\documentclass[aps,nofootinbib,showkeys,noshowpacs,preprintnumbers,amsmath,amssymb]{revtex4}
\pdfoutput=1

\def\be{\begin{equation}}
\def\ee{\end{equation}}
\def\ba{\begin{eqnarray}}
\def\ea{\end{eqnarray}}
\def\bea{\begin{eqnarray}}
\def\eea{\end{eqnarray}}
\def\bes{\begin{subequations}}
\def\ees{\end{subequations}}
\def\btd{{\widetilde d}}
\newcommand{\tk}{{\widetilde k}}
\newcommand{\tK}{{\widetilde K}}
\newcommand{\tD}{{\widetilde {\cal D}}}
\newcommand{\A}{{\mathcal{A}}}
\newcommand{\tA}{{\widetilde {\mathcal{A}}}}
\newcommand{\ta}{{\widetilde a}}
\newcommand{\tal}{{\widetilde \alpha}}

\newcommand{\td}{{\widetilde d}}

\newcommand{\MSbar}{\overline{\rm MS}}  
\newcommand{\cM}{{\cal M}}
\usepackage{graphics}
\usepackage{graphicx}
\usepackage{dcolumn}
\usepackage{bm}
\usepackage{epsfig}
\usepackage{graphicx,color}
\usepackage{multirow}

\begin{document}


\title{Lattice-motivated QCD coupling and hadronic contribution to muon $g-2$}

\author{Gorazd Cveti\v{c}$^1$}
\author{Reinhart K\"ogerler$^2$}

\affiliation{$^1$Department of Physics, Universidad T{\'e}cnica Federico Santa Mar{\'\i}a, Casilla 110-V, Valpara{\'\i}so, Chile\\ 
  $^2$Department of Physics, Universit\"at Bielefeld, 33501 Bielefeld, Germany}

\date{\today}

\begin{abstract}
  We present an updated version of a QCD coupling which fulfills various physically motivated conditions: at high momenta it practically coincides with the perturbative QCD (pQCD) coupling; at intermediate momenta it reproduces correctly the physics of the semihadronic tau decay; and at very low momenta it is suppressed as suggested by large-volume lattice calculations. An earlier presented analysis is updated here in the sense that the Adler function, in the regime $|Q^2| \lesssim 1 \ {\rm GeV}^2$, is evaluated by a renormalon-motivated resummation method. This Adler function is then used here in the evaluation of the quantities related with the semihadronic (strangeless) $\tau$-decay spectral functions, including Borel-Laplace sum rules in the (V+A)-channel. The analysis is then extended to the evaluation of the hadronic vacuum polarization contribution to the muon anomalous magnetic moment, $a_{\mu}^{\rm had(1)}$, where we include in the Adler function the V-channel higher-twist OPE terms which are regulated in the infrared (IR) by mass parameters which are expected to be $\lesssim 1$ GeV. The correct value of $a_{\mu}^{\rm had(1)}$ can be reproduced with the mentioned IR-regulating mass parameters if the value of the condensate $\langle O_4 \rangle_{\rm V+A}$ is positive (and thus the gluon condensate value is positive). This restriction and the requirement of the acceptable quality of the fits to the various mentioned sum rules then lead us to the restriction $0.1171 < \alpha_s(M_Z^2;\MSbar) < 0.1180$.
 \end{abstract}

\maketitle

\section{Introduction}
\label{sec:intr}

In this work, we present an updated construction of a QCD coupling $\A(Q^2)$ [the analog of the underlying pQCD coupling $a(Q^2) \equiv \alpha_s(Q^2)/\pi$] which was developed in Ref.~\cite{3dAQCD}. The construction is based on a specific parametrization of the behaviour of the discontinuity (spectral) function $\rho_{\A}(\sigma) \equiv {\rm Im} \ \A(-\sigma-i \epsilon)$ in the low-$\sigma$ regime (where deviations from pQCD are expected) in terms of three Dirac-delta functions. The parameters of the model are determined by a set of three groups of conditions: (I) at high momenta ($|Q^2| > 1 \ {\rm GeV}^2$) the coupling $\A(Q^2)$ practically coincides with the underlying pQCD coupling $a(Q^2)$; (II) at intermediate momenta ($|Q^2| \sim 1 \ {\rm GeV}^2$) the coupling reproduces the correct values of the semihadronic $\tau$-lepton decay ratio $r_{\tau} \approx 0.20$; (III) at very low momenta ($|Q^2| < 1 \ {\rm GeV}^2$) the coupling is suppressed, $\A(Q^2) \sim Q^2$ when $Q^2 \to 0$, as suggested by a natural interpretation of the large-volume lattice results. The obtained coupling $\A(Q^2)$ turns out to have no Landau singularities, i.e., it has singularities only along the negative semiaxis in the complex $Q^2$-plane (we denote: $q^2 \equiv (q^0)^2 - {\vec q}^2 \equiv - Q^2$).\footnote{Various QCD couplings free of Landau singularities (holomorphic couplings) have been constructed in the literature. Well known among them is the Minimal Analytic coupling [usually named: Analytic Perturbation Theory (APT)], cf.~Refs.~\cite{APT,KS,BMS}; for reviews and applications of APT cf.~Refs.~\cite{revAPT} and Refs.~\cite{APTappl1,APTappl1b,APTappl2,APTappl3}, respectively. For other forms of holomorphic couplings, cf.~Refs.~\cite{APTappl1b,Nest2,Webber,Boucaud,Alekseev,mes2,CV12,1danQCD,2danQCD,anOPE,anOPE2,Brod,Brod2,ArbZaits,Shirkovmass,KKS,Luna1,Luna2,Nest1,Pelaez,Siringo,NestBook}. Some of them \cite{Boucaud,mes2,ArbZaits,Luna2,Siringo} fulfill the relation $\A(Q^2)=0$ at $Q^2=0$, i.e., the same relation that is fulfilled by the coupling $\A(Q^2)$ considered in this work. For reviews of a variety of holomorphic couplings, cf.~Refs.~\cite{GCrev,Brodrev}. For some of such couplings, mathematical packages for their evaluation also exist \cite{BK,ACprogr}. Most of such holomorphic couplings are constructed with dispersive approach (as is the case also for the coupling considered in this work). On the other hand, related dispersive approaches can be applied also directly to spacelike QCD observables \cite{MSS,MagrGl,mes2,DeRafael,MagrTau,Nest3a,Nest3b,NestBook}.}
In comparison with the construction in Ref.~\cite{3dAQCD}, the leading-twist (i.e., $D=0$) Adler function ${\cal D}(Q^2)_{(D=0)}$ is evaluated, at $|Q^2| \lesssim 1 \ {\rm GeV}^2$, with a renormalon-motivated resummation method of Ref.~\cite{renmod}\footnote{In Ref.~\cite{3dAQCD}, the basis of evaluation of ${\cal D}(Q^2)_{(D=0)}$ was the knowledge of the first four terms in the (truncated) perturbation expansion of ${\cal D}(Q^2)_{(D=0)}$ and the  absence of Landau singularities of $\A(Q^2)$.} which takes into account in an adequate manner the main renormalon structure of ${\cal D}(Q^2)_{(D=0)}$ and the absence of Landau singularities of $\A(Q^2)$. The Borel-Laplace sum rules of the $\tau$-lepton spectral function in the full (V+A) channel are also performed here to obtain the values of the first condensates, as in Ref.~\cite{3dAQCD}, but ${\cal D}(Q^2)_{(D=0)}$ in these sum rules is evaluated with the mentioned renormalon-motivated resummation. The obtained values of the $D=4, 6$ condensates in the full (V+A) channel then enable us to obtain the V-channel values of the corresponding condensates. This then permits us to construct an OPE for the V-channel Adler function ${\cal D}(Q^2)_{\rm V}$, where the $D=4, 6$ terms include the obtained consensates and are regulated in the IR-regime ($|Q^2| < 1 \ {\rm GeV}^2$) with regulator masses ${\cal M}_{D}$. The obtained ${\cal D}(Q^2)_{\rm V}$ is then used for the evaluation of the hadronic vacuum polarization contribution to the muon anomalous magnetic moment, $a_{\mu}^{\rm had(1)}$. The obtained value of $a_{\mu}^{\rm had(1)}$ is heavily influenced by the behaviour of the coupling $\A(Q^2)$ and  ${\cal D}(Q^2)_{(D=0)}$ at very low momenta $|Q^2| \sim m^2_{\mu}$ ($\lesssim 10^{-2} \ {\rm GeV}^2$), and by the IR-regulating masses ${\cal M}_{D}$ ($D=4,6$). So an interesting question arises in this ($\A$)QCD framework: does the condition of reproduction of the correct value of $a_{\mu}^{\rm had(1)}$ give us positive squared masses  ${\cal M}_{D}^2>0$; and if so, are the obtained values $\lesssim 1 \ {\rm GeV}^2$, i.e., typical nonperturbative QCD scales, as expected physically? We point out that ${\cal M}_{D}^2>0$ implies that the OPE terms with $D>0$ ($D=4, 6$) of ${\cal D}(Q^2)_{\rm V}$ are free of Landau singularities (and not just the $D=0$ term).

In Sec.~\ref{sec:constr} we resume the construction of the considered QCD coupling $\A(Q^2)$. In Sec.~\ref{sec:renmod} we recapitulate the renormalon-motivated resummation method and how it is applied to the leading-twist Adler function ${\cal D}(Q^2)_{(D=0)}$; additional details are given in Appendix \ref{app:renmod}. In Sec.~\ref{sec:Ares} we present the numerical results for the parameters of the coupling $\A(Q^2)$, for various QCD reference strength values $\alpha_s(M_Z^2;\MSbar)$ and various values of $D=0$ $\tau$-lepton decay ratio $r_{\tau}^{(D=0)}$ ($r_{\tau}^{(D=0)} \approx 0.20$). In Sec.~\ref{sec:BSR} we present the analysis of the Borel-Laplace sum rules with the OPAL and ALEPH data, and extract the values of the $D=4, 6$ condsensates $\langle \mathcal{O}_D \rangle_{{\rm V+A}}$ of the (V+A)-channel Adler function. We also perform the $r_{\tau}$-consistency checks by using OPAL and ALEPH data. In Sec.~\ref{sec:amu} we deduct from the obtained values of $\langle \mathcal{O}_D \rangle_{{\rm V+A}}$ the values of the V-channel condensates $\langle \mathcal{O}_D \rangle_{{\rm V}}$. Then we finally apply there all the obtained results for the evaluation of the hadronic vacuum polarization contribution to the muon anomalous magnetic moment, $a_{\mu}^{\rm had(1)}$, and extract the values of the IR-regulator squared masses ${\cal M}^2_{D}$ ($D=4,6$). We also interpret all the results obtained in this work. In Sec.~\ref{sec:summ} we present a summary of the presented work. \textcolor{black}{A short version with the main features of this work was published in Ref.~\cite{shortv}.}

\section{Construction of the coupling $\A(Q^2)$}
\label{sec:constr}

As mentioned, we will follow mainly Ref.~\cite{3dAQCD} in the construction of the QCD coupling $\A(Q^2)$, which is the analog of the usual (underlying) pQCD coupling $a(Q^2) \equiv \alpha_s(Q^2)/\pi$ in the same renormalization scheme. The coupling $\A(Q^2)$ should be valid also for the low-momentum regime where the number of active flavours is taken to be $N_f=3$.

The following dispersion integral form  for the (underlying) pQCD coupling $a(Q^2)$ is obtained by application of the Cauchy theorem to the integrand  $a(Q^{'2})/(Q^{'2} - Q^2)$:
\be a (Q^2) = \frac{1}{\pi} \int_{\sigma= - {Q^2_{\rm br}} - \eta}^{\infty}
\frac{d \sigma {\rho}_a (\sigma) }{(\sigma + Q^2)}
   \qquad (\eta \to +0),
\label{adisp} \ee
where $\rho_a (\sigma) \equiv {\rm Im} \ a (Q^{'2}=-\sigma - i \epsilon)$ is the discontinuity (spectral) function of $a(Q^{'2})$ along its cut. The integration in Eq.~(\ref{adisp}) is applied along the entire cut of $a(Q^{'2})$ in the complex $Q^{'2}$-plane: $-\infty < Q^{'2}<Q^2_{\rm br}$, where $Q^2_{\rm br} >0$ is the branching point. We note that $0 < Q^{'2}<Q^2_{\rm br}$ is the Landau cut of the pQCD coupling  $a(Q^{'2})$; usually we have $Q^2_{\rm br} \sim 0.1$-$1 \ {\rm GeV}^2$.

On the other hand, the coupling $\A(Q^2)$ is defined by the analogous form of the dispersion integral
\be
\A(Q^2) = \frac{1}{\pi} \int_{\sigma=M^2_{\rm thr}}^{\infty} \frac{d \sigma \rho_{\A}(\sigma)}{(\sigma + Q^2)} .
\label{Adisp1} \ee
The corresponding spectral function $\rho_{\A}(\sigma) \equiv {\rm Im} \; \A(Q^2=-\sigma - i \varepsilon)$ is specified in the following way: In the high energy regime $ \rho_{\A}(\sigma)$ is considered to coincide with that of the underlying pQCD coupling (in the same renormalization scheme)
\be
\rho_{\A}(\sigma) =\rho_a (\sigma) \qquad {\rm for} \; \sigma \geq M_0^2.
\label{pQCDonset} \ee
Here, $M_0^2 \sim 1$-$10 \ {\rm GeV}^2$ can be regarded as pQCD onset-scale. On the other hand, the branching (threshold) point $Q^{'2}=-M^2_{\rm thr}$ of the cut of $\A(Q^{'2})$ may be different from $Q^2_{\rm br}$; it is expected to be comparable to the light meson scales $\sim 0.01 \ {\rm GeV}^2$, and a hope is that $-M^2_{\rm thr}$ turns out to be negative in the construction of $\A(Q^{2})$ [if $-M^2_{\rm thr}$ is negative, then $\A(Q^2)$ has no Landau singularities]. In the low-energy regime ($\sigma < M_0^2$) the spectral function is unknown, it is expected to differ from the pQCD expression $\rho_a(\sigma)$ and would contain an IR-contribution
\be
\Delta \A_{\rm IR}(Q^2) = \frac{1}{\pi} \int_{\sigma=M^2_{\rm thr}}^{M_0^2} \frac{d \sigma \rho_{\A}(\sigma)}{(\sigma + Q^2)} ,
\label{AIRdisp} \ee
which is a priori unknown. We parametrize this quantity as a nearly-diagonal Pad\'e  $[M-1/M](Q^2)$,\footnote{These approximants work very well \cite{Peris} for spacelike QCD quantities ${\cal D}(Q^2)$ such as current correlators.} and we take $M=3$
\be
\Delta \A_{\rm IR}(Q^2) = \frac{\sum_{n=1}^{2} A_n Q^{2n}}{\sum_{n=1}^3 B_n Q^{2n}}
=  \sum_{j=1}^{3} \frac{{\cal F}_j}{Q^2 + M_j^2}.
\label{PFM1M}
\ee
Combining these terms, the considered coupling $\A(Q^2)$ has the form
\be
\A(Q^2)= \sum_{j=1}^{3} \frac{{\cal F}_j}{Q^2 + M_j^2} + \frac{1}{\pi} \int_{\sigma=M_0^2}^{\infty} \frac{d \sigma \rho_{a}(\sigma)}{(\sigma + Q^2)},
\label{Adisp2} \ee
and the corresponding spectral function, in vew of Eqs.~(\ref{Adisp1}) and (\ref{Adisp2}), can be written as
\be
\rho_{\A}(\sigma) = \pi \sum_{j=1}^{3} {\cal F}_j \delta(\sigma - M_j^2) + \Theta(\sigma - M_0^2) \rho_a(\sigma).
\label{rhoA} \ee
When we consider the underlying pQCD coupling $a(Q^2)$ as already chosen (determined), our ansatz for the new coupling $\A(Q^2)$ has seven parameters: $(M_j^2, {\cal F}_j)$ ($j=1,2,3$) and the pQCD-onset scale $M_0^2$. It is expected that we will have $0 < M_1^2 < M_2^2 < M_3^2 < M_0^2$.

Due to its construction, however, the coupling $\A(Q^2)$ will differ from the underlying ($N_f=3$) pQCD coupling $a(Q^2)$ by nonperturbative (NP) contributions. The coupling will fulfill several physically-motivated conditions. Since pQCD is known to give correct results for QCD quantities ${\cal D}(Q^2)$ at large $|Q^2| > 1 \ {\rm GeV}^2$, we will require that the coupling $\A(Q^2)$ agree to a large precision with the (underlying) pQCD coupling $a(Q^2)$ in that regime
\be
\A(Q^2) - a(Q^2) \sim \left( \frac{\Lambda^2_{\rm QCD}}{Q^2} \right)^{\cal N}
\qquad (|Q^2| > 1 \ {\rm GeV}^2, \; {\cal N}=5).
\label{diffAa}
\ee
Here, $\Lambda^2_{\rm QCD} \sim 0.1 \ {\rm GeV}^2$, and ${\cal N}$ is a (large) positive integer. The larger is ${\cal N}$, the better is the agreement between $\A$ and $a$ at large $|Q^2| > 1 \ {\rm GeV}^2$. We will take ${\cal N}=5$; it turns out that this choice represents a set of four conditions (by default we would have: ${\cal N}=1$). We refer to Ref.~\cite{3dAQCD} for more details on this aspect and for the explicit form of these conditions.

Another condition comes from the requirement that the leading-twist ($D=0$) part of the $\tau$-lepton semihadronic decay width gives the value (approximately) known by the experiments $r_{\tau}^{(D=0)} \approx 0.20$, where the theoretical expression for this quantity is
\be
r^{(D=0)}_{\tau, {\rm th}} = \frac{1}{2 \pi} \int_{-\pi}^{+ \pi}
d \phi \ (1 + e^{i \phi})^3 (1 - e^{i \phi}) \
d(Q^2=m_{\tau}^2 e^{i \phi})_{(D=0)} \; \left( \approx 0.20 \right).
\label{rtaucont}
\ee
Here, $d(Q^2)_{(D=0)} = {\cal D}(Q^2)_{(D=0)} - 1$ is the massless Adler function whose perturbation expansion is $d_{(D=0)} = a + {\cal O}(a^2)$. This condition can be regarded as originating from the intermediate (i.e., moderately low) momentum regime, $|Q^2| \sim m_{\tau}^2 \sim 1 \ {\rm GeV}^2$.

In addition to these hitherto five conditions, we impose two conditions coming from the deep infrared (IR) regime $|Q^2| \lesssim 0.1 \ {\rm GeV}^2$ and are motivated by the large-volume lattice calculations \cite{LattcoupNf02a,LattcoupNf02b,Lattcoupb,Lattcoupc}: as a function of positive $Q^2$ the coupling $\A(Q^2)$ has a maximum at
\be
\A(Q^2) = \A_{\rm max} \quad {\rm for} \; Q^2 \approx 0.135 \ {\rm GeV}^2,
\label{Amax} \ee
and at $Q^2 \to 0$ it is suppressed by $\sim Q^2$ behaviour
\be
\A(Q^2) \sim Q^2 \qquad {\rm for} \; Q^2 \to 0.
\label{A00} \ee
We recall that within QCD the running coupling $a(Q^2)$ can be generally related to the renormalization functions via
\be
a (Q^2)  =  a (\Lambda^2) \frac{Z_{\rm gl}^{(\Lambda)}(Q^2) Z_{\rm gh}^{(\Lambda)}(Q^2)^2}{Z_1 ^{(\Lambda)}(Q^2)^2} \qquad (\Lambda^2_{\rm QCD} < |Q^2| < \Lambda^2),
\label{alatt}
\ee  
where $Z_1$ is the dressing function of the gluon-ghost-ghost vertex, which in the Langau gauge is constant $Z_1^{(\Lambda)}(Q^2)=1$ to all orders \cite{Taylor}. Therefore, within a lattice approach to QCD a QCD-lattice coupling can be defined \cite{LattcoupNf02a,LattcoupNf02b,Lattcoupb,Lattcoupc}
\be
\A_{\rm latt}(Q^2)  \equiv  \A_{\rm latt}(\Lambda^2) Z_{\rm gl}^{(\Lambda)}(Q^2) Z_{\rm gh}^{(\Lambda)}(Q^2)^2 \qquad (0 < |Q^2| < \Lambda^2).
\label{Alatt}
\ee
Within the mentioned large-volume  lattice calculations \cite{LattcoupNf02a,LattcoupNf02b,Lattcoupb,Lattcoupc},\footnote{Somewhat different but qualitatively similar behaviour [$\A_{\rm latt}(Q^2) \to 0$ when $Q^2 \to 0$] is obtained when defining a (lattice) coupling which involves the lattice-calculated 3-gluon Green function \cite{Latt3gluon}.} the Landau gauge dressing functions $Z_{\rm gl}^{(\Lambda)}(Q^2)$ and $Z_{\rm gh}^{(\Lambda)}(Q^2)^2$ were calculated for low positive $Q^2 < 1 \ {\rm GeV}^2$, in  the lattice MiniMOM (MM) renormalization scheme \cite{MiniMOM,BoucaudMM,CheRet}.\footnote{Interestingly, it was shown in Ref.~\cite{AKGCR} that the $\beta$-function factorisation of the conformal symmetry breaking contribution to the generalised Crewther relation is preserved in the MM scheme in the Landau gauge.} It turns out that the product (\ref{Alatt}) has the property Eq.~(\ref{A00}), \textcolor{black}{i.e., $\A_{\rm latt}(Q^2) \sim Q^2$ when $Q^2 \to 0$. This is so because the  mentioned large-volume lattice results indicate that in the deep IR regime the Landau ghost propagator behaves as a tree-level propagator $\sim 1/Q^2$ [thus the dressing function $Z_{\rm gh}^{(\Lambda)}(Q^2) \to {\rm const}$ when $Q^2 \to 0$] and the Landau gauge gluon propagator goes to a nonvanishing constant [thus the dressing function $Z_{\rm gl}^{(\Lambda)}(Q^2) \sim Q^2$ there]. Similar results are obtained for the mentioned dressing functions with the decoupling solution of the Dyson-Schwinger equation (DSE) approach \cite{DSEdecoup}.
Furthermore, the mentioned lattice results \cite{LattcoupNf02a,LattcoupNf02b} show that the product (\ref{Alatt})} achieves the maximum as written in Eq.~(\ref{Amax}) once the momenta in the lattice MM scheme are rescaled to the usual $\MSbar$-like scale, $Q^2 = Q^2_{\rm latt} (\Lambda^2_{\MSbar}/\Lambda_{\rm MM})^2$. Stated otherwise, the conditions (\ref{Amax})-(\ref{A00}) mean that our constructed coupling $\A(Q^2)$ qualitatively agrees with the lattice coupling (\ref{Alatt}) in the (deep) IR regime $Q^2 < 1 \ {\rm GeV}^2$.  For further explanation of the relations (\ref{Amax})-(\ref{A00}) and (\ref{alatt})-(\ref{Alatt}), we refer to Ref.~\cite{3dAQCD}.

The seven aforementioned requirements can be used to determine the seven parameters of the coupling $\A(Q^2)$. There is also an eighth, more hidden, parameter which determines the strength of the underlying pQCD coupling $a(Q^2)$; in our approach, this strength parameter will be represented by the value of $\alpha_s(M_Z^2;\MSbar)$ (at $N_f=5$); the recent world average \cite{PDG2019} is  $\alpha_s(M_Z^2;\MSbar)=0.1179 \pm 0.0010$. The underlying ($N_f=3$) pQCD coupling $a(Q^2)$ is in the same MM renormalization scheme in which the large-volume lattice calculations were performed \cite{LattcoupNf02a,LattcoupNf02b,Lattcoupb,Lattcoupc}, with the 3-loop and 4-loop beta coefficients $\beta_j$ ($j=2,3$) coinciding with the MM scheme coefficients \cite{MiniMOM} (cf.~also \cite{BoucaudMM,CheRet})\footnote{The MM scheme in pQCD sense is known up to 4-loop level.}
\be
\left( \frac{d a(Q^2)}{d \ln Q^2} = \right) \beta_{\rm MM}(a) = - \beta_0 a^2 - \beta_1 a^3 - \beta_2({\rm MM}) a^4 - \beta_3({\rm MM}) a^5 + {\cal O}(a^5),
\label{betaMM} \ee
where the first two coefficients  $\beta_0=9/4$ and $\beta_1 = 4$ are universal, and $\beta_2({\rm MM})=20.9183$ and  $\beta_3({\rm MM})=160.771$ (all for $N_f=3$).\footnote{For comparison, the ($N_f=3$) $\MSbar$ values are $\beta_2(\MSbar)=10.0599$ and $\beta_3(\MSbar)=47.2281$.}
We point out, however, that the renormalization scheme of the considered underlying coupling $a(Q^2)$, and of our coupling $\A(Q^2)$, has the squared momenta rescaled as earlier mentioned, from the lattice MM scale to the usual $\MSbar$-like scale: $Q^2 = Q^2_{\rm latt.} ({\Lambda}_{\MSbar}/{\Lambda_{\rm MM}})^2$ [at $N_f=0$ this means: $Q^2 \approx Q^2_{\rm latt.}/1.90^2$].\footnote{The maximum value of $\A_{\rm latt}(Q^2)$ in the MM scheme for $N_f=0$ \cite{LattcoupNf02a} and $N_f=2$  \cite{LattcoupNf02b} is at about $Q^2_{\rm latt} \approx 0.45 \ {\rm GeV}^2$, and this then corresponds in the $\MSbar$-like rescaling (LMM) to $Q^2 \approx 0.45/1.90^2 \ {\rm GeV}^2 \approx 0.125 \ {\rm GeV}^2$ and $0.45/1.85^2 \ {\rm GeV}^2 \approx 0.131 \ {\rm GeV}^2$, respectively. In our construction of $\A(Q^2)$, we have $N_f=3$ and we take the maximum at $Q^2 = 0.135 \ {\rm GeV}^2$, Eq.~(\ref{Amax}).}
We call this rescaled scheme the Lambert MiniMOM (LMM) scheme, and it differs from the $\MSbar$ scheme only due to the changed $\beta_j$ coefficients ($j \geq 2$). We recall that the construction of the coupling $\A(Q^2)$ is based on the dispersion integral (\ref{Adisp2}) with the discontinuity (spectral) function $\rho_{\A}(\sigma)$ equal at large energy scales to the spectral function of the underlying pQCD coupling $\rho_a(\sigma)$, Eqs.~(\ref{pQCDonset}) and (\ref{rhoA}). In practice, we need for an efficient evaluation of the dispersion integral (\ref{Adisp2}) an efficient and precise evaluation of the integrand function $\rho_a (\sigma) \equiv {\rm Im} \ a (Q^{'2}=-\sigma - i \epsilon)$, i.e., of $a (Q^{'2})$ in the complex $Q^{'2}$-plane and even close to the cut. This is obtained if we take for the pQCD $\beta$-function a specific form of Pad\'e which, on one hand, allows for an explicit solution of the RGE $d a/d \ln Q^2 = \beta(a)$ \cite{GCIK} and, on the other hand, when expanded it agrees with the MM-$\beta$-function Eq.~(\ref{betaMM}) up to the known terms $\sim a^5$
\be
\frac{d a(Q^2)}{d \ln Q^2}  =
\beta(a(Q^2))
\equiv
- \beta_0 a(Q^2)^2 \frac{ \left[ 1 + a_0 c_1 a(Q^2) + a_1 c_1^2 a(Q^2)^2 \right]}{\left[ 1 - a_1 c_1^2 a(Q^2)^2 \right] \left[ 1 + (a_0-1) c_1 a(Q^2) + a_1 c_1^2 a(Q^2)^2 \right]} \ ,
\label{beta} \ee
where $c_j \equiv \beta_j/\beta_0$ and
\be
a_0  =  1 + \sqrt{c_3({\rm MM})/c_1^3}, \quad 
a_1 = c_2({\rm MM})/c_1^2 +   \sqrt{c_3({\rm MM})/c_1^3} .
\label{a0a1} \ee
Here, $c_j({\rm MM}) \equiv \beta_j({\rm MM})/\beta_0$ ($j=2,3$) are taken in the MM scheme (with $N_f=3$). It turns out that the expansion of the $\beta$-function of the RGE (\ref{beta}) up to $\sim a(Q^2)^5$ reproduces the four-loop polynomial MM-scheme $\beta$-function (\ref{betaMM}). As shown in Ref.~\cite{GCIK}, the RGE (\ref{beta}) has explicit solution in terms of the Lambert functions $W_{\mp 1}(z)$
\be
a(Q^2) = \frac{2}{c_1}
\left[ - \sqrt{\omega_2} - 1 - W_{\mp 1}(z) + 
\sqrt{(\sqrt{\omega_2} + 1 + W_{\mp 1}(z))^2 
- 4(\omega_1 + \sqrt{\omega_2})} \right]^{-1},
\label{a4l} \ee
where $\omega_1= c_2({\rm MM})/c_1^2$, $\omega_2=c_3({\rm MM})/c_1^3$,  $Q^2 = |Q^2| \exp(i \phi)$. The Lambert function $W_{-1}$ is used when $0 \leq \phi < \pi$, and $W_{+1}$ when $-\pi \leq \phi < 0$. The argument $z = z(Q^2)$ in $W_{\pm 1}(z)$ is
\be
z \equiv z(Q^2) =
-\frac{1}{c_1 e}
\left( \frac{\Lambda_L^2}{Q^2} \right)^{\beta_0/c_1} \ ,
\label{zexpr} \ee
where $\Lambda_L$ is a scale which we call the Lambert scale ($\Lambda_L \sim \Lambda_{\rm QCD}$).\footnote{In Ref.~\cite{GCIK} for $z$ the expression without the factor $1/( c_1 e)$ was used, which just redefines the Lambert scale $\Lambda_L$. We point out that the expression Eq.~(\ref{a4l}) can be used as an explicit solution for any chosen values of $c_2$ and $c_3$ \cite{GCIK}, not just those of the MM scheme.}
The form of Eq.~(\ref{a4l}) is very convenient for the evaluation of $a(Q^2)$, and thus of $\rho_a(\sigma) \equiv {\rm Im} \ a (Q^{'2}=-\sigma - i \epsilon)$, in Mathematica software which has the evaluation of the Lambert functions $W_{\mp}(z)$ implemented in a stable manner.\footnote{\textcolor{black}{We wish to point out that when $\beta$-function $\beta(a)$ has a general Pad\'e form, an explicit solution $a=a(Q^2)$ in general does not exist; it exists only for some specific sets of Pad\'e forms [such as Eq.~(\ref{beta})] as pointed out in Ref.~\cite{GCIK}.}}

The Lambert scale $\Lambda_L$ can be fixed once we have chosen the reference value for the strength parameter $\alpha_s(M_Z^2;\MSbar)$. The algorithm relating the values of $\Lambda_L$ (at $N_f=3$, in the LMM scheme) and $\alpha_s(M_Z^2;\MSbar)$ (at $N_f=5$, in the $\MSbar$ scheme) is described in Ref.~\cite{3dAQCD}, and we refer to it for details. For example, the value of $\Lambda_L=0.11100$ GeV (in LMM) corresponds to $\alpha_s(M_Z^2;\MSbar)=0.1177$. We point out that we use in the present work for the RGE running of pQCD coupling $a(Q^2;\MSbar)$ in the $\MSbar$ scheme (from $M_Z^2$ down to $Q^2 \sim 1 \ {\rm GeV}^2$) everywhere the five-loop $\MSbar$ $\beta$-function \cite{5lMSbarbeta} and the corresponding four-loop quark threshold matching \cite{4lquarkthresh} at the scales $Q^2_{\rm thr} = \kappa {\bar m}_q^2$ with $\kappa=2$, and ${\bar m}_q \equiv {\bar m}_q( {\bar m}_q^2)$ equal to $4.2$ GeV ($q=b$) and $1.27$ GeV ($q=c$). In Ref.~\cite{3dAQCD}, the use of four-loop $\MSbar$ $\beta$-function \cite{4lMSbarbeta} with the corresponding three-loop  quark threshold matching \cite{3lquarkthresh} was made (with $\kappa=2$).\footnote{
For this reason, for example, for $\alpha_s(M_Z^2;\MSbar)=0.1181$ the value of $\Lambda_L=0.1136$ GeV was obtained in \cite{3dAQCD}, while here  $\Lambda_L=0.1130$ GeV.}

\section{Renormalon-motivated resummation}
\label{sec:renmod}

The evaluation of the leading-twist ($D=0$) contribution to $r_{\tau}$ in $\A$QCD is reduced to the evaluation of the leading-twist ($D=0$) contribution of the massless Adler function, $d(Q^2)_{(D=0)}$, in $\A$QCD along the contour $|Q^2|=m^2_{\tau}$ in the $Q^2$-complex plane, according to Eq.~(\ref{rtaucont}).\footnote{The Adler function $d(Q^2)_{(D=0)}$ is the logarithmic derivative of the quark current correlator $\Pi_{(D=0)}$: $d(Q^2)_{(D=0)} = - 2 \pi^2 (d/d \ln Q^2) \Pi(Q^2)_{(D=0)} - 1$.}
The perturbation expansion of $d(Q^2)_{(D=0)}$ is known up to $\sim a^4$ \cite{d1,d2,d3}\footnote{In the $\MSbar$ scheme (with $N_f=3$), the values of $(d_1,d_2,d_3)$ are:  ($1.63982$, $6.37101$, $49.0757$). On the other hand, in the LMM scheme (with $N_f=3$), the corresponding values are significantly different: ($1.63982$, $1.54508$, $8.01658$).}
\bes
\label{dD0TPS}
\bea
d(Q^2)_{(D=0)}^{{\rm pt};[4]} & = & a(Q^2) + \sum_{n=1}^{3} d_n a(Q^2)^{n+1}
\label{dD0pt}
\\
& = & a(Q^2) + \sum_{n=1}^{3} \td_n \ta_{n+1}(Q^2),
\label{dD0lpt}
\eea \ees
where in the last relation the power series was reorganized in terms of the logarithmic derivatives
\be
\ta_{n+1}(Q^2) \equiv \frac{(-1)^n}{n! \beta_0^n} \left( \frac{d}{d \ln Q^2} \right)^n a(Q^2) \qquad (n=1,2,\ldots).
\label{tan} \ee
The connections between $a^{n+1}$ and $\ta_{n+1}$ are obtained via the RGE (\ref{betaMM}) (in a general scheme, not just in LMM), and have the form
\bes
\label{tananrel}
\bea
\ta_{n+1} &=& a^{n+1} + \sum_{m \geq 1} k_m(n+1) \; a^{n+1+m},
\label{tanan}
\\
a^{n+1} & = & \ta_{n+1} + \sum_{m \geq 1} \tk_m(n+1) \; \ta_{n+1+m},
\label{antan} \eea \ees
with appropriate constants $k_m(n+1)$ and $\tk_m(n+1)$. This leads to the relations  of the following form between the coefficients $\td_n$ and the original coefficients $d_n$:
\be
\td_n = d_n + \sum_{s=1}^{n-1} \tk_s(n+1-s) \; d_{n-s}.
\label{tdndn} \ee
All the coefficients $k_m(p)$ and $\tk_m(p)$ are independent of the (physical) scale $Q^2$, but do depend on the scheme-characterizing beta coefficients $\beta_j$ ($j \geq 2$). Specifically,
\bes
\label{tdns}
\bea
\td_1 &=& d_1, \qquad \td_2 = d_2 - \left( \frac{\beta_1}{\beta_0} \right) d_1,
\label{td1td2} \\
\td_3 & = & d_3 - \frac{5}{2} \left( \frac{\beta_1}{\beta_0} \right) d_2 + \left[ -\left( \frac{\beta_2}{\beta_0} \right) +  \frac{5}{2} \left( \frac{\beta_1}{\beta_0} \right)^2 \right] d_1,
\label{td3} \eea \ees
etc. In Ref.~\cite{3dAQCD}, we had evaluated the truncated series (\ref{dD0lpt})  by the necessary $\A$QCD replacements
\be
a \mapsto \A, \quad
\ta_{n+1} \mapsto \tA_{n+1}  \equiv \frac{(-1)^n}{n! \beta_0^n} \left( \frac{d}{d \ln Q^2} \right)^n \A(Q^2) \qquad (n=1,2,\ldots),
\label{tatA}
\ee
leading to
\be
d(Q^2)_{(D=0)}^{\A{\rm QCD};[4]}  =  \A(Q^2) + \sum_{n=1}^{3} \td_n \; \tA_{n+1}(Q^2).
\label{dD0TPSA}
\ee
In the present work, however, we will extend the evaluation of the Adler function $d(Q^2)_{(D=0)}$ to include a renormalon-motivated estimate of the infinite sequence of the higher-order coefficients $\td_n$ ($\leftrightarrow d_n$) with $n=4,5,\ldots$ as explained in Ref.~\cite{renmod}.\footnote{\textcolor{black}{For a review of renormalon physics cf.~Ref.~\cite{BenekeRev}; and for some other recent developments, cf.~Refs.~\cite{Rens1,Rens2,Rens3,Rens4}.}}
In doing so, we are motivated by the observation that the precise connection between the (analytically improved) coupling $\A$ and the experimental value for the (moderately-low-energy) quantity $r_{\tau}$ is crucial for the optimal construction of $\A$. The resummed Adler function can
be reexpressed in terms of integrals for $d(Q^2)_{(D=0)}$ involving the coupling $a$ and the characteristic functions $G_d^{(\pm)}$ and $G_d^{\rm (SL)}$
\be
d(Q^2)_{(D=0)}^{\rm pt; res} =  \int_0^1 \frac{dt}{t} G_d^{(-)}(t) a(t e^{-\tK} Q^2) + \int_1^\infty \frac{dt}{t} G_d^{(+)}(t) a(t e^{-\tK} Q^2) + \int_0^{1} \frac{dt}{t} G^{\rm (SL)}_d(t) \left[ a(t e^{-\tK} Q^2) - a(e^{-\tK} Q^2) \right] ,
\label{drespQCD}
\ee
where, in the LMM scheme, the obtained characteristic functions have the form
\bes
\label{Gds}
\bea
G_d^{(-)}(t) & = &  \pi t^2 \left[{\td}_{2,1}^{\rm IR} - {\td}_{3,2}^{\rm IR} t \ln t  \right], 
\label{Gdmi}
\\
G_d^{(+)}(t) &=&  \frac{\pi}{t} { \td}_{1,2}^{\rm UV} \ln t,
\label{Gdpl}
\\
G_d^{\rm (SL)}(t) &=&  - { \tal} { \td}_{2,1}^{\rm IR} \frac{\pi t^2}{\ln t},
\label{GdSL}
\eea
\ees
and the obtained numerical values of the rescaling parameter $\tK$ and of the residue-like parameters ${\td}_{i,j}$ are (in LMM): ${ \td}_{2,1}^{\rm IR}=-1.831$,  ${ \td}_{3,2}^{\rm IR}=11.05$, ${ \td}_{1,2}^{\rm UV}=0.005885$, ${ \tal}=-0.14$; ${ \tK}=-0.7704$. These parameters appear in the Borel-Laplace transform ${\rm B}[\btd](u)$ of an auxiliary quantity $\btd(Q^2; \mu^2)$ related to the Adler function $d(Q^2)_{(D=0)}$ (in the one-loop approximation these two quantities would coincide). The resummed expression for the Adler function in the considered $\A$QCD framework is obtained from the expression (\ref{drespQCD}) by the simply replacing the (underlying) pQCD coupling $a$ there by $\A$
\be
d(Q^2)_{(D=0)}^{\A;{\rm res}} =  \int_0^1 \frac{dt}{t} G_d^{(-)}(t) \A(t e^{-\tK} Q^2) + \int_1^\infty \frac{dt}{t} G_d^{(+)}(t) \A(t e^{-\tK} Q^2) + \int_0^{1} \frac{dt}{t} G^{\rm (SL)}_d(t) \left[ \A(t e^{-\tK} Q^2) - \A(e^{-\tK} Q^2) \right] ,
\label{dresA}
\ee
We refer to Appendix \ref{app:renmod} for the main steps in this construction, and to Ref.~\cite{renmod} for full details.

 It turns out that the evaluation of the integrals in Eq.~(\ref{drespQCD}) for $Q^2>0$ at $t \to 0$, is ambiguous because the pQCD coupling $a(t \exp(-\tK) Q^2)$ has a cut-discontinuity at $0 < t \exp(-\tK) Q^2 < Q^2_{\rm br}$ (Landau singularities). However, since the coupling $\A(Q^{'2})$ in $\A$QCD has no Landau singularities, the evaluation of the expression (\ref{dresA}) is unambiguous even for $Q^2>0$.

\section{The extracted parameters of the coupling $\A(Q^2)$}
\label{sec:Ares}

The seven requirements explained in Sec.~\ref{sec:constr} then give us the values of the seven parameters of the considered $\A$-coupling: ${\cal F}_j, M_j^2$ ($j=1,2,3$) and $M_0^2$. This means that for any chosen values of the ``strength'' constant $\alpha_s(M_Z^2;\MSbar)$ ($\leftrightarrow$ Lambert scale $\Lambda_L$) and of the $\tau$-decay ratio $r_{\tau}^{(D=0)}$ ($\approx 0.20$), we are then able to obtain the values of the seven parameters, which are written in the dimensionless form
\be
f_k \equiv \frac{{\cal F}_k}{\Lambda_L^2}   \quad (k=1,2,3), \quad
s_j \equiv \frac{M_j^2}{\Lambda_L^2}   \quad (j=0,1,2,3) \ .
\label{not}
\ee
In practice, the five parameters $s_j$ ($j=2,3$) and $f_k$ ($k=1,2,3$) are expressed as functions of the parameters $s_0$ and $s_1$ via the five requirements (\ref{diffAa}) and (\ref{A00}). Then, for a given chosen value of $\alpha_s(M_Z^2;\MSbar)$, the positive values of the two parameters $s_0$ and $s_1$ (with $s_0 < s_1 < s_2 < s_3$) are varied so that the other two requirements [Eqs.~(\ref{rtaucont}) and (\ref{Amax})] are met: (I) a chosen value value of $r_{\tau}^{(D=0)}$  ($ \approx 0.20$) is achieved and, simultaneously, (II) the coupling $\A(Q^2)$ achieves for positive $Q^2$ the maximum at $Q^2=0.135 \ {\rm GeV}^2$.

When the chosen values of $\alpha_s(M_Z^2;\MSbar)$ are higher and those of $r_{\tau}^{(D=0)}$ ($\approx 0.20$) are lower, it turns out that the mentioned two requirements [(I) and (II)] can be met only in the limit $s_1 \to s_2-0$. In such cases, the spectral function $\rho_{\A}(\sigma)$ and the coupling $\A(Q^2)$ then attain the following limiting form [cf.~Eqs.~(\ref{rhoA}) and (\ref{Adisp2})]:
\bes \label{2d1dp1dpp}
\bea
\frac{1}{\pi} \rho_{\A}(\sigma) &=& \sum_{j=1}^2 {\cal F}_j \delta(\sigma - M_j^2) +{\cal F}_1^{(1)} \delta'(\sigma - M_1^2) +{\cal F}_1^{(2)} \delta^{\prime \prime}(\sigma - M_1^2) + \frac{1}{\pi}\Theta(\sigma - M_0^2) \rho_a(\sigma),
\label{rhoAalt} \\
\A(Q^2) &=& \sum_{j=1}^{2} \frac{{\cal F}_j}{(Q^2 + M_j^2)} +  \frac{{\cal F}_1^{(1)}}{(Q^2 + M_j^2)^2}  + \frac{2 {\cal F}_j^{(2)}}{(Q^2 + M_j^2)^3} + \frac{1}{\pi} \int_{\sigma=M_0^2}^{\infty} \frac{d \sigma \rho_{a}(\sigma)}{(\sigma + Q^2)},
\label{Adisp2alt} \eea
\ees
where the primes at the Dirac-delta functions denote derivatives. In this case, the new dimensionless parameters are
\be
f_1^{(1)}= \frac{{\cal F}_1^{(1)}}{\Lambda_L^4}, \quad
f_1^{(2)}= \frac{{\cal F}_1^{(2)}}{\Lambda_L^6}.
\label{notalt} \ee
Effectively, we can always regard the chosen values of $\alpha_s(M_Z^2;\MSbar)$ and  $r_{\tau}^{(D=0)}$ as the input parameters which then determine the $\A$-coupling. In Table \ref{tabres3d} we present solutions for various chosen values of these input parameters $\alpha_s(M_Z^2;\MSbar)$ and $r_{\tau}^{(D=0)}$. The range of the values $\alpha_s(M_Z^2;\MSbar)$ is the interval of the recent world average values $\alpha_s(M_Z^2;N_f=5;\MSbar)=0.1179 \pm 0.0010$ \cite{PDG2019}.
 \begin{table}
   \caption{The dimensionless parameters of the coupling $\A(Q^2)$: $s_j \equiv M_j^2/\Lambda_L^2$ ($j=0,1,2,3$); $f_j \equiv {\cal F}_j/\Lambda_L^2$ ($j=1,2,3$) [cf.~Eqs.~(\ref{not}) and (\ref{Adisp2})] for various chosen values of $\alpha_s(M_Z^2;\MSbar)$ ($\leftrightarrow \Lambda_{L}$) and $r^{(D=0)}_{\tau, {\rm th}}$. Included is also the value of $\pi \A(Q^2)$ at the local maximum (for positive $Q^2$) $Q^2 = 0.1350 \ {\rm GeV}^2$. At higher values of $\alpha_s(M_Z^2;\MSbar)$ or lower values of  $r_{\tau}^{(D=0)}$, the values of the alternative needed parametrization Eq.~(\ref{notalt}) [cf.~Eq.~(\ref{Adisp2alt})] are given.}
\label{tabres3d}  
\begin{ruledtabular}
  \begin{tabular}{ll|rrrrrrr|r}
${\overline \alpha}_s(M_Z^2)$ [$\Lambda_L$ (GeV)] &  $r^{(D=0)}_{\tau, {\rm th}}$ & $s_0$ & $s_1$  & $s_2$ & $s_3$ & $f_1$ & $f_2$ & $f_3$  & $\pi \A_{\rm max}$
  \\
  \hline
 $0.1184 \; [0.114517]$ & 0.203 & 642.399 & 4.63154 & 16.5725 & 466.910 & -3.97366 & 12.8485 & 5.20916 & 0.92452
 \\
 $0.1181 \; [0.113003]$ & 0.201 & 673.556 & 3.71461 & 21.3864 & 490.509 & -2.27773 & 11.3625 & 5.35749 & 0.87219
 \\
 $0.1181 \; [0.113003]$ & 0.203 &  755.931 & 2.29898 & 33.3151 & 552.949 & -0.911199 & 10.5375 & 5.74122 & 0.78240
\\
 \hline
 $0.1180 \; [0.112500]$ & 0.202 & 755.040 & 2.35204 & 33.1302 & 552.268 & -0.937899 & 10.5583 & 5.73732 & 0.78157
 \\
 $0.1180 \; [0.112500]$ & 0.203 & 829.491 & 1.7491 & 43.3075 & 608.703 & -0.568146 & 10.6616 & 6.07637 & 0.722972
\\ 
 $0.1179 \; [0.111999]$ & 0.201 & 753.964 & 2.40862 & 32.9178 & 551.446 & -0.967082 & 10.5803 & 5.73256 & 0.78090
 \\
$0.1179 \; [0.111999]$ & 0.202 & 823.902 & 1.81634 & 42.5024 & 604.46 & -0.598431 & 10.6567 & 6.05134 & 0.72555
 \\
 $0.1178 \; [0.111500]$ & 0.200 & 752.710 & 2.46899 & 32.6783 & 550.489 & -0.998984 & 10.6039 & 5.72699 & 0.78037
\\
 $0.1178 \; [0.111500]$ & 0.201 & 818.918. & 1.88306 & 41.7757 & 600.676 & -0.628746 & 10.6555 & 6.02901 & 0.72772
 \\
  $0.1177 \; [0.111001]$ & 0.199 & 751.287 & 2.53325 & 32.4130 & 549.404 & -1.03377 & 10.6293 & 5.72064 & 0.77997
 \\
$0.1177 \; [0.111001]$ & 0.200 & 814.291 & 1.95043 & 41.0942 & 597.163 & -0.659738 & 10.6572 & 6.00827 & 0.72964
  \\
 $0.1176 \; [0.110504]$ & 0.199 & 809.917 & 2.01909 & 40.4438 & 593.841 & -0.691761 & 10.6615 & 5.98866 & 0.73140
\\
 $0.1175 \; [0.110008]$ & 0.197 & 747.883 & 2.6754 & 31.7959 & 546.811 & -1.11385 & 10.6869 & 5.70537 & 0.77963
\\
$0.1175 \; [0.110008]$ & 0.198 & 805.718 & 2.08956 & 39.8139 & 590.652 & -0.725134 & 10.6682 & 5.96983 & 0.73303
\\
$0.1175 \; [0.110008]$ & 0.1988 & 910.653 & 1.48732 & 54.005 & 670.205 & -0.418181 & 11.0113 & 6.43904 & 0.66892
\\
$0.1174 \; [0.109513]$ & 0.197 & 801.637 & 2.1623 & 39.1966 & 587.552 & -0.760166 & 10.6772 & 5.95151 & 0.73457
\\
$0.1173 \; [0.109019]$ & 0.196 & 797.636 & 2.23771 & 38.5866 & 584.513 & -0.797146 & 10.6886 & 5.93354 & 0.73605
\\
$0.1173 \; [0.109019]$ & 0.197 & 904.897 & 1.5672 & 53.1420 & 665.829 & -0.445743 & 11.0034 & 6.41399 & 0.66981
\\
$0.1172 \; [0.108527]$ & 0.195 & 793.679 & 2.31620 & 37.9789 & 581.506 & -0.836399 & 10.7025 & 5.91576 & 0.73750
\\
$0.1172 \; [0.108527]$ & 0.196 & 883.815 & 1.69725 & 50.2540 & 649.839 & -0.501017 & 10.9297 & 6.32083 & 0.68000
\\
$0.1171 \; [0.108036]$ & 0.195 & 870.166 & 1.80233 & 48.359 & 639.485 & -0.54636 & 10.8909 & 6.26031 & 0.68650
\\
$0.1170 \; [0.107546]$ & 0.194 & 859.428 & 1.89941 & 46.8507 & 631.337 & -0.588937 & 10.8669 & 6.2126 & 0.69151
\\
$0.1169 \; [0.107058]$ & 0.193 & 850.281 & 1.99365 & 45.5518 & 624.396 & -0.631008 & 10.8520 & 6.17189 & 0.69570
\\
\hline
   ${\overline \alpha}_s(M_Z^2)$ [$\Lambda_L$ (GeV)] &  $r^{(D=0)}_{\tau, {\rm th}}$ & $s_0$ & $s_1$  & $s_2$ & $f_1$ & $f_1^{(1)}$ & $f_1^{(2)}$ & $f_2$  & $\pi \A_{\rm max}$
\\
\hline
$0.1181 \; [0.113003]$ & 0.199 & 614.709 & 10.1622 & 445.893 & 8.68391 & -71.7103 & -127.796 & 5.07754 & 0.96237
\\
$0.1184 \; [0.114517]$ & 0.201 & 587.16 & 11.6459 & 424.893 & 8.48770 & -27.1570 & -483.952 & 4.94869 & 1.01338
\\
$0.1184 \; [0.114517]$ & 0.199 & 523.647 & 13.4469 & 375.641 & 7.98996 & 62.3606 & -1246.69 & 4.67979 & 1.08931
\\
$0.1189 \; [0.117067]$ & 0.201 & 434.498 & 14.2200 & 304.382 & 7.15111 & 174.204 & -2092.06 & 4.39556 & 1.21831
\end{tabular}
\end{ruledtabular}
\end{table}
In all the displayed cases, we can arrive from one of them to another adjacent case by continuous variation of the input parameter values of $\alpha_s(M_Z^2;\MSbar)$ and  $r^{(D=0)}_{\tau, {\rm th}}$.
Later in the analysis it will become clear why we displayed, at a given value of $\alpha_s(M_Z^2;\MSbar)$, the specific values or (narrow) ranges of values of  $r_{\tau}^{(D=0)}$ as given in Table \ref{tabres3d}. For $\alpha_s(M_Z^2;\MSbar) \leq 0.1180$, the displayed cases always include the one with the largest possible value of $r^{(D=0)}_{\tau, {\rm th}}$ (with the precision of 0.001) within our ($3 \delta$ $\A$QCD) approach of the construction of $\A$-coupling. For example, when $\alpha_s(M_Z^2;\MSbar)= 0.1179$, the value $r^{(D=0)}_{\tau, {\rm th}}=0.202$ can be achieved, but not the value $r^{(D=0)}_{\tau, {\rm th}}=0.203$; when $\alpha_s(M_Z^2;\MSbar)= 0.1172$, the value $r^{(D=0)}_{\tau, {\rm th}}=0.195$ can be achieved, but not the value  $r^{(D=0)}_{\tau, {\rm th}}=0.196$. Only in the case of $\alpha_s(M_Z^2;\MSbar)= 0.1175$ we increased the value of  $r^{(D=0)}_{\tau, {\rm th}}$ to its maximal value with the precision of $0.0001$, i.e.,  $r^{(D=0)}_{\tau, {\rm th}}=0.1988$ can be achieved but not $r^{(D=0)}_{\tau, {\rm th}}=0.1989$. We recall that in all these adjustments, we required that the coupling $\A(Q^2)$ at positive $Q^2$ achieve its maximum at $Q^2=0.1350 \ {\rm GeV}^2$ (as suggested by the aforementioned lattice calculations).

The numerical implementation of the couplings of Table \ref{tabres3d} is available in the Mathematica code from the web page www.gcvetic.usm.cl.\footnote{E.g., the case with the input parameters  $\alpha_s(M_Z^2;\MSbar)=0.1177$ and $r^{(D=0)}_{\tau, {\rm th}}=0.200$ is in the program 3dAQCDrt0200al01177N.m; and the other cases are in the programs with analogous names there. The programs also contain information on how to use them in Mathematica software.} 

\begin{figure}[htb] 
\begin{minipage}[b]{.49\linewidth}
  \centering\includegraphics[width=85mm]{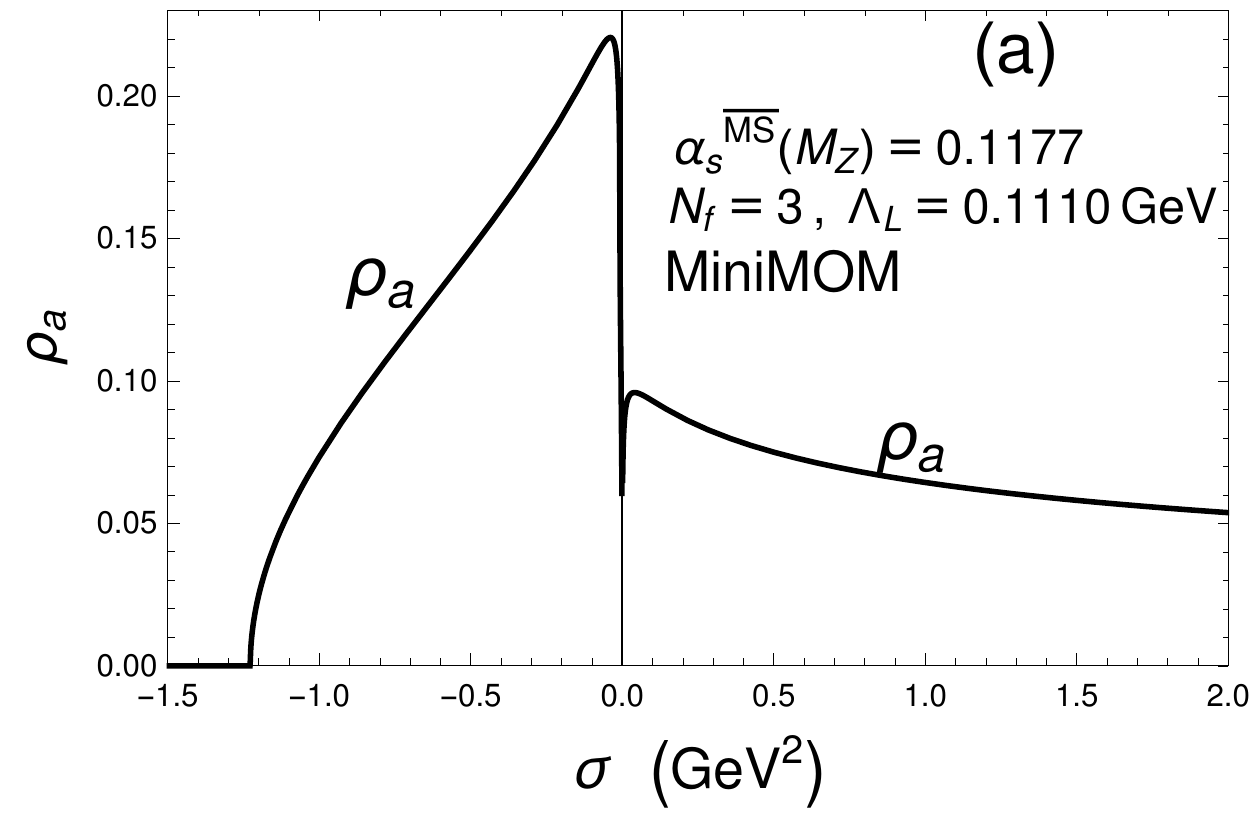}
  \end{minipage}
\begin{minipage}[b]{.49\linewidth}
  \centering\includegraphics[width=85mm]{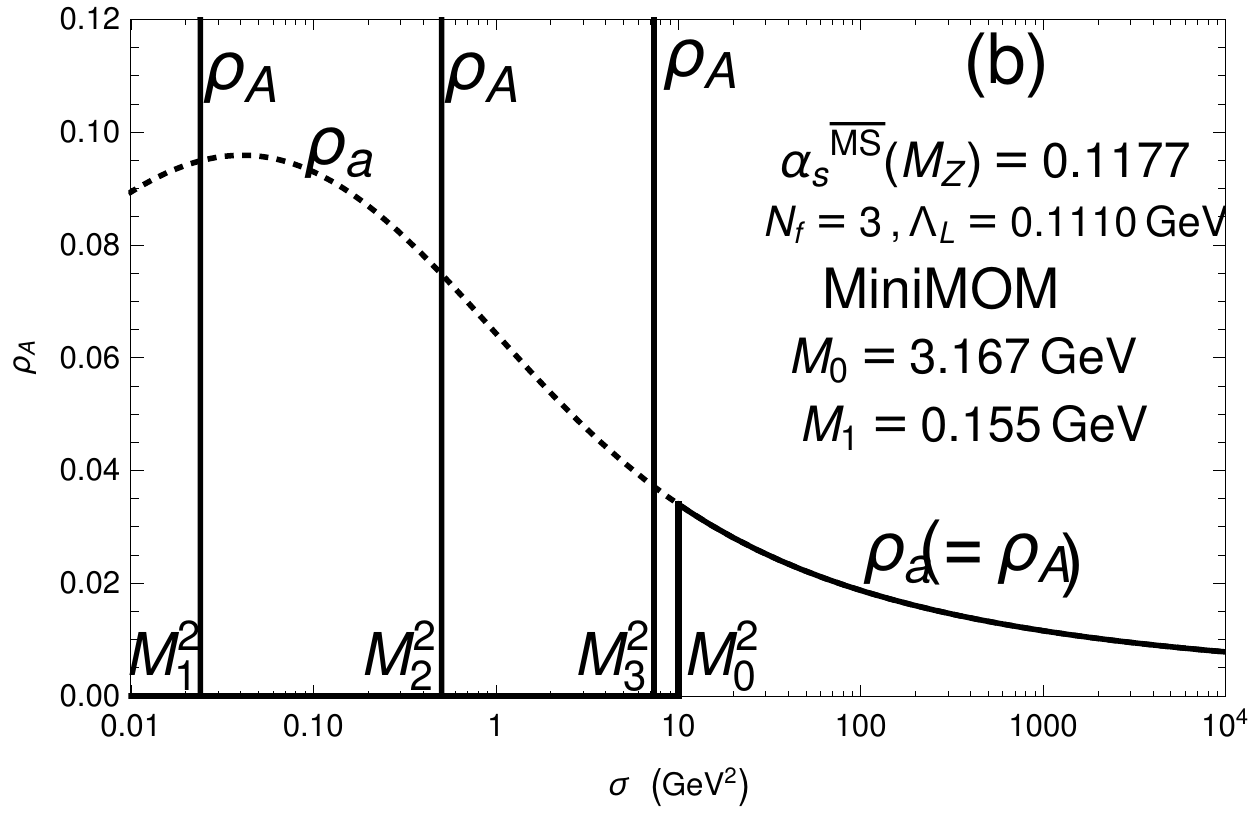}
\end{minipage}
\vspace{-0.2cm}
\caption{\footnotesize  (a) The pQCD spectral function $\rho_a (\sigma) = {\rm Im} \; a (Q^2=-\sigma - i \epsilon)$ in the 4-loop LMM scheme ($\sigma$ is on linear scale); (b) $\rho_{\A}(\sigma) =  {\rm Im} \; \A(Q^2=-\sigma - i \epsilon)$ (where: $\sigma > 0$ is on logarithmic scale). The delta function at $M_1^2$ is in fact negative, but is shown as positive for convenience.}
\label{Figrho}
\end{figure}

\begin{figure}[htb] 
   \centering\includegraphics[width=90mm,height=60mm]{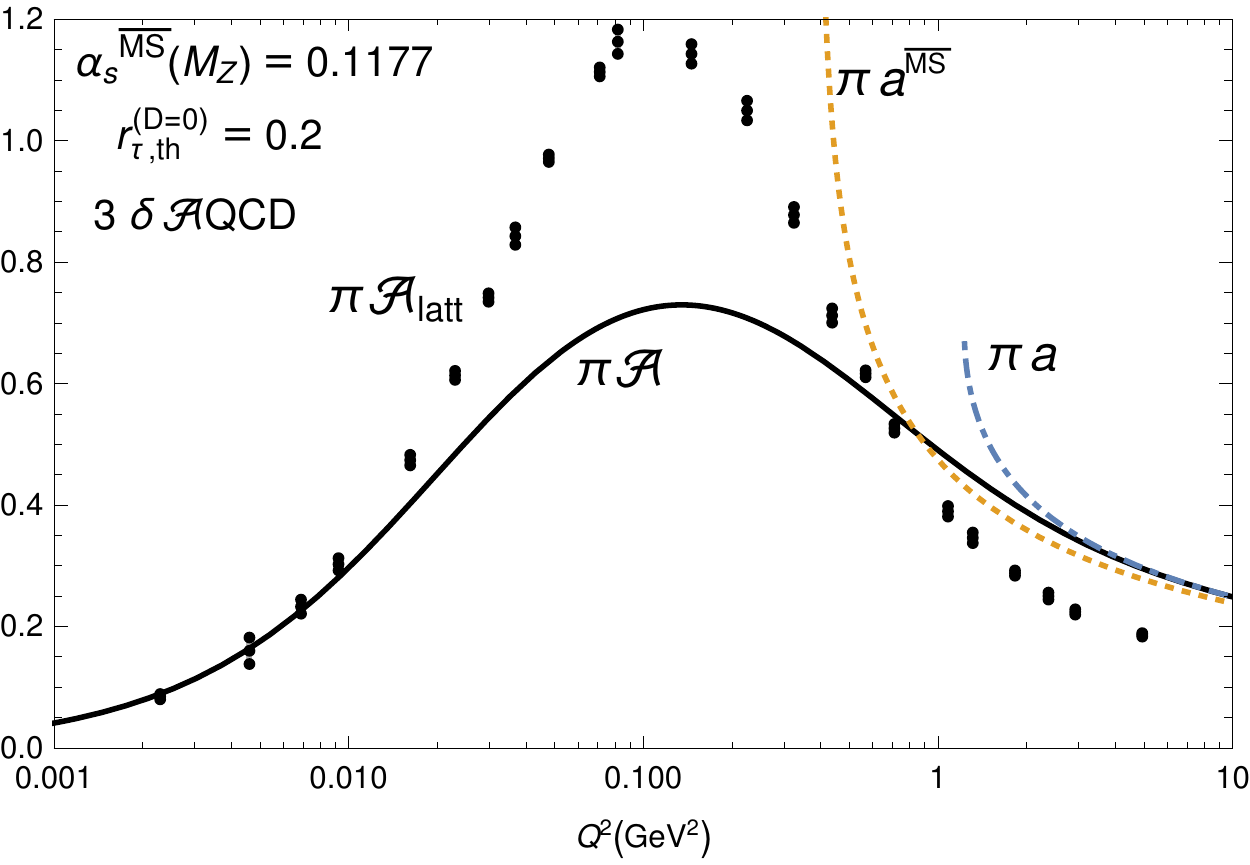}
\vspace{-0.2cm}
\caption{\footnotesize The considered $N_f=3$ coupling $\pi \A$ (solid curve), at positive $Q^2$. For comparison, the underlying LMM pQCD coupling $\pi a$ (dot-dashed curve) and $\MSbar$ pQCD coupling $\pi a^{\MSbar}$ (dotted curve) are included. Further, the large-volume lattice results $\pi \A_{\rm latt}$ \cite{LattcoupNf02a} are shown (points with bars); for them, the momenta $Q^2$ were rescaled from the lattice MM to the LMM scheme: $Q^2=Q^2_{\rm latt} (\Lambda_{\MSbar}/\Lambda_{\rm MM})^2 \approx Q^2_{\rm latt}/1.9^2$. The branching points of the Landau cuts for $a$ and $a^{\MSbar}$ are $Q^2_{\rm br}=1.228 \ {\rm GeV}^2$ and $0.393 \ {\rm GeV}^2$, respectively; $a$ is finite at its branching point.}
\label{FigAa}
 \end{figure}

For the choice of the input parameters $\alpha_s(M_Z^2;\MSbar)=0.1177$ and $r^{(D=0)}_{\tau, {\rm th}}=0.200$, the mass parameters $M_j^2 = s_j \Lambda_{L}^2$ and the residue parameters ${\cal F}_j = f_j \Lambda_L^2$ are\footnote{\label{ftrestrunc} If, for the parameters of Eqs.~(\ref{M0123}), we evaluated $r^{(D=0)}_{\tau, {\rm th}}$ Eq.~(\ref{rtaucont}) with the Adler function taken as the truncated series Eq.~(\ref{dD0TPSA}) (as in Ref.~\cite{3dAQCD}), instead of the renormalon-motivated resummation Eq.~(\ref{dresA}), we would obtain  $r^{(D=0)}_{\tau, {\rm th}}=0.1988$ (instead of $r^{(D=0)}_{\tau, {\rm th}}=0.2000$).}
\bes
\label{M0123}
\bea
M_0^2 &= & 10.033  \ {\rm GeV}^2 \; (M_0 \approx 3.167 \ {\rm GeV}); 
\label{M0}
\\
M_1^2 &=& 0.0240 \ {\rm GeV}^2 \; (M_1 \approx 0.155 \ {\rm GeV}), \quad {\cal F}_1  =  -0.00813 \ {\rm GeV}^2,
\label{M1}
\\
M_2^2&=&0.506  \ {\rm GeV}^2 \; (M_2 \approx 0.712 \ {\rm GeV}), \quad {\cal F}_2 = 0.1313 \ {\rm GeV}^2,
\label{M2}
\\
M_3^2 &=& 7.358  \ {\rm GeV}^2 \; (M_3 \approx 2.713 \ {\rm GeV}), \quad {\cal F}_3 = 0.0740 \ {\rm GeV}^2.
\label{M3}
\eea
\ees
In Figs.~\ref{Figrho} and \ref{FigAa} we present the spectral function $\rho_{\A}(\sigma)$ and the coupling $\pi \A(Q^2)$ (for $Q^2>0$), respectively, for the choice $\alpha_s(M_Z^2;\MSbar)=0.1177$ and $r^{(D=0)}_{\tau, {\rm th}}=0.200$.
 In Fig.~\ref{Figrho}(a) the spectral function $\rho_a(\sigma)$ of the underlying pQCD coupling $a$ is presented (for $\sigma>0$ and $\sigma<0$), while in Fig.~\ref{Figrho}(b) the resulting $\rho_{\A}(\sigma)$ is presented ($\sigma>0$). In Fig.~\ref{FigAa} we include, for comparison, the underlying pQCD coupling $\pi a(Q^2)$ (i.e., in the LMM scheme) and the pQCD coupling in the $\MSbar$ scheme. We further include in Fig.~\ref{FigAa} the resulting lattice coupling $\pi \A_{\rm latt}(Q^2)$ at $N_f=0$ \cite{LattcoupNf02a}, where we rescaled $Q^2$ from the MM scheme of Ref.~\cite{LattcoupNf02a} to the LMM scheme, cf.~discussion in Sec.~\ref{sec:constr}. At $Q^2 \to 0$ the coupling $\A(Q^2)$ behaves as $k Q^2$, with $k \approx 13.6 \ {\rm GeV}^2$, agreeing qualitatively with the lattice coupling $\A_{\rm latt}(Q^2)$. The couplings $\A$ and $\A_{\rm latt}$ both achieve the local maximum (for positive $Q^2$) at $Q^2=0.135 \ {\rm GeV}^2$. The maximal value of $\A(Q^2)$ is lower than that of $\A_{\rm latt}(Q^2)$; however, the height of the maximum depends significantly on the chosen reference value $\alpha_s(M_Z^2; \MSbar)$ in our approach and in the lattice calculations. Further, we note that at large $Q^2 > 1 {\rm GeV}^2$, the large-volume lattice results are unreliable \cite{Sternb}.

\begin{figure}[htb] 
\centering\includegraphics[width=90mm,height=60mm]{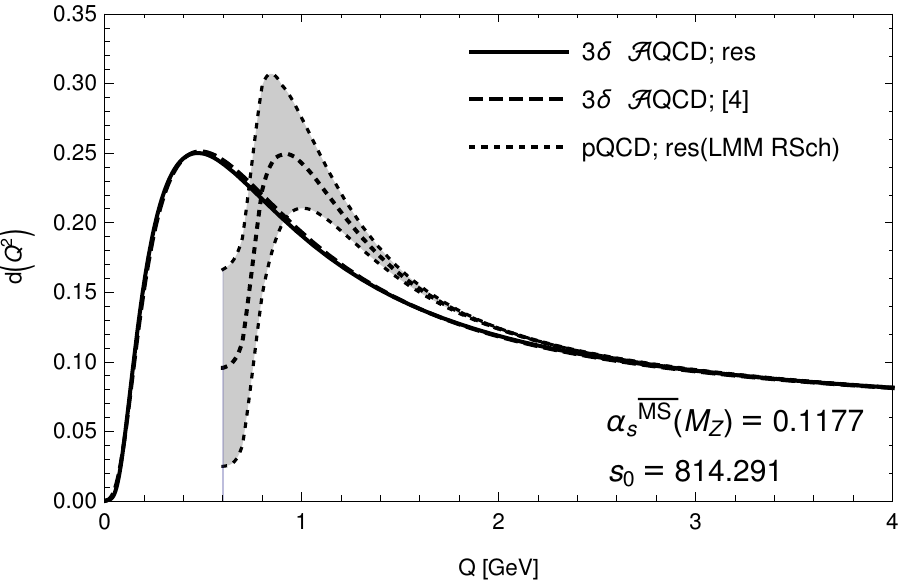}
\vspace{-0.2cm}
\caption{\footnotesize The Adler function $d(Q^2)_{(D=0)}$ as a function of $Q \equiv \sqrt{Q^2}$ for positive $Q^2$, in the considered $3 \delta$ $\A$QCD: evaluated as the resummation Eq.~(\ref{dresA}) (solid curve); as the truncated (with four terms, [4]) series Eq.~(\ref{dD0TPSA}) (dashed curve). Included are the resummation results Eqs.~(\ref{drespQCDPV})-(\ref{drespQCDIm}) using the underlying pQCD coupling (dotted curves, grey band).}
\label{FigdposQ2}
 \end{figure}

 In Fig.~\ref{FigdposQ2} we show the behaviour of the Adler function $d(Q^2)_{(D=0)}$ for positive $Q^2$, when it is evaluated with the $\A$-coupling with the renormalon-motivated  resummation Eq.~(\ref{dresA}) and with the truncated series Eq.~(\ref{dD0TPSA}), both for the $\A$-coupling with the parameter values Eqs.~(\ref{M0123}) and $\alpha_s(M_Z^2;\MSbar)=0.1177$ (i.e., $\Lambda_L=0.1110$ GeV). 

 We include in the Figure the renormalon-motivated resummation using the underlying pQCD coupling $a(Q^2)$, cf.~Eq.~(\ref{drespQCD}), and fixing the ambiguity from the mentioned Landau cut by taking the principal value (PV)
 \bea
\lefteqn{ 
  d(Q^2; {\rm PV})_{(D=0)}^{\rm pt; res} =  {\rm Re} {\Bigg \{} \int_0^1 \frac{dt}{t} G_d^{(-)}(t) a(t e^{-\tK} Q^2 + i \epsilon) + \int_1^\infty \frac{dt}{t} G_d^{(+)}(t) a(t e^{-\tK} Q^2+ i \epsilon)
}
\nonumber\\
&&
+ \int_0^{1} \frac{dt}{t} G^{\rm (SL)}_d(t) \left[ a(t e^{-\tK} Q^2+ i \epsilon) - a(e^{-\tK} Q^2 + i \epsilon) \right] {\Bigg \}} \qquad (Q^2>0, \epsilon \to +0).
\label{drespQCDPV}
\eea
The grey band around the central pQCD curve illustrates the uncertainty (ambiguity) $\delta d$ of this pQCD result, where $\delta d$ is
\bea
\lefteqn{ 
  \delta d(Q^2)_{(D=0)}^{\rm pt; res} =  \pm \frac{1}{\pi} {\rm Im} {\Bigg \{} \int_0^1 \frac{dt}{t} G_d^{(-)}(t) a(t e^{-\tK} Q^2 + i \epsilon) + \int_1^\infty \frac{dt}{t} G_d^{(+)}(t) a(t e^{-\tK} Q^2+ i \epsilon)
}
\nonumber\\ &&
+ \int_0^{1} \frac{dt}{t} G^{\rm (SL)}_d(t) \left[ a(t e^{-\tK} Q^2+ i \epsilon) - a(e^{-\tK} Q^2 + i \epsilon) \right] {\Bigg \}} \qquad (Q^2>0, \epsilon \to +0).
\label{drespQCDIm}
\eea
We deduce from Fig.~\ref{FigdposQ2} that the (resummed) pQCD approach fails in the regime $Q < 2$ GeV, the main reason being the Landau singularities (cut) of the pQCD coupling. On the other hand, the truncated series (\ref{dD0TPSA}) gives results apparently very close to those of the resummed expression (\ref{dresA}). This results then in a minor deviation ($\sim 0.001$) of the evaluated $r_{\tau}$ ratio, cf.~footnote \ref{ftrestrunc}. Nonetheless, as we will see later in this work, even such apparently minor deviations of $r_{\tau}$ will play a significant role in the determinations of nonperturbative parameters such as condensate values and IR-regulating masses.

\section{Borel-Laplace sum rules of the $\tau$-lepton decay data}
\label{sec:BSR}

\subsection{Borel-Laplace sum rules: description of the method}
\label{subs:des}

In this Section, we will extract values of the condensates with dimension $D=4$ and $6$ of the (V+A)-channel current-current correlator (Adler function) in our approach. These values will be extracted by applying Borel-Laplace sum rules to the corresponding spectral function measured by the OPAL and ALEPH Collaborations. The approach is following closely the sum rule analysis presented in Ref.~\cite{3dAQCD}, only that this time the $D=0$ (leading-twist) contribution of the Adler function will not be evaluated as truncated series (\ref{dD0TPSA}) but as the renormalon-motivated resummation (\ref{dresA}). Therefore, we will not go into all the technical details of this extraction, but will refer to Ref.~\cite{3dAQCD} for details.

The central quantities appearing in these sum rules are the quark current-current corelators
\be
\Pi_{\rm{J}, \mu\nu}(q) =  i \int  d^4 x \; e^{i q \cdot x} 
\langle T J_{\mu}(x) J_{\nu}(0)^{\dagger} \rangle
=  (q_{\mu} q_{\nu} - g_{\mu \nu} q^2) \Pi_{\rm J}^{(1)}(Q^2)
+ q_{\mu} q_{\nu} \Pi_{\rm J}^{(0)}(Q^2),
\label{PiJ}
\ee
where J=V,A, and $Q^2 \equiv -q^2$ is the squared momentum transfer. The quark currents are (for J=V) $J_{\mu} = {\overline u} \gamma_{\mu} d$ and (for J=A) $J_{\mu} = {\overline u} \gamma_{\mu} \gamma_5 d$. The sum rules will be applied to the (strangeless) (V+A)-channel; the corresponding polarization function is
\be
\Pi(Q^2) \equiv \Pi_{V+A}(Q^2) =
\Pi^{(1)}_{V}(Q^2) +  \Pi^{(1)}_{A}(Q^2) + \Pi^{(0)}_{A}(Q^2),
\label{Pi1}
\ee
where the term $\Pi^{(0)}_{V}(Q^2)$ is neglected because ${\rm Im} \Pi^{(0)}_{V}(-\sigma - i \epsilon) \propto (m_d-m_u)^2$. The corrections ${\cal O}(m_{u,d}^2)$ and ${\cal O}(m_{u,d}^4)$ will not be included because they are numerically negligible.

\begin{figure}[htb] 
\centering\includegraphics[width=70mm]{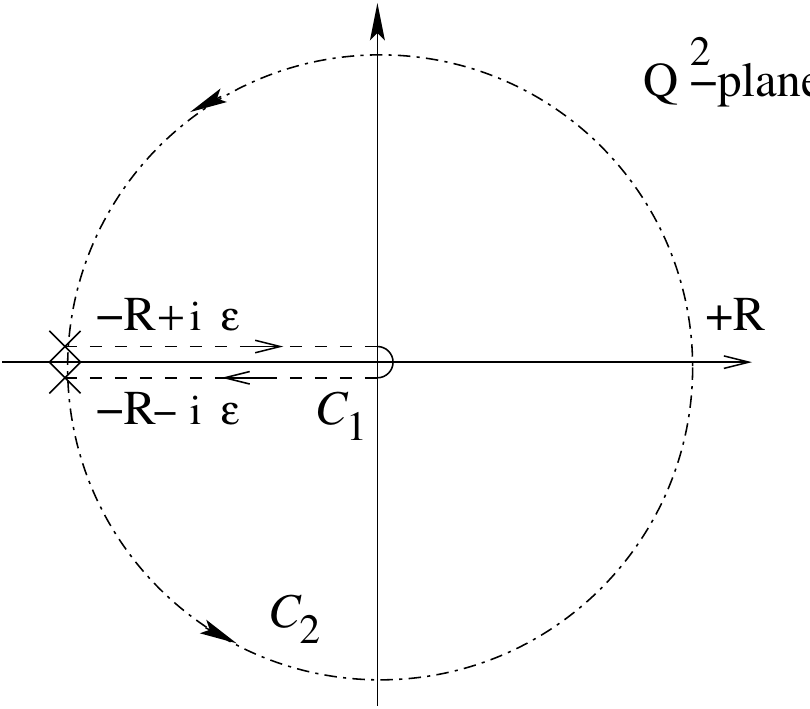}
\caption{\footnotesize The closed contour $C_1+C_2$ for integration of $g(Q^2) \Pi(Q^2)$. Here, the contour radius is $R=\sigma_{\rm max}$ ($\leq m_{\tau}^2$).}
\label{Figcont}
 \end{figure}

\begin{figure}[htb] 
\begin{minipage}[b]{.49\linewidth}
  \centering\includegraphics[width=85mm]{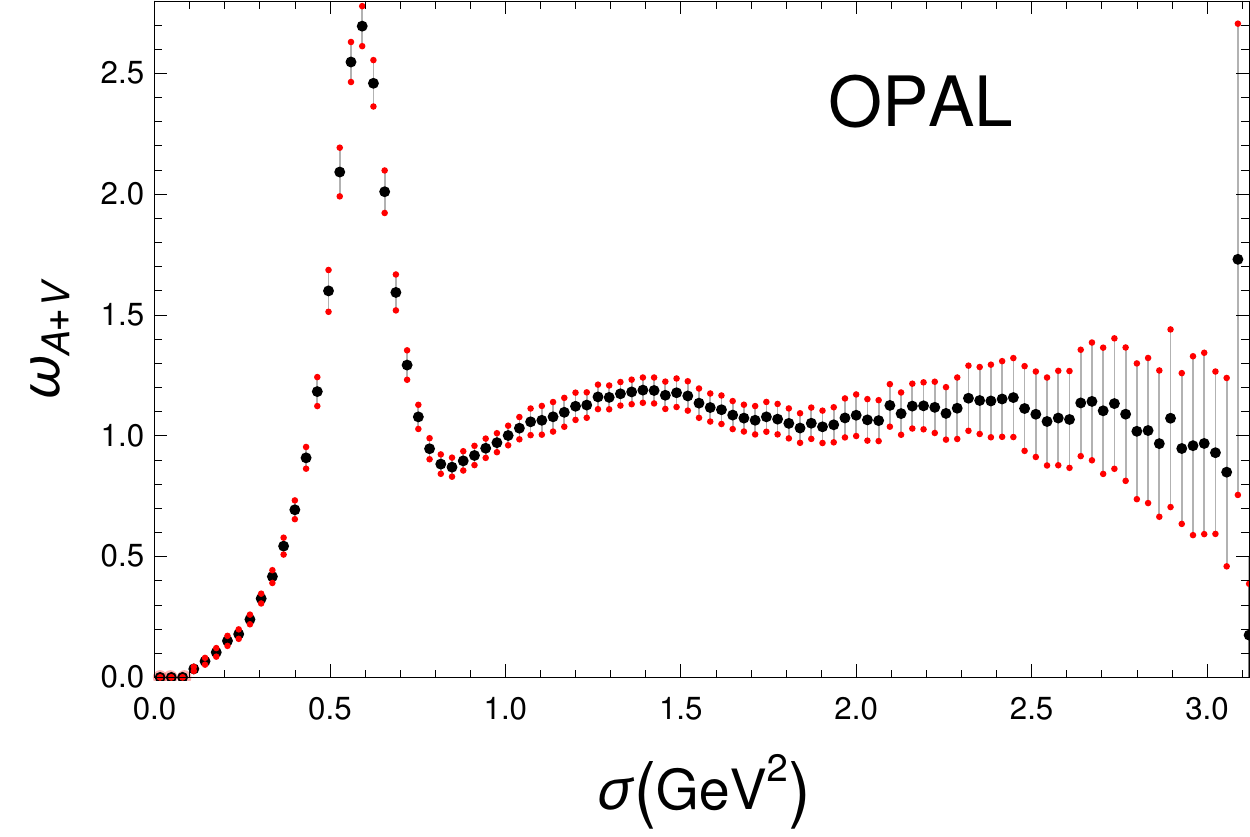}
  \end{minipage}
\begin{minipage}[b]{.49\linewidth}
  \centering\includegraphics[width=85mm]{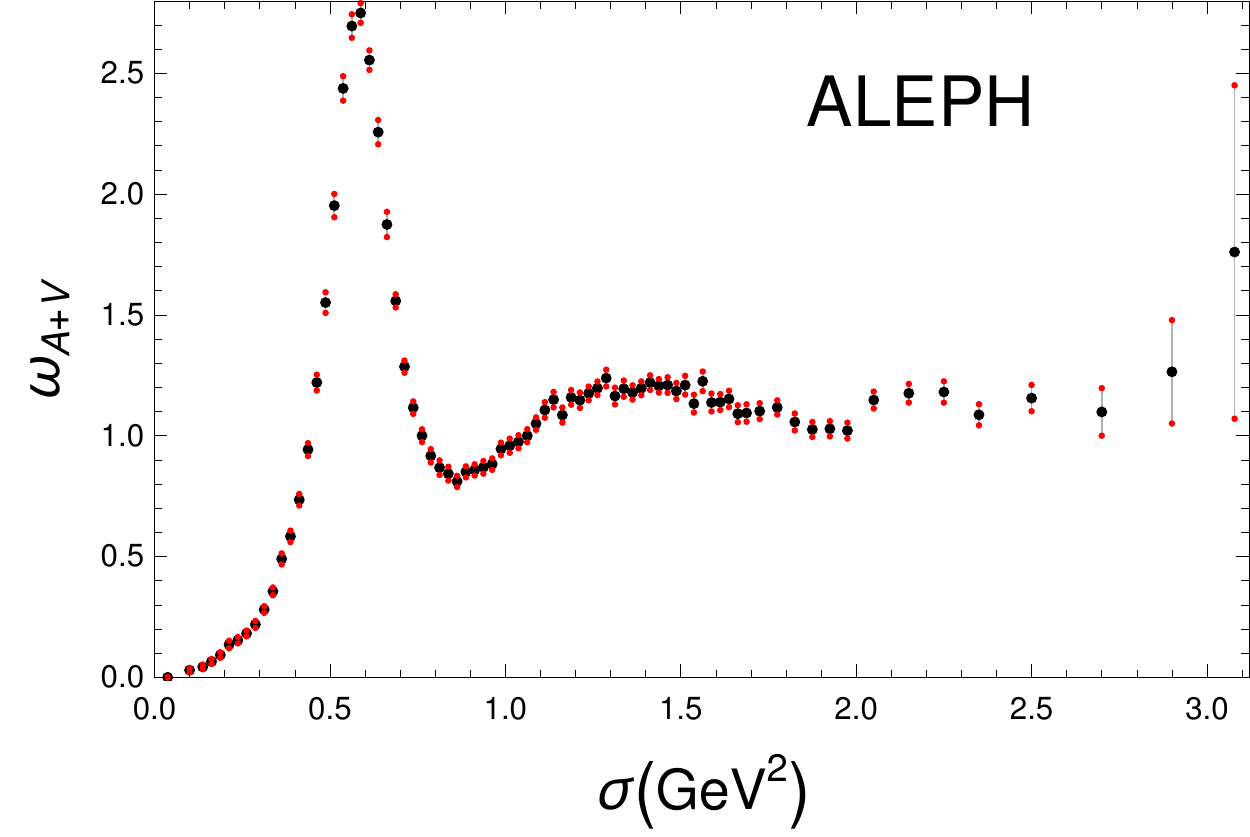}
\end{minipage}
\vspace{-0.2cm}
\caption{\footnotesize  The spectral function $\omega(\sigma)$ for the (V+A)-channel, measured by OPAL (left-hand) and by ALEPH Collaboration (right-hand), without the pion peak contribution. We will take $\sigma_{\rm max}=3.136$ for OPAL and $2.80 \ {\rm GeV}^2$ for ALEPH.}
\label{FigOmega}
\end{figure}

By the general principles of Quantum Field Theory, the correlator is a holomorphic (analytic) function in the complex $Q^2$-plane, with the exception of the the negative semiaxis, $Q^2  \in \mathbb{C} \backslash (-\infty, -m_{\pi}^2)$. If we multiply $\Pi(Q^2)$ with any function $g(Q^2)$ which is holomorphic in the $Q^2$-plane, then the application of the Cauchy theorem to the integration contour presented in Fig.~\ref{Figcont} gives the sum rule
\bes
\label{Cauchysr1}
\bea
\oint_{C_1+C_2} d Q^2 g(Q^2) \Pi(Q^2) & = & 0 \; \Rightarrow
\label{Cauchy} \\
\int_0^{\sigma_{\rm max}} d \sigma g(-\sigma) \omega_{\rm (exp)}(\sigma)  &=&
-i \pi  \oint_{|Q^2|=\sigma_{\rm max}}
d Q^2 g(Q^2) \Pi_{\rm (th)}(Q^2) ,
\label{sr1} \eea \ees
where $\omega(\sigma)$ is the spectral function of $\Pi(Q^2)$ (along the cut)
\be
\omega(\sigma) \equiv 2 \pi \; {\rm Im} \ \Pi(Q^2=-\sigma - i \epsilon) \ ,
\label{om1}
\ee
In the sum rule (\ref{sr1}), in practice the left-hand side is determined by the measured values, and the right-hand side by the theoretical values.
 The spectral function $\omega(\sigma)$ was measured in the $\tau$-lepton semihadronic decays by the OPAL \cite{OPAL,PerisPC1,PerisPC2} and ALEPH Collaboration \cite{ALEPH2,DDHMZ,ALEPHfin,ALEPHwww}. These experimental results are presented in Figs.~\ref{FigOmega}.
In addition to these measured contributions, the pion peak contribution $\delta \omega_{\pi}(\sigma)= 2 \pi^2 f_{\pi}^2 \delta(\sigma - m_{\pi}^2)$ (where $f_{\pi}=0.1305$ GeV) must be added. The largest value of $\sigma$ in the case of OPAL is  $\sigma_{\rm max}=3.136 \ {\rm GeV}^2$ (which is close to $m^2_{\tau}=3.157 \ {\rm GeV}^2$). On the other hand, the uncertainties of the last few bins of the ALEPH data are very large, so we decided to take in the case of ALEPH $\sigma_{\rm max}=2.80 \ {\rm GeV}^2$.

The correlator function has an Operator Product Expansion (OPE)
\be
\Pi_{\rm th}(Q^2) =  - \frac{1}{2 \pi^2} \ln(Q^2/\mu^2) + 
\Pi_{\rm th}(Q^2)_{(D=0)}
+ \sum_{n \geq 2} \frac{ \langle O_{2n} \rangle_{V+A}}{(Q^2)^n}
  \left( 1 + {\cal C}_n a(Q^2) \right) ,
\label{OPE1}
\ee
where $\langle O_{2n} \rangle_{V+A}$ are condensates (vacuum expectation values) of local operators of dimension $D=2 n$ ($\geq 4$). The terms ${\cal C}_n a(Q^2)$ in the Wilson coefficients turn out to be negligible. Therefore, the OPE of the (full) Adler function ${\cal D}(Q^2)$, which is the logarithmic derivative of the correlator, is
\bea
{\cal D}(Q^2) &\equiv&  - 2 \pi^2 \frac{d \Pi_{{\rm th}}(Q^2)}{d \ln Q^2} 
 =  1 + d(Q^2)_{(D=0)} + 2 \pi^2 \sum_{n \geq 2} \frac{ n \langle O_{2n} \rangle_{V+A}}{(Q^2)^n}.
\label{Adlfull}
\eea
If we apply integration by parts on the right-hand side of the sum rule (\ref{sr1}), we obtain
\be
\int_0^{\sigma_{\rm max}} d \sigma g(-\sigma) \omega_{\rm exp}(\sigma)  =
-\frac{i}{2 \pi}   \oint_{|Q^2|=\sigma_{\rm max}}
\frac{d Q^2}{Q^2} {\cal D}(Q^2) G(Q^2) ,
\label{sr2}
\ee
where $G(Q^2)$ is such that $d G(Q^2)/d Q^2 = g(Q^2)$ and $G(-\sigma_{\rm max})=0$
\be
G(Q^2)= \int_{-\sigma_{\rm max}}^{Q^2} d Q^{'2} g(Q^{'2}).
\label{GQ2} \ee
The Borel-Laplace sum rules are those where
\be
g(Q^2) = \frac{1}{M^2} \exp \left( \frac{Q^2}{M^2} \right),
\label{gQ2} \ee
and $M^2$ is a chosen complex Borel scale, and for convenience (see later) we equate only the real parts of the integrals
\be
{\rm Re} B_{\rm exp}(M^2) =  {\rm Re} B_{\rm th}(M^2) \ ,
\label{sr}
\ee
where
\be
B_{\rm exp}(M^2) = \frac{1}{M^2} \int_0^{\sigma_{\rm max}} d \sigma \exp(-\sigma/M^2) \omega_{\rm exp}(\sigma)
\label{Bexp} \ee
and
\bes
\label{Bth}
\bea
B_{\rm th}(M^2)  & = & - \frac{i}{2 \pi}
\int_{\phi=-\pi}^{\pi} \frac{d Q^2}{Q^2} {\cal D}(Q^2)
\left[ \exp( Q^2/M^2) -  \exp( -\sigma_{\rm max}/M^2) \right] {\big |}_{Q^2 = \sigma_{\rm max} \exp(i \phi)}
\label{Btha}
\\
& = &  \left( 1 - \exp(-\sigma_{\rm max}/M^2) \right)
+ B_{\rm th}(M^2)_{(D=0)} + 2 \pi^2 \sum_{n \geq 2}
 \frac{ \langle O_{2n} \rangle_{V+A}}{ (n-1)! \; (M^2)^n} \ .
\label{Bthb}
\eea \ees
In Eq.~(\ref{Bthb}), the OPE expression (\ref{Adlfull}) was taken into account,
and the $D=0$ part of $B_{\rm th}(M^2)$ is
\bea
B_{\rm th}(M^2)_{(D=0)} &=&
\frac{1}{2 \pi}\int_{-\pi}^{\pi}
d \phi \; d(Q^2\!=\!\sigma_{\rm max} e^{i \phi})_{(D=0)} \left[ 
\exp \left( \frac{\sigma_{\rm max} e^{i \phi}}{M^2} \right) -
\exp \left( - \frac{\sigma_{\rm max}}{M^2} \right) \right] \ .
\label{BD0}
\eea

\subsection{Borel-Laplace sum rules: numerical results in V+A channel}
\label{subs:num}

As mentioned, in contrast to the sum rule analysis in Ref.~\cite{3dAQCD}, we now use in the expression (\ref{BD0}) for $d(Q^2)_{(D=0)}$ the renormalon-motivated resummation (\ref{dresA}) [and not the truncated series (\ref{dD0TPSA})]. Further, the $D$($\equiv 2n$) $=2$ contribution in the OPEs (\ref{Adlfull}) and (\ref{BD0}) is negligible because the effects ${\cal O}(m^2_{u,d})$ are negligible as mentioned earlier.

In practice, we include in the OPE only two terms beyond the leading-twist, i.e., $D=4$ and $D=6$ terms. One reason for this is practical. Namely, if $M^2=|M^2| \exp(i \Psi)$, it is straightforward to see that the contribution of $D=2 n$ to ${\rm Re} B_{\rm th}(M^2)$ is proportional to $\cos( i D \Psi/2)$. Therefore, when $\Psi=\pi/6$, the condensates with $D=6,18,\ldots$ do not contribute; when $\Psi=\pi/4$, the condensates with $D=4, 12, \ldots$ do not contribute to ${\rm Re} B_{\rm th}(M^2)$. Hence, when we have only two $D>0$ terms ($D=4,6$), the choice $\Psi=\pi/6$ eliminates the $D=6$ term and we have only one parameter $\langle O_{4} \rangle_{V+A}$ to fit; the choice $\Psi=\pi/4$ eliminates the $D=4$ term and we have only one parameter $\langle O_{6} \rangle_{V+A}$ to fit. The other practical reason to include only two $D>0$ terms is that in Eq.~(\ref{Bthb}) the sum over the condensates is proportional to $\langle O_{2n} \rangle_{V+A}/((n-1)! M^{2n})$; since we will have in our fit procedure $|M^2| \sim 1 \ {\rm GeV}^2$, the terms with higher $n$ are suppressed by the effect of the inverse factorial $(n-1)!$ provided that the values of the condensates $\langle O_{2n} \rangle_{V+A}$ (in units of ${\rm GeV}^{2 n}$) do not increase significantly with increasing $n$. We will see that the latter is true for $n=2, 3$ in some of the considered cases of interest.

We wish to point out yet another observation about the Borel-Laplace transforms $B(M^2)$: as seen in Eq.~(\ref{Bexp}), lower values of $|M^2|$ give more weight to lower values of $\sigma$, i.e., more weight to the IR-regime.

When we perform the fits in the way described in Ref.~\cite{3dAQCD}, at $\Psi=\pi/6$ and $\Psi=\pi/4$, we obtain, for each chosen value of the input parameters $\alpha_s(M_Z^2;\MSbar)$ and $r_{\tau, {\rm th}}^{(D=0)}$, a specific central value and the experimental variation, separately for the OPAL and the ALEPH data. For example, in the case of the input parameters $\alpha_s(M_Z^2;\MSbar)=0.1177$ and $r_{\tau, {\rm th}}^{(D=0)}=0.200$ we obtain [for comparison, we also give the corresponding values obtained by the same approach but using $\MSbar$ pQCD $a(Q^2;\MSbar)$ instead of $\A(Q^2)$]
\bes
\label{resOPAL}
\bea
\langle O_4 \rangle_{{\rm V+A}}^{\rm (OPAL)} & = & (+3.3 \pm 2.9) \times 10^{-4} \ {\rm GeV}^4
\label{O4OPAL}
\\ 
\Rightarrow  \; \langle a G G \rangle^{\rm (OPAL)} &=&  (+3.97 \pm 1.71) \times 10^{-3} \ {\rm GeV}^4 ,
\label{aGGOPAL}
\\
\langle O_6 \rangle_{{\rm V+A}}^{\rm (OPAL)} & = & (+6.3 \pm 3.3) \times 10^{-4} \ {\rm GeV}^6.
\label{O6OPAL}
\eea
\bea
\langle O_4 \rangle_{{\rm V+A},\MSbar}^{\rm (OPAL)} & = & (+19.2 \pm 2.9) \times 10^{-4} \ {\rm GeV}^4, 
\label{O4MSOPAL}
\\
\langle O_6 \rangle_{{\rm V+A},\MSbar}^{\rm (OPAL)} & = & (-48.6 \pm 3.3) \ 10^{-4} {\rm GeV}^6. 
\label{O6MSOPAL}
\eea
\ees
\bes
\label{resALEPH}
\bea
\langle O_4 \rangle_{{\rm V+A}}^{\rm (ALEPH)} & = & (+2.2 \pm 1.2) \times 10^{-4} \ {\rm GeV}^4
\label{O4ALEPH}
\\ 
\Rightarrow  \; \langle a G G \rangle^{\rm (ALEPH)} &=&  (+3.32 \pm 0.74) \times 10^{-3} \ {\rm GeV}^4 ,
\label{aGGALEPH}
\\
\langle O_6 \rangle_{{\rm V+A}}^{\rm (ALEPH)} & = & (+8.5 \pm 1.4) \times 10^{-4} \ {\rm GeV}^6.
\label{O6ALEPH}
\eea
\bea
\langle O_4 \rangle_{{\rm V+A},\MSbar}^{\rm (ALEPH)} & = & (+15.5 \pm 1.2) \times 10^{-4} \ {\rm GeV}^4, 
\label{O4MSALEPH}
\\
\langle O_6 \rangle_{{\rm V+A},\MSbar}^{\rm (ALEPH)} & = & (-41.5 \pm 1.4) \ 10^{-4} {\rm GeV}^6. 
\label{O6MSALEPH}
\eea
\ees
These values were obtained by the same kind of fit as in Ref.~\cite{3dAQCD}, and we refer for details to that reference, including the determination of the above (experimental) uncertainties. It turns out that these uncertainties displayed in Eqs.~(\ref{resOPAL})-(\ref{resALEPH}) remain practically unchanged when the input values of  $\alpha_s(M_Z^2;\MSbar)$ and $r_{\tau, {\rm th}}^{(D=0)}$ are varied as in Table \ref{tabres3d}. We note that in the ALEPH case the experimental uncertainties are significantly smaller, because the values of the covariance matrix of the measured spectral function values $\omega_{\rm exp}(\sigma)$ are significantly smaller in the ALEPH case than in the OPAL case (cf.~Appendix C of \cite{3dAQCD} for details). 

In Table \ref{tabO4O6} we display the extracted central values of the condensates for various cases of the input parameters $\alpha_s(M_Z^2;\MSbar)$ and $r_{\tau, {\rm th}}^{(D=0)}$, i.e., for the cases appearing in Table \ref{tabres3d}.
\begin{table}
\caption{The central extracted values of the condensates $\langle O_D \rangle_{V+A}$ ($D=4,6$), in units of ${\rm GeV}^D$, for various values of the input parameters  $\alpha_s(M_Z^2;\MSbar)$ and $r_{\tau, {\rm th}}^{(D=0)}$ (in the same order as in Table \ref{tabres3d}). The associated uncertainties are $\delta \langle O_D \rangle_{V+A}({\rm exp})$ given in Eqs.~(\ref{resOPAL})-(\ref{resALEPH}).}
\label{tabO4O6}  
\begin{ruledtabular}
  \begin{tabular}{ll|rr|rr}
    ${\overline \alpha}_s(M_Z^2)$ &  $r^{(D=0)}_{\tau, {\rm th}}$ & $\langle O_4 \rangle_{V+A}^{({\rm OPAL})}$ & $\langle O_4 \rangle_{V+A}^{({\rm ALEPH})}$ & $\langle O_6 \rangle_{V+A}^{({\rm OPAL})}$ & $\langle O_6 \rangle_{V+A}^{({\rm ALEPH})}$ 
\\
\hline
$0.1184$ & 0.203 & -0.00098 & -0.00107 & +0.00133 & +0.00155
 \\
$0.1181$ & 0.201 & -0.00059 & -0.00068 & +0.00106 & +0.00128
\\
 $0.1181$ & 0.203 &  -0.00023 & -0.00033 & +0.00104 & +0.00126
 \\
 \hline
 $0.1180$ & 0.202 &  -0.00015 & -0.00026 & +0.00095 & +0.00118
 \\
 $0.1180$ & 0.203 &  +0.00018 & +0.00008 & +0.00085 & +0.00107
 \\
 $0.1179$ & 0.201 &  -0.00008 & -0.00018 & +0.00087 & +0.00109
 \\
$0.1179$ & 0.202 &  +0.00023 & +0.00012 & +0.00078 & +0.00100
 \\ 
 $0.1178$ & 0.200 & -0.00000 & -0.00011 & +0.00078 & +0.00100
 \\
 $0.1178$ & 0.201 & +0.00028 & +0.00017 & +0.00070 & +0.00093
 \\
 $0.1177$ & 0.199 & +0.00007 & -0.00004 & +0.00069 & +0.00091
 \\
$0.1177$ & 0.200 & +0.00033 & +0.00022 & +0.00063 & +0.00085
 \\
$0.1176$ & 0.199 & +0.00038 & +0.00027 & +0.00055 & +0.00077
 \\
$0.1175$ & 0.197 & +0.00021 & +0.00010 & +0.00051 & +0.00074
 \\
$0.1175$ & 0.198 & +0.00044 & +0.00033 & +0.00047 & +0.00069
 \\
 $0.1175$ & 0.1988 & +0.00088 & +0.00077 & +0.00029 & +0.00051
 \\ 
$0.1174$ & 0.197 & +0.00049 & +0.00039 & +0.00039 & +0.00061
 \\
 $0.1173$ & 0.196 & +0.00055 & +0.00044 & +0.00031 & +0.00053
 \\
  $0.1173$ & 0.197 & +0.00098 & +0.00087 & +0.00015 & +0.00037
 \\
 $0.1172$ & 0.195 & +0.00061 & +0.00050 & +0.00023 & +0.00045
 \\
 $0.1172$ & 0.196 & +0.00096 & +0.00085 & +0.00011 & +0.00034
 \\
$0.1171$ & 0.195 & +0.00097 & +0.00086 & +0.00006 & +0.00028
 \\
$0.1170$ & 0.194 & +0.00099 & +0.00088 & -0.00000 & +0.00022
 \\  
$0.1169$ & 0.193 & +0.00103 & +0.00092 & -0.00007 & +0.00015
 \\   
\hline
$0.1181$ & 0.199 &  -0.00082 & -0.00091 & +0.00100 & +0.00121
\\
$0.1184$ & 0.201 &  -0.00121 & -0.00130 & +0.00126 & +0.00147
\\
$0.1184$ & 0.199 &  -0.00143 & -0.00153 & +0.00119 & +0.00140
\\
$0.1189$ & 0.201 &  -0.00223 & -0.00233 & +0.00157 & +0.00179 
\end{tabular}
\end{ruledtabular}
\end{table}

The numerical results are somewhat different from those obtained in Ref.~\cite{3dAQCD}, primarily (a) due to different values of $\alpha_s(M_Z^2;\MSbar) = 0.1179 \pm 0.0010$, \cite{PDG2019};\footnote{In \cite{3dAQCD} we used $\alpha_s(M_Z^2;\MSbar) = 0.1185 \pm 0.0004$.} and (b) due to the already mentioned different (improved) evaluation of the Adler function $d(Q^2)_{(D=0)}$ Eq.~(\ref{dresA}) as opposed to the truncated series (\ref{dD0TPSA}). To a lesser degree, the differences appear due to the earlier mentioned use of the five-loop $\MSbar$ beta function \cite{5lMSbarbeta} and the corresponding four-loop quark threshold matching \cite{4lquarkthresh}, as opposed to the four-loop $\MSbar$ beta function and the corresponding three-loop quark threshold matching.

\begin{figure}[htb] 
\begin{minipage}[b]{.49\linewidth}
  \centering\includegraphics[width=88mm]{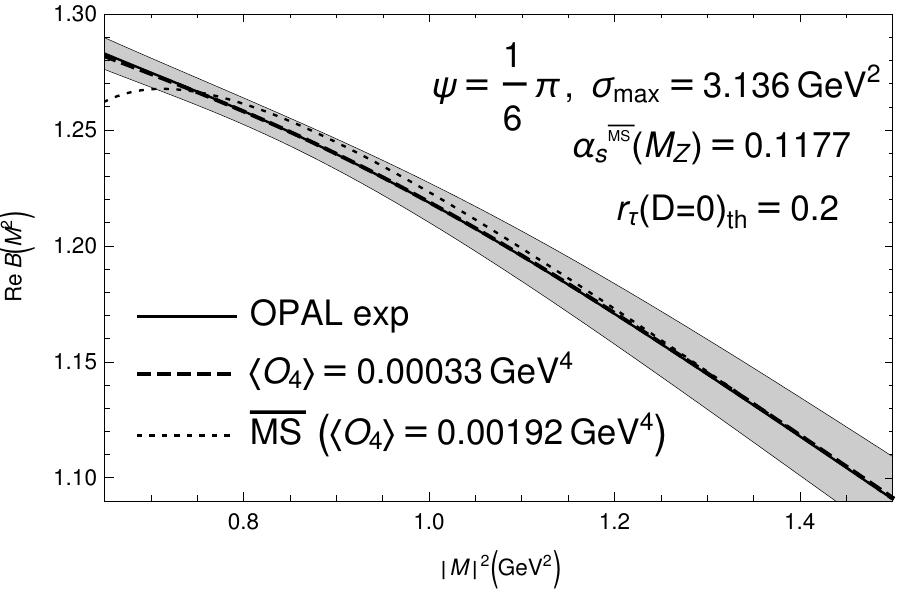}
  \end{minipage}
\begin{minipage}[b]{.49\linewidth}
  \centering\includegraphics[width=88mm]{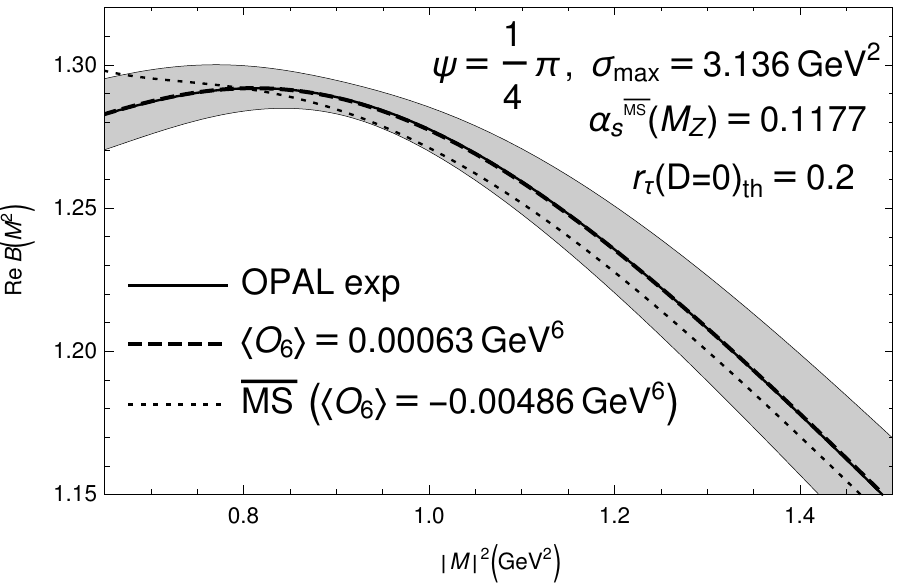}
\end{minipage}
\vspace{-0.2cm}
\caption{\footnotesize Borel-Laplace transforms ${\rm Re} B(M^2)$ along the rays $M^2 = |M^2| \exp(i \Psi)$ with $\Psi=\pi/6$ (left-hand side) and $\Psi = \pi/4$ (right-hand side), as a function of $|M^2|$, fitted to the OPAL data. The values of ${\rm Re} B(M^2)$ obtained from the (OPAL) experimental results are represented as the grey band, the central experimental values as thick solid curve.}
\label{BTOPPi6Pi4}
\end{figure}
\begin{figure}[htb] 
  \centering\includegraphics[width=90mm]{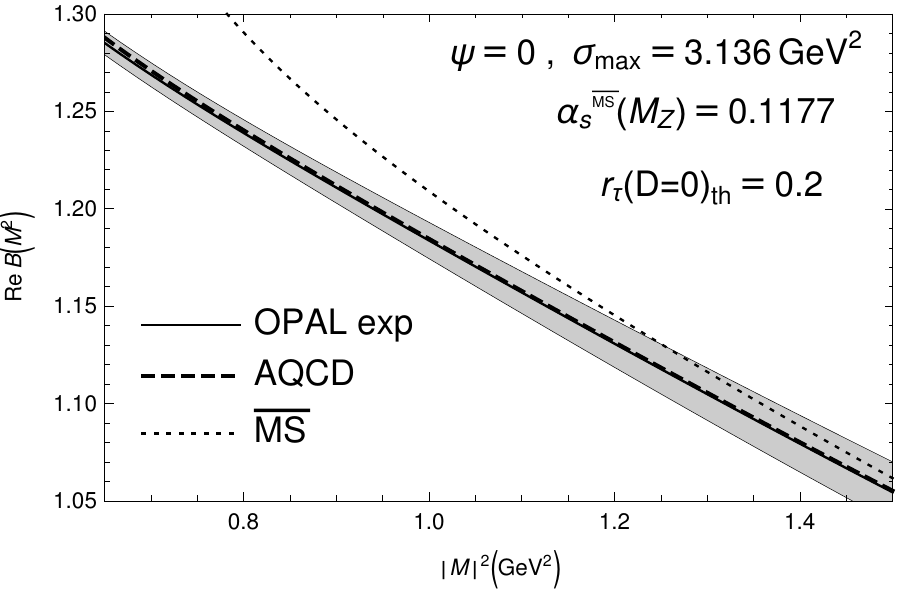}
\vspace{-0.3cm} 
 \caption{\footnotesize  Analogous to the previous Figures \ref{BTOPPi6Pi4}, but now the Borel-Laplace transforms $B(M^2)$ are for real $M^2 > 0$.}
\label{BTOP0}
 \end{figure}
We present in Figs.~\ref{BTOPPi6Pi4} the results of the Borel-Laplace sum rule fit to the OPAL data for $M^2=|M^2| \exp(i \Psi)$ with $\Psi=\pi/6$ and $\pi/4$, for the central case of $\alpha_s(M_Z^2;\MSbar) = 0.1177$ and $r_{\tau, {\rm th}}^{(D=0)}=0.200$. Using the central condensate values obtained through these fits, in Fig.~\ref{BTOP0} we present the results of the sum rule for $\Psi=0$, where the theoretical curve now presents the prediction (not fit).

\begin{figure}[htb] 
\begin{minipage}[b]{.49\linewidth}
  \centering\includegraphics[width=88mm]{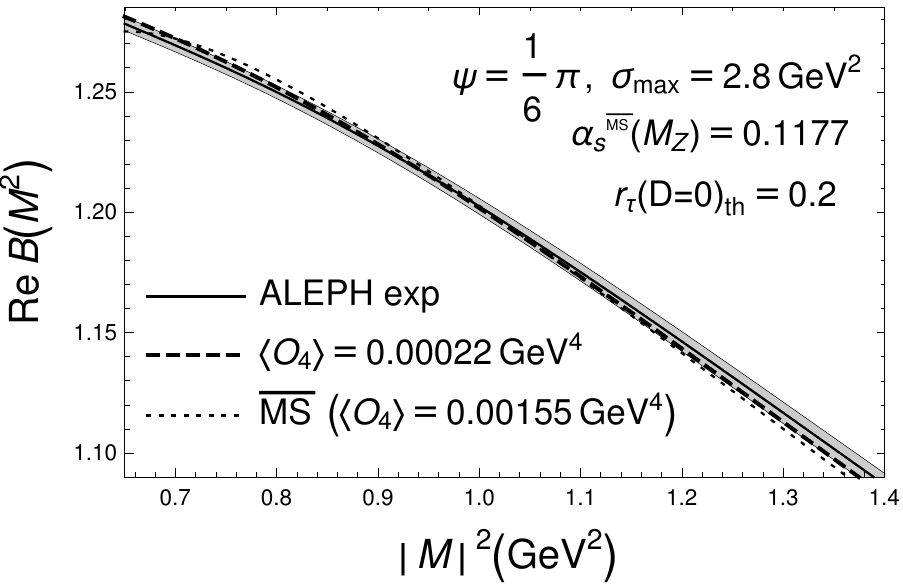}
  \end{minipage}
\begin{minipage}[b]{.49\linewidth}
  \centering\includegraphics[width=88mm]{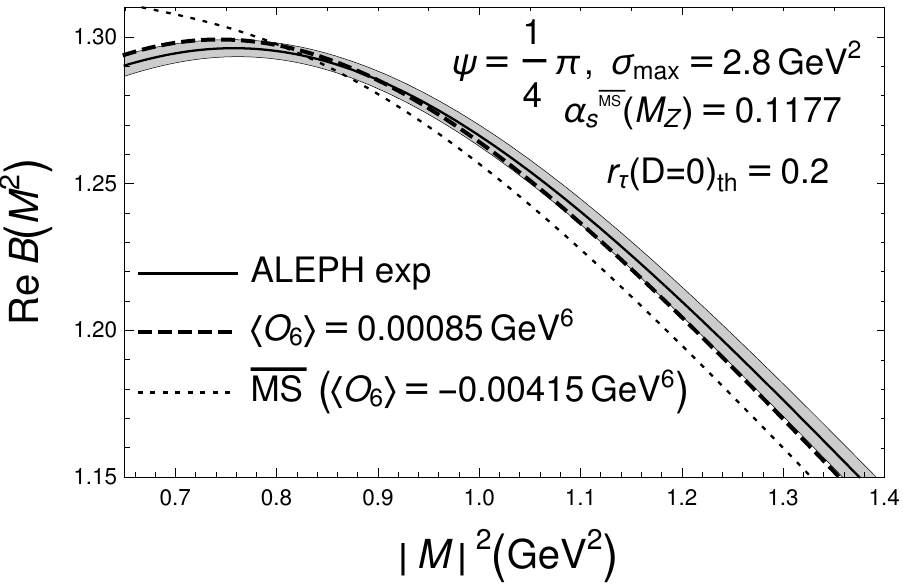}
\end{minipage}
\vspace{-0.2cm}
\caption{Borel-Laplace transforms ${\rm Re} B(M^2)$ along the rays $M^2 = |M^2| \exp(i \Psi)$ with $\Psi=\pi/6$ (left-hand) and $\Psi = \pi/4$ (right-hand), as a function of $|M^2|$, fitted to the ALEPH data.  The values of ${\rm Re} B(M^2)$ obtained from the (ALEPH) experimental results are represented as the grey band, the central experimental values as thick solid curve.}
\label{BTALPi6Pi4}
\end{figure}

\begin{figure}[htb] 
  \centering\includegraphics[width=90mm]{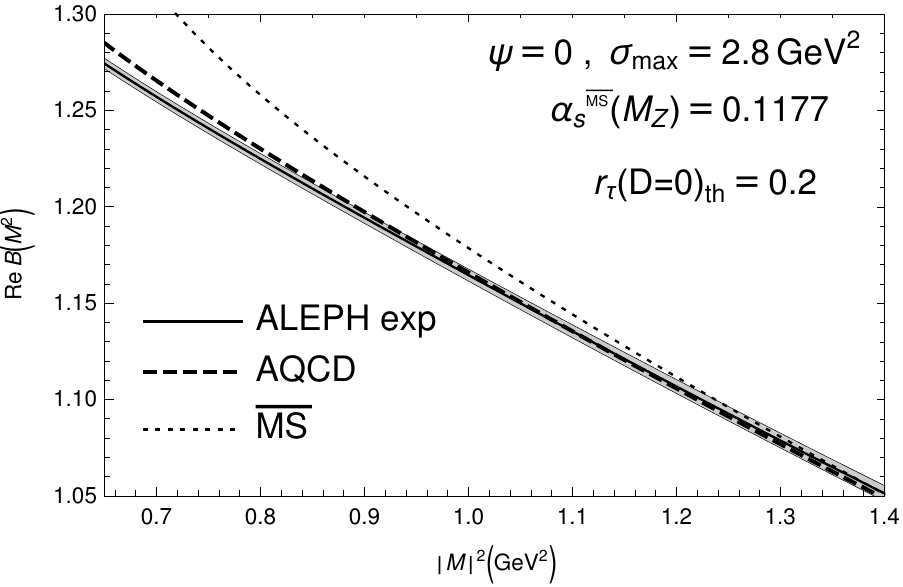}
  \vspace{-0.3cm} 
 \caption{Analogous to the previous Figures \ref{BTALPi6Pi4}, but now the Borel-Laplace transforms $B(M^2)$ are for real $M^2 > 0$.}
\label{BTAL0}
 \end{figure}

In Figs.~\ref{BTALPi6Pi4} and \ref{BTAL0} we present the analogous curves when the ALEPH data are used instead\footnote{\textcolor{black}{Figs.~\ref{BTALPi6Pi4} and \ref{BTAL0} were presented also in the short version of this work Ref.~\cite{shortv}.}}

All Figs.~\ref{BTOPPi6Pi4}-\ref{BTAL0} include, for comparison, the results for the $\MSbar$ case, i.e., when in the $D=0$ Adler function Eq.~(\ref{dresA}) we replace $\A(Q^{'2}) \mapsto a(Q^{'2};\MSbar)$ (with $N_f=3$) corresponding to the strength value $\alpha_s(M_Z^2;\MSbar)=0.1177$, i.e., with the best fit condensate values Eqs.~(\ref{O4MSOPAL})-(\ref{O6MSOPAL}) and (\ref{O4MSALEPH})-(\ref{O6MSALEPH}).

As can be seen from Figs.~\ref{BTOPPi6Pi4}-\ref{BTAL0}, the fit results with $\A$QCD are considerably better than those with $\MSbar$ pQCD coupling. This can be seen also by evaluating the $\chi^2$ quality parameters (which were minimized by the fit). The parameters $\chi^2$ were evaluated by dividing the considered interval $0.65 \ {\rm GeV}^2 \leq |M^2| \leq 1.50 \ {\rm GeV}^2$ in $n=85$ equally long intervals
\be
\chi^2(\Psi) = \frac{1}{n} \sum_{\alpha=0}^n \left( {\rm Re} B_{\rm th} (M_{\alpha}^2) - {\rm Re} B_{\rm exp} (M_{\alpha}^2) \right)^2 ,
\label{chi2}
\ee
where $M_{\alpha}^2 = |M_{\alpha}|^2 \exp(i \Psi)$ with $\Psi$ fixed ($\pi/6$; $\pi/4$; $0$) and $|M_{\alpha}|^2$ are the $n+1$ ($=86$) equidistant points in the mentioned $|M|^2$ interval $[0.65,1.50] \ {\rm GeV}^2$: $|M_0|^2=0.65 \ {\rm GeV}^2$, $|M_1|^2=0.66 \ {\rm GeV}^2$, ...,$|M_{85}|^2=1.50 \ {\rm GeV}^2$. ${\rm Re} B_{\rm exp} (M_{\alpha}^2)$ in Eq.~(\ref{chi2}) are the values using the central experimental data (the central experimental solid curves in Figs.~\ref{BTOPPi6Pi4}-\ref{BTAL0}).\footnote{In Figs.~\ref{BTALPi6Pi4}-\ref{BTAL0} we displayed the curves in the shorter $|M|^2$-interval $[0.65,1.40] \ {\rm GeV}^2$, for better visibility, although the fits were performed in the mentioned wider interval $[0.65,1.50] \ {\rm GeV}^2$.}
\begin{table}
\caption{The values of the quality parameters $\chi^2(\Psi)$ for $\A$QCD and $\MSbar$ pQCD, for OPAL and ALEPH data, for $\Psi=\pi/6$, $\pi/4$ and $\Psi=0$. For the theoretical curves, the input strength parameter $\alpha_s(M_Z^2;\MSbar)=0.1177$ was used, and for $\A$QCD in addition the input parameter value $r_{\tau, {\rm th}}^{(D=0)}=0.200$ was taken. The corresponding parameters of the experimental deviations are included for comparison.}
\label{tabchi2c}
 \begin{ruledtabular}
   \begin{tabular}{l|lll|lll|lll}
  data & $\chi^2(\pi/6)$ & $\chi^2(\pi/6)_{\MSbar}$ & $\chi^2(\pi/6)_{{\rm exp}}$ &  $\chi^2(\pi/4)$ & $\chi^2(\pi/4)_{\MSbar}$ & $\chi^2(\pi/4)_{{\rm exp}}$ &  $\chi^2(0)$ & $\chi^2(0)_{\MSbar}$ & $\chi^2(0)_{{\rm exp}}$
  \\
\hline
OPAL & $1.7 \times 10^{-7}$ & $2.4 \times 10^{-5}$ & $1.4 \times 10^{-4}$ & $4.4 \times 10^{-8}$ & $5.1 \times 10^{-5}$ & $2.0 \times 10^{-4}$ & $1.4 \times 10^{-6}$ &  $1.6 \times 10^{-3}$ & $1.2 \times 10^{-4}$
\\
ALEPH & $8.9 \times 10^{-6}$ & $2.7 \times 10^{-5}$ & $1.4 \times 10^{-5}$ & $2.3 \times 10^{-5}$ & $1.7 \times 10^{-4}$ & $2.0 \times 10^{-5}$  & $1.6 \times 10^{-5}$ & $6.3 \times 10^{-4}$ & $1.2 \times 10^{-5}$
\end{tabular}
\end{ruledtabular}
\end{table}  
In Table \ref{tabchi2c} we present the values of this best fit quality parameter for OPAL and ALEPH data for the input values [$\alpha_s(M_Z^2;\MSbar) = 0.1177$ and $r_{\tau, {\rm th}}^{(D=0)}=0.200$]. There, we include the experimental values $\chi^2(\Psi)_{{\rm exp}}$ which were obtained by applying  the expression (\ref{chi2}) to the difference between the central experimental curve and the upper (or lower) edge of the experimental band presented in Figs.~\ref{BTOPPi6Pi4}-\ref{BTAL0}.\footnote{The experimental bands in Figs.~\ref{BTOPPi6Pi4}-\ref{BTAL0} are obtained by the use of the full covariance matrix of the OPAL and ALEPH data, respectively, cf.~Appendix C of Ref.~\cite{3dAQCD}.} We recall that the values of $\chi^2(\Psi)$ for $\Psi=0$ are not obtained by the fit, but by the prediction when using for ${\rm Re} B_{\rm th}(M^2)$ for $M^2>0$ the condensate values obtained from the fits of the cases $\Psi=\pi/6$ and $\pi/4$. In Table  \ref{tabchi2c} we include the best fit quality parameters for the case when the $\MSbar$ pQCD coupling ${\bar a}$ is used.

In Tables \ref{tabchi2OP} and \ref{tabchi2AL} we show the best fit quality parameters $\chi^2(\Psi)$ for the various considered cases of the input parameters $\alpha_s(M_Z^2;\MSbar)$ and $r_{\tau, {\rm th}}^{(D=0)}$, for OPAL and ALEPH data, respectively. In these Tables, we include the ratios
\be
{\rm rat}_{\chi^2}(\Psi) \equiv \frac{\chi^2(\Psi)}{\chi^2(\Psi)_{{\rm exp}}},
\label{ratPsi} \ee
which can be considered, in comparisons of the different cases, as the relative quality fit parameters.
\begin{table}
  \caption{The values of the Borel-Laplace best fit quality parameters $\chi^2(\Psi)$ for $\A$QCD, for OPAL data, for $\Psi=\pi/6$, $\pi/4$ and $\Psi=0$. The values are for various input parameters  $\alpha_s(M_Z^2;\MSbar)$ and $r_{\tau, {\rm th}}^{(D=0)}$ (in the same order as in Table \ref{tabres3d}). Included are also the values of the relative quality fit ratios ${\rm rat}_{\chi^2}(\Psi)$, cf.~Eq.~(\ref{ratPsi}).}
\label{tabchi2OP}
 \begin{ruledtabular}
   \begin{tabular}{ll|rr|rr|rr}
 ${\overline \alpha}_s(M_Z^2)$ &  $r^{(D=0)}_{\tau, {\rm th}}$ &  $\chi^2(\pi/6)$ & ${\rm rat}_{\chi^2}(\pi/6)$ &  $\chi^2(\pi/4)$ & ${\rm rat}_{\chi^2}(\pi/4)$ &  $\chi^2(0)$ & ${\rm rat}_{\chi^2}(0)$
 \\
\hline
$0.1184$ & 0.203 & $9.6 \times 10^{-7}$ & 0.007 & $3.5 \times 10^{-7}$ & 0.002 & $4.6 \times 10^{-6}$ & 0.039
 \\
$0.1181$ & 0.201 & $1.5 \times 10^{-7}$ & 0.001 & $1.7 \times 10^{-7}$ & 0.001 & $2.4 \times 10^{-6}$ & 0.020
\\
 $0.1181$ & 0.203 & $1.3 \times 10^{-7}$ & 0.001 & $1.0 \times 10^{-6}$ & 0.005 & $4.7 \times 10^{-6}$ & 0.039 
 \\
 \hline
 $0.1180$ & 0.202 &  $6.5 \times 10^{-7}$ & 0.005 & $2.5 \times 10^{-7}$ & 0.001 & $3.5 \times 10^{-6}$ & 0.029
 \\
 $0.1180$ & 0.203 & $2.4 \times 10^{-6}$ & 0.017 & $2.9 \times 10^{-6}$ & 0.014 & $4.2 \times 10^{-6}$ & 0.035
 \\
 $0.1179$ & 0.201 & $2.2 \times 10^{-7}$ & 0.002 & $3.0 \times 10^{-8}$ & 0.000 & $2.5 \times 10^{-6}$ & 0.021
 \\
$0.1179$ & 0.202 &  $1.4 \times 10^{-6}$ & 0.010 & $1.3 \times 10^{-6}$ & 0.006 & $3.1 \times 10^{-6}$ & 0.026
 \\ 
 $0.1178$ & 0.200 & $2.3 \times 10^{-8}$ & 0.000 & $3.5 \times 10^{-7}$ & 0.002 & $1.7 \times 10^{-6}$ & 0.014
 \\
 $0.1178$ & 0.201 & $6.2 \times 10^{-7}$ & 0.005 & $3.3 \times 10^{-7}$ & 0.002 & $2.1 \times 10^{-6}$ & 0.018
 \\
 $0.1177$ & 0.199 & $6.8 \times 10^{-8}$ & 0.000 & $1.2 \times 10^{-6}$ & 0.006 & $1.0 \times 10^{-6}$ & 0.009
 \\
$0.1177$ & 0.200 & $1.7 \times 10^{-7}$ & 0.001 & $4.4 \times 10^{-8}$ & 0.000 & $1.4 \times 10^{-6}$ & 0.011
 \\
$0.1176$ & 0.199 & $7.1 \times 10^{-9}$ & 0.000 & $3.9 \times 10^{-7}$ & 0.002 & $7.8 \times 10^{-7}$ & 0.006
 \\
$0.1175$ & 0.197 & $7.1 \times 10^{-7}$ & 0.005 & $4.0 \times 10^{-6}$ & 0.020 & $2.9 \times 10^{-7}$ & 0.002
 \\
$0.1175$ & 0.198 & $1.2 \times 10^{-7}$ & 0.001 & $1.4 \times 10^{-6}$ & 0.007 & $4.0 \times 10^{-7}$ & 0.003
 \\
 $0.1175$ & 0.1988 & $2.9 \times 10^{-7}$ & 0.002 & $1.9 \times 10^{-7}$ & 0.001 & $4.1 \times 10^{-7}$ & 0.003
 \\ 
$0.1174$ & 0.197 & $5.0 \times 10^{-7}$ & 0.004 & $2.9 \times 10^{-6}$ & 0.014 & $2.1 \times 10^{-7}$ & 0.002
 \\
 $0.1173$ & 0.196 & $1.1 \times 10^{-6}$ & 0.008 & $5.1 \times 10^{-6}$ & 0.025 & $2.1 \times 10^{-7}$ & 0.002
 \\
  $0.1173$ & 0.197 & $3.8 \times 10^{-8}$ & 0.000 & $7.0 \times 10^{-7}$ & 0.003 & $9.7 \times 10^{-9}$ & 0.000
 \\
 $0.1172$ & 0.195 & $2.1 \times 10^{-6}$ & 0.015 & $7.8 \times 10^{-6}$ & 0.038 & $4.1 \times 10^{-7}$ & 0.003
 \\
 $0.1172$ & 0.196 & $4.2 \times 10^{-7}$ & 0.003 & $2.4 \times 10^{-6}$ & 0.012 & $4.9 \times 10^{-8}$ & 0.000
 \\
$0.1171$ & 0.195 & $1.2 \times 10^{-6}$ & 0.008 & $4.8 \times 10^{-6}$ & 0.023 & $2.8 \times 10^{-7}$ & 0.002
 \\
$0.1170$ & 0.194 & $2.2 \times 10^{-6}$ & 0.016 & $7.8 \times 10^{-6}$ & 0.038 & $7.0 \times 10^{-7}$ & 0.006
 \\  
$0.1169$ & 0.193 & $3.5 \times 10^{-6}$ & 0.026 & $1.1 \times 10^{-5}$ & 0.057 & $1.3 \times 10^{-6}$ & 0.011
 \\   
\hline
$0.1181$ & 0.199 & $8.1 \times 10^{-8}$ & 0.001 & $1.9 \times 10^{-6}$ & 0.009 & $7.5 \times 10^{-7}$ & 0.006   
\\
$0.1184$ & 0.201 & $1.3 \times 10^{-7}$ & 0.001 & $3.6 \times 10^{-7}$ & 0.002 & $2.0 \times 10^{-6}$ & 0.017   
\\
$0.1184$ & 0.199 & $1.1 \times 10^{-7}$ & 0.001 & $2.4 \times 10^{-6}$ & 0.012 & $4.8 \times 10^{-7}$ & 0.004    
\\
$0.1189$ & 0.201 & $8.8 \times 10^{-8}$ & 0.001 & $9.0 \times 10^{-7}$ & 0.004 & $1.3 \times 10^{-6}$ & 0.010 
\end{tabular}
\end{ruledtabular}
\end{table}
\begin{table}
 \caption{The same as in Table \ref{tabchi2OP}, but now for ALEPH data.}
 \label{tabchi2AL}
\begin{ruledtabular}
  \begin{tabular}{ll|rr|rr|rr}
 ${\overline \alpha}_s(M_Z^2)$ &  $r^{(D=0)}_{\tau, {\rm th}}$ &  $\chi^2(\pi/6)$ & ${\rm rat}_{\chi^2}(\pi/6)$ &  $\chi^2(\pi/4)$ & ${\rm rat}_{\chi^2}(\pi/4)$ &  $\chi^2(0)$ & ${\rm rat}_{\chi^2}(0)$
 \\
\hline
$0.1184$ & 0.203 & $8.3 \times 10^{-6}$ & 0.58 & $2.3 \times 10^{-5}$ & 1.15 & $2.5 \times 10^{-5}$ & 2.14
 \\
$0.1181$ & 0.201 & $1.1 \times 10^{-5}$ & 0.80 & $3.0 \times 10^{-5}$ & 1.15 & $2.1 \times 10^{-5}$ & 1.83
\\
 $0.1181$ & 0.203 &  $5.5 \times 10^{-6}$ & 0.39 & $1.5 \times 10^{-5}$ & 0.76 & $2.2 \times 10^{-5}$ & 1.90
 \\
 \hline
 $0.1180$ & 0.202 &  $7.0 \times 10^{-6}$ & 0.50 & $1.9 \times 10^{-5}$ & 0.95 & $2.0 \times 10^{-5}$ & 1.76
 \\
 $0.1180$ & 0.203 & $3.7 \times 10^{-6}$ & 0.26 & $1.1 \times 10^{-5}$ & 0.52 & $1.9 \times 10^{-5}$ & 1.64
 \\
 $0.1179$ & 0.201 &  $8.8 \times 10^{-6}$ & 0.62 & $2.4 \times 10^{-5}$ & 1.17 & $1.9 \times 10^{-5}$ & 1.65
  \\
$0.1179$ & 0.202 &  $5.2 \times 10^{-6}$ & 0.37 & $1.4 \times 10^{-5}$ & 0.70 & $1.8 \times 10^{-5}$ & 1.54
 \\ 
 $0.1178$ & 0.200 & $1.1 \times 10^{-5}$ & 0.76 & $2.9 \times 10^{-5}$ & 1.41 & $1.8 \times 10^{-5}$ & 1.55
 \\
 $0.1178$ & 0.201 & $6.9 \times 10^{-6}$ & 0.49 & $1.8 \times 10^{-5}$ & 0.90 & $1.7 \times 10^{-5}$ & 1.45
 \\
 $0.1177$ & 0.199 & $1.3 \times 10^{-5}$ & 0.93 & $3.4 \times 10^{-5}$ & 1.68 & $1.7 \times 10^{-5}$ & 1.47
 \\
$0.1177$ & 0.200 & $8.9 \times 10^{-6}$ & 0.63 & $2.3 \times 10^{-5}$ & 1.14 & $1.6 \times 10^{-5}$ & 1.37
 \\
$0.1176$ & 0.199 & $1.1 \times 10^{-5}$ & 0.78 & $2.8 \times 10^{-5}$ & 1.40 & $1.5 \times 10^{-5}$ & 1.31
 \\
$0.1175$ & 0.197 & $1.8 \times 10^{-5}$ & 1.31 & $4.7 \times 10^{-5}$ & 2.30 & $1.6 \times 10^{-5}$ & 1.37
 \\
$0.1175$ & 0.198 & $1.4 \times 10^{-5}$ & 0.96 & $3.4 \times 10^{-5}$ & 1.69 & $1.5 \times 10^{-5}$ & 1.27
 \\
 $0.1175$ & 0.1988 & $7.9 \times 10^{-6}$ & 0.56 & $2.0 \times 10^{-5}$ & 0.98 & $1.1 \times 10^{-5}$ & 0.97
 \\ 
$0.1174$ & 0.197 & $1.6 \times 10^{-5}$ & 1.15 & $4.1 \times 10^{-5}$ & 2.01 & $1.4 \times 10^{-5}$ & 1.23
 \\
 $0.1173$ & 0.196 & $1.9 \times 10^{-5}$ & 1.36 & $4.8 \times 10^{-5}$ & 2.36 & $1.4 \times 10^{-5}$ & 1.22
 \\
  $0.1173$ & 0.197 & $1.2 \times 10^{-5}$ & 0.85 & $3.7 \times 10^{-5}$ & 1.46 & $1.1 \times 10^{-5}$ & 0.93
 \\
 $0.1172$ & 0.195 & $2.3 \times 10^{-5}$ & 1.60 & $5.5 \times 10^{-5}$ & 2.73 & $1.4 \times 10^{-5}$ & 1.22
 \\
 $0.1172$ & 0.196 & $1.6 \times 10^{-5}$ & 1.10 & $3.8 \times 10^{-5}$ & 1.87 & $1.1 \times 10^{-5}$ & 0.98
 \\
$0.1171$ & 0.195 & $1.9 \times 10^{-5}$ & 1.35 & $4.6 \times 10^{-5}$ & 2.28 & $1.2 \times 10^{-5}$ & 1.02
 \\
$0.1170$ & 0.194 & $2.3 \times 10^{-5}$ & 1.61 & $5.5 \times 10^{-5}$ & 2.70 & $1.2 \times 10^{-5}$ & 1.06
 \\  
$0.1169$ & 0.193 & $2.7 \times 10^{-5}$ & 1.89 & $6.4 \times 10^{-5}$ & 3.15 & $1.3 \times 10^{-5}$ & 1.11
 \\   
\hline
$0.1181$ & 0.199 & $1.6 \times 10^{-5}$ & 1.11 & $4.2 \times 10^{-5}$ & 2.06 & $1.8 \times 10^{-5}$ & 1.54
\\
$0.1184$ & 0.201 & $1.1 \times 10^{-5}$ & 0.81 & $3.2 \times 10^{-5}$ & 1.57 & $2.0 \times 10^{-5}$ & 1.72
\\
$0.1184$ & 0.199 &  $1.6 \times 10^{-5}$ & 1.11 & $4.3 \times 10^{-5}$ & 2.10 & $1.6 \times 10^{-5}$ & 1.42
\\
$0.1189$ & 0.201 & $1.2 \times 10^{-5}$ & 0.84 & $3.4 \times 10^{-5}$ & 1.67 & $1.7 \times 10^{-5}$ & 1.49
\end{tabular}
\end{ruledtabular}
\end{table}
For example, we can consider ${\rm rat}_{\chi^2}(\Psi) \leq 2$ as a fit of acceptably good quality. The results in Table \ref{tabchi2OP} imply that the fit quality is in general very good for the OPAL data. However, for ALEPH data the fit quality is not always acceptably good, especially for the $\Psi=\pi/4$ Borel-Laplace sum rule.\footnote{When comparing the values of ${\rm rat}_{\chi^2}(\Psi)$ of ALEPH with OPAL data, we should also keep in mind that the experimental band width for Borel-Laplace transforms is in the OPAL case significantly wider than in the ALEPH case: $\chi^2(\pi/6)_{\rm exp}$, $\chi^2(\pi/4)_{\rm exp}$ and $\chi^2(0)_{\rm exp}$ are $1.38 \times 10^{-4}$, $2.03 \times 10^{-4}$, $1.20 \times 10^{-4}$ for OPAL data, respectively; and  $1.42 \times 10^{-5}$, $2.03 \times 10^{-5}$, $1.16 \times 10^{-5}$ for ALEPH data, respectively.}
  We can deduce from the results for ${\rm rat}_{\chi^2}(\pi/4)$ for ALEPH data in Table \ref{tabchi2AL} that the value of the parameter $r_{\tau, {\rm th}}^{(D=0)}$ must increase in order to decrease the value of ${\rm rat}_{\chi^2}(\pi/4)$ for ALEPH data.

We can now evaluate the values of the condensates obtained from combined OPAL and ALEPH data. In the next Section we will apply specific versions of the described $\A$QCD to the evaluation of an emblematic low-energy observable $a_{\mu}^{\rm had(1)}$, the hadronic vacuum polarization contribution to the anomalous magnetic moment of muon. $\A$QCD and information about its condensate values $\langle O_D \rangle_{V+A}$ ($D=4,6$), which will be used in the evaluation of this new observable, are determined by the ``input'' parameter values $\alpha_s(M_Z^2;\MSbar)$ and $r_{\tau,{\rm th}}^{(D=0)}$. Therefore, we will consider the values of the condensates $\langle O_D \rangle_{V+A}$ ($D=4,6$) from combined OPAL and ALEPH data for each of the specific cases given in Table \ref{tabres3d}. The central value of the condensate  $\langle O_D \rangle_{V+A}$ is taken as the arithmetic mean\footnote{According to Eq.~(\ref{avO}), we gave equal weight to the condensate values extracted from the OPAL and ALEPH data. Namely, although the experimental uncertainties of ALEPH data are smaller than those of OPAL data, the quality of the fit of our theoretical $\A$QCD curves to the central experimental curves is better in the case of OPAL data.} 
\be
\langle O_D \rangle = \frac{1}{2} \left( \langle O_D^{\rm (OPAL)} \rangle + \langle O_D^{\rm (ALEPH)} \rangle \right).
\label{avO} \ee
The final uncertainty $\sigma_O$ will then be obtained from quadrature of half of the difference and the experimental uncertainties of each of them
\bes
\label{sigavO}
\bea
\sigma_O & = & \left[ \left(\sigma_O^{\rm (OP-AL)}\right)^2 + \left( \sigma_{O,{\rm exp}} \right)^2 \right]^{1/2},
\label{sigavOa}
\\
{\rm where:} \;\; \sigma_O^{\rm (OP-AL)} & = & \frac{1}{2} {\big |} \langle O_D^{\rm (OPAL)} \rangle - \langle O_D^{\rm (ALEPH)} \rangle {\big |},
\quad \sigma_{O,{\rm exp}}=\frac{1}{2} \left[ (\sigma_{O,{\rm OP, exp}})^2 + (\sigma_{O,{\rm AL, exp}})^2 \right]^{1/2}.
\label{sigavOb} \eea \ees
The values of $\sigma_{O,{\rm OP, exp}}$ and $\sigma_{O,{\rm AL, exp}}$ are given in Eqs.~(\ref{resOPAL})-(\ref{resALEPH}), and they give [for fixed specific values of input parameters $\alpha_s(M_Z^2;\MSbar)$ and $r_{\tau,{\rm th}}^{(D=0)}$]
\bea
\sigma_{O4,{\rm exp}} = 0.00016 \ {\rm GeV}^4, \quad
\sigma_{O6,{\rm exp}} = 0.00018 \ {\rm GeV}^6.
\label{sigexp}
\eea
Further, $\sigma_{O4}^{\rm (OP-AL)}=0.00005 \ {\rm GeV}^4$ and  $\sigma_{O6}^{\rm (OP-AL)}=0.00011 \ {\rm GeV}^6$. These values, as well as the values in Eq.~(\ref{sigexp}), are approximately independent of the input values $\alpha_s(M_Z^2;\MSbar)$ and $r_{\tau,{\rm th}}^{(D=0)}$. Hence, we obtain
\be
\sigma_{O4}=0.00016 \ {\rm GeV}^4, \quad \sigma_{O6}=0.00021 \ {\rm GeV}^6,
\label{sigmaO} \ee
and both these values are independent of the input parameters, up to the digits displayed in Eq.~(\ref{sigmaO}).
This then gives, for the input parameters $\alpha_s(M_Z^2;\MSbar)=0.1177$ and $r_{\tau,{\rm th}}^{(D=0)}=0.200$, the combined OPAL+ALEPH values
\bes
\label{O11770200}
\bea
\langle O_4 \rangle_{V+A} &=&-0.00028 \pm 0.00016\ {\rm GeV}^4,
\label{O411770200} \\
\langle O_6 \rangle_{V+A} &=& +0.00074 \pm 0.00021 \ {\rm GeV}^6.
\label{O611770200}
\eea \ees
For the other choices of the input parameters, only the central values in Eqs.~(\ref{O11770200}) change accordingly, and can be obtained from the arithmetic averages of the OPAL and ALEPH values in Table \ref{tabO4O6}, cf.~Eq.~(\ref{avO}).

\subsection{Borel-Laplace sum rules: consistency checks with $r_{\tau}$}
\label{subs:cons}

Consistency checks of the obtained results for $\langle O_6 \rangle_{V+A}$ can be made by comparing the theoretical and the experimental values of the $r_{\tau}^{(D=0, \sigma_{\rm max})}$ quantity, where the effective mass $m^2_{\tau}$ ($=3.157 \ {\rm GeV}^2$) in Eq.~(\ref{rtaucont}) is replaced by the lower values $\sigma_{\rm max}=3.136 \ {\rm GeV}^2$ in the case of OPAL data and $\sigma_{\rm max}=2.80 \ {\rm GeV}^2$ in the case of ALEPH data. The theoretical and experimental expressions are
\bes
\label{rtau}
\bea
r^{(D=0, \sigma_{\rm max})}_{\tau, {\rm th}} &=& \frac{1}{2 \pi} \int_{-\pi}^{+ \pi}
d \phi \ (1 + e^{i \phi})^3 (1 - e^{i \phi}) \
d(Q^2=\sigma_{\rm max} e^{i \phi})_{(D=0)},
\label{rtauth} \\
r^{(D=0, \sigma_{\rm max})}_{\tau,{\rm exp}} &=&
\left[ 2 \int_{0}^{\sigma_{\rm max}} \frac{d \sigma}{\sigma_{\rm max}} \left( 1 - \frac{\sigma}{\sigma_{\rm max}} \right)^2 \left( 1 + 2  \frac{\sigma}{m_{\tau}^2} \right) \omega_{\rm exp}(\sigma) - 1 \right]
+ 12 \pi^2 \frac{\langle O_6 \rangle_{{\rm V+A}}}{\sigma^3_{\rm max}}
\label{rtauexp}
\eea \ees
For example, for the case of $\alpha_s(M_Z^2,\MSbar)=0.1177$ and $r^{(D=0)}_{\tau, {\rm th}}$ ($\equiv r^{(D=0, m_{\tau}^2)}_{\tau, {\rm th}}$) $=0.200$ we obtain\footnote{The uncertainty of $r^{(D=0, \sigma_{\rm max})}_{\tau, {\rm exp}}$ is obtained by using the full covariance matrix  of $\omega_{\rm exp}(\sigma)$ values in the integral in Eq.~(\ref{rtauexp}), and the uncertainty $\delta \langle O_6 \rangle_{V+A}({\rm exp})$ [cf.~Eqs.~(\ref{O6OPAL}) and (\ref{O6ALEPH})] in the last term in Eq.~(\ref{rtauexp}), and adding them in quadrature.}
\bes
\label{rt11770200}
\bea
r^{(D=0, \sigma_{\rm max})}_{\tau, {\rm th}}({\rm OPAL}) & = & 0.20072,
\qquad
r^{(D=0, \sigma_{\rm max})}_{\tau,{\rm exp}}({\rm OPAL}) = 0.20122 \pm 0.00586,
\label{rt11770200OP}
\\
r^{(D=0, \sigma_{\rm max})}_{\tau, {\rm th}}({\rm ALEPH}) & = & 0.21329,
\qquad
r^{(D=0, \sigma_{\rm max})}_{\tau,{\rm exp}}({\rm ALEPH}) = 0.21068 \pm 0.00271.
\label{rt11770200AL} \eea \ees
The main uncertainty in the experimental values above ($\pm 0.00572$ in the case of OPAL; $\pm 0.00260$ in the case of ALEPH) comes from the integral involving $\omega_{\rm exp}(\sigma)$. Eqs.~(\ref{rt11770200}) show that in the case of the input strength value $\alpha_s(M_Z^2;\MSbar)=0.1177$, the input value $r^{(D=0}_{\tau, {\rm th}}=0.200$ is consistent with the (OPAL and ALEPH) experimental data. In Table \ref{tabrtau} we present these results for all the ``input'' cases of $\A$QCD of Table \ref{tabres3d}, and include the deviation factors ('dev')
\bes
\label{dev}
\bea
{\rm dev}_{r_{\tau}}({\rm OPAL}) & \equiv & \left[ r^{(D=0, \sigma_{\rm max})}_{\tau, {\rm th}}({\rm OPAL}) - r^{(D=0, \sigma_{\rm max})}_{\tau,{\rm exp}}({\rm OPAL}) \right]/0.00586 \qquad \quad \;  (\sigma_{\rm max}=3.136 \ {\rm GeV}^2),
\label{devOP}
\\
{\rm dev}_{r_{\tau}}({\rm ALEPH}) & \equiv & \left[ r^{(D=0, \sigma_{\rm max})}_{\tau, {\rm th}}({\rm ALEPH}) - r^{(D=0, \sigma_{\rm max})}_{\tau,{\rm exp}}({\rm ALEPH}) \right]/0.00271  \qquad (\sigma_{\rm max}=2.800 \ {\rm GeV}^2).
\label{devAL}
\ea \ees
 \begin{table}
   \caption{Comparison of the theoretical and experimental values of $r_{\tau}^{(D=0,\sigma_{\rm max})}$ for various values of the input parameters $\alpha_s(M_Z^2;\MSbar)$ and $r_{\tau, {\rm th}}^{(D=0)}$ ($=r_{\tau}^{(D=0,m^2_{\tau})}$), in the same order as in Table \ref{tabres3d}. The cases of OPAL data ($\sigma_{\max}=3.136 \ {\rm GeV}^2$) and ALEPH data ($\sigma_{\max}=2.80 \ {\rm GeV}^2$) are given in separate columns. Included are the deviation factors Eqs.~(\ref{devOP})-(\ref{devAL}).}
\label{tabrtau}  
\begin{ruledtabular}
  \begin{tabular}{ll|rrr|rrr}
  ${\overline \alpha}_s(M_Z^2)$ &  $r^{(D=0)}_{\tau, {\rm th}}$ & OPAL: $\;\; r^{(D=0, \sigma_{\rm max})}_{\tau, {\rm th}}$ & $r^{(D=0, \sigma_{\rm max})}_{\tau,{\rm exp}}$ & ${\rm dev}_{r_{\tau}}$ & ALEPH: $\;\; r^{(D=0, \sigma_{\rm max})}_{\tau, {\rm th}}$ & $r^{(D=0, \sigma_{\rm max})}_{\tau,{\rm exp}}$ & ${\rm dev}_{r_{\tau}}$
\\
\hline
$0.1184$ & 0.203 & 0.20379 & 0.20391 & -0.02 & 0.21778 & 0.21443 & 1.24
 \\
$0.1181$ & 0.201 & 0.20177 & 0.20288 & -0.19 & 0.21540 & 0.21300 & 0.89
 \\
$0.1181$ & 0.203 &  0.20376 & 0.20282 & 0.16 & 0.21702 & 0.21291 & 1.52
 \\
\hline
$0.1180$ & 0.202 & 0.20275 & 0.20248 & 0.05 & 0.21594 & 0.21244 & 1.29
 \\
$0.1180$ & 0.203 & 0.20373 & 0.20208 & 0.28 & 0.21636 & 0.21189 & 1.65
 \\
$0.1179$ & 0.201 &  0.20175 & 0.20214 & -0.07 & 0.21486 & 0.21196 & 1.07
 \\
$0.1179$ & 0.202 & 0.20273 & 0.20180 & 0.16 & 0.21535 & 0.21149 & 1.42
 \\
$0.1178$ & 0.200 & 0.20074 & 0.20180 & -0.18 & 0.21379 & 0.21149 & 0.85
\\
$0.1178$ & 0.201 & 0.20172 & 0.20152 & 0.04 & 0.21432 & 0.21109 & 1.19
\\
$0.1177$ & 0.199 & 0.19974 & 0.20146 & -0.29 & 0.21271 & 0.21101 & 0.63
\\ 
$0.1177$ & 0.200 & 0.20072 & 0.20122 & -0.09 & 0.21329 & 0.21068 & 0.96
\\
$0.1176$ & 0.199 & 0.19972 & 0.20092 & -0.21 & 0.21225 & 0.21025 & 0.74
\\ 
$0.1175$ & 0.197 & 0.19773 & 0.20078 & -0.52 & 0.21056 & 0.21006 & 0.18
\\
$0.1175$ & 0.198 & 0.19872 & 0.20062 & -0.32 & 0.21121 & 0.20983 & 0.51
\\
$0.1175$ & 0.1988 & 0.19948 & 0.19992 & -0.08 & 0.21113 & 0.20885 & 0.84
\\
$0.1174$ & 0.197 & 0.19771 & 0.20031 & -0.44 & 0.21016 & 0.20940 & 0.28
\\
$0.1173$ & 0.196 & 0.19671 & 0.20000 & -0.56 & 0.20912 & 0.20896 & 0.06
\\
$0.1173$ & 0.197 & 0.19767 & 0.19939 & -0.29 & 0.20929 & 0.20810 & 0.44
\\
$0.1172$ & 0.195 & 0.19571 & 0.19969 & -0.68 & 0.20806 & 0.20853 & -0.17
\\
$0.1172$ & 0.196 & 0.19668 & 0.19925 & -0.44 & 0.20842 & 0.20790 & 0.19
\\
$0.1171$ & 0.195 & 0.19568 & 0.19904 & -0.57 & 0.20747 & 0.20761 & -0.05
\\
$0.1170$ & 0.194 & 0.19468 & 0.19880 & -0.70 & 0.20649 & 0.20727 & -0.29
\\
$0.1169$ & 0.193 & 0.19368 & 0.19854 & -0.83 & 0.20550 & 0.20690 & -0.52
\\
\hline
$0.1181$ & 0.199 &  0.19976 & 0.20264 & -0.49 & 0.21334 & 0.21264 & 0.26
\\
$0.1184$ & 0.201 &  0.20177 & 0.20363 & -0.32 & 0.21566 & 0.21405 & 0.59
\\
$0.1184$ & 0.199 &  0.19976 & 0.20337 & -0.62 & 0.21346 & 0.21367 & -0.08
\\
$0.1189$ & 0.201 &  0.20177 & 0.20486 & -0.53 & 0.21581 & 0.21577 & 0.01
\end{tabular}
\end{ruledtabular}
\end{table}
 For a given pair of values of the input parameters  $\alpha_s(M_Z^2;\MSbar)$ and  $r_{\tau,{\rm th}}^{(D=0)}$ ($\equiv r_{\tau,{\rm th}}^{(D=0,m_{\tau}^2)}$), we will require $|{\rm dev}_{r_\tau}| < 1.5$ for both OPAL and ALEPH data as the condition of passing the $r_{\tau}$-consistency check.
 
 Comparison of these results in Table \ref{tabrtau} for $\alpha_s(M_Z^2;\MSbar)=0.1177$ suggests that both values $r_{\tau,{\rm th}}^{(D=0)} \approx 0.199$ and $0.200$ pass the above $r_{\tau}$-consistency check, both for OPAL and ALEPH data; but in the next Section, we will see that $r_{\tau,{\rm th}}^{(D=0)} =0.200$ is preferred. And in the case of other $\alpha_s(M_Z^2;\MSbar)$ values, we will see that only restricted intervals of values $r_{\tau,{\rm th}}^{(D=0)}$ are preferred (of width $\approx 0.001$ or less).

\section{Hadronic vacuum polarization contribution to muon $g-2$}
\label{sec:amu}

After having constructed the $\A$ coupling, and determined the condensate values $\langle O_D \rangle_{V+A}$ from the physics of semihadronic decays of $\tau$ lepton, we now turn to a very-low-energy QCD observable, namely the leading order hadronic vacuum polarization (had(1)) contribution to the anomalous magnetic moment of $\mu$ lepton,  $(g_{\mu}/2-1)^{\rm had(1)} \equiv a_{\mu}^{\rm had(1)}$. This quantity can be deduced experimentally to a high precision from the available measurements of the cross section $e^+ e^- \to \gamma^{\ast} \to$ hadrons.\footnote{\textcolor{black}{In Ref.~\cite{AKR} this cross section was used to extract the Adler function ${\cal D}_V(Q^2)$. For theoretical high order perturbative contributions to this cross section, see e.g.~Refs.~\cite{ANR}.}} The recent values of $a_{\mu}^{\rm had(1)}$ extracted in this way are \cite{Davier:2019can}\footnote{\label{amuSM} The full SM value is $10^{10} \times a_{\mu}^{\rm (SM)} = 11 \ 659 \ 181.0 \ (4.3)$ \cite{amurev} [these authors took $10^{10} \times a_{\mu}^{\rm had(1)} = 693.1(4.0)$], while the directly measured ('dir.exp') value of the full $a_{\mu}$ is \cite{Brook} $10^{10} \times a_{\mu}^{\rm dir.exp} = 11 \ 659 \ 208.9 \ (6.3)$, which is $3.7 \sigma$ higher. The BMW Collaboration \cite{Borsanyi:2020mff} obtained recently from lattice calculation significantly higher values for the had(1)-contribution, $10^{10} \times a_{\mu; {\rm latt}}^{\rm had(1)} = 712.4 \pm 4.5$  \cite{Borsanyi:2020mff}, which would indicate that the deviation between $a_{\mu}^{\rm dir.exp}$ and $a_{\mu}^{\rm (SM)}$ reduces to about $1 \sigma$ and that no new physics beyond QCD (beyond SM) is needed for the explanation of the deviation. However, the results of lattice calculation by another group \cite{Lehner:2020crt} indicates that higher statistical uncertainties appear in $a_{\mu; {\rm latt}}^{\rm had(1)} $ than assumed in \cite{Borsanyi:2020mff}. This would then ease tension with the results Eq.~(\ref{amuhad1}) based on the measurements of  $e^+ e^- \to \gamma^{\ast} \to$ hadrons, but restore the tension between $a_{\mu}^{\rm dir.exp}$ and $a_{\mu}^{\rm (SM)}$. For some early estimates on $a_{\mu}^{\rm had(1)}$ see Refs.~\cite{amuHT,amuBR}. For some of the works on the prospects of specific new physics to explain the difference $a_{\mu}^{\rm dir.exp}-a_{\mu}^{\rm (SM)}$, we refer to Refs.~\cite{LPQ,SF,CSK}. The impact of $a_{\mu}^{\rm had(1)}$ on the global electroweak (EW) fit was investigated in \cite{Pass,CHMM}.}
\be
10^{10}   \times a_{\mu; {\rm exp}}^{\rm had(1)} \approx 694 \pm 4 \ .
\label{amuhad1}
\ee
The theoretical evaluation of this quantity involves the current-current correlation function $\Pi_{V}(Q^2) = \Pi^{(1)}_{V}(Q^2)$ [cf.~Eqs.~(\ref{PiJ})-(\ref{Pi1})]. For this reason, the full perturbative $+$ nonperturbative V-channel Adler function ${\cal D}_V(Q^2)$ should be evaluated
\bea
{\cal D}_{\rm V}(Q^2) &\equiv&  - 4 \pi^2 \frac{d \Pi_{\rm V}(Q^2)}{d \ln Q^2} 
= d(Q^2)_{D=0} + {\cal D}_{\rm V}(Q^2)^{\rm (NP)}
\nonumber\\
 &=&  d(Q^2)_{(D=0)} + 1 + 2 \pi^2 \sum_{n \geq 2}
 \frac{ n 2 \langle O_{2n} \rangle_{\rm V}}{(Q^2)^n}.
\label{DV}
\eea
The normalization here was chosen in the usual way, so that the perturbative ($D=0$) part is the same as in the (V+A)-channel case, and for the condensates we have the simple relations $\langle O_D \rangle_{V \pm A} = \langle O_D \rangle_{V} \pm \langle O_D \rangle_{A}$ [cf.~Eq.~(\ref{Adlfull})]. We recall that $D=0$ part is evaluated in the same way as previously, Eq.~(\ref{dresA}).

\begin{figure}[htb] 
\centering\includegraphics[width=70mm]{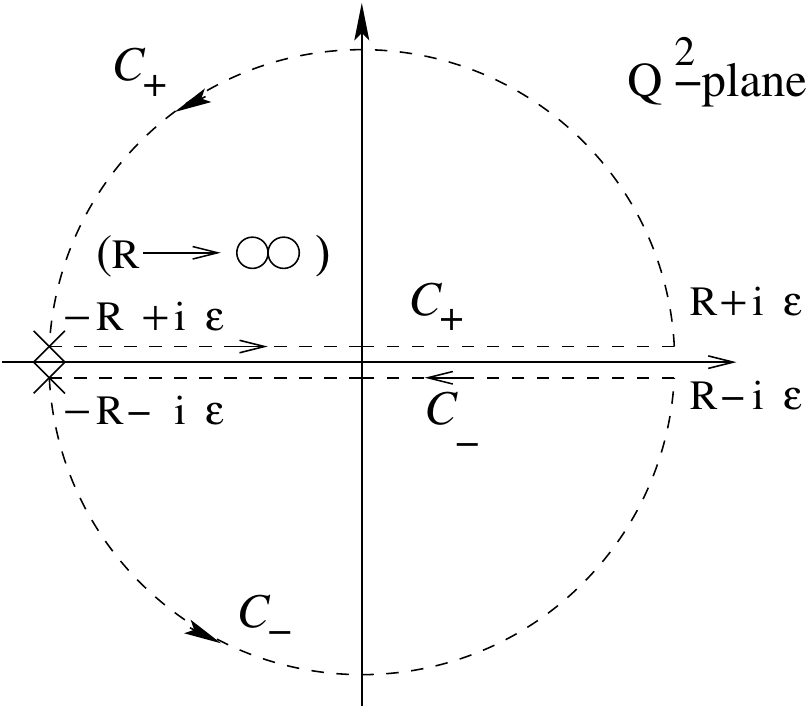}
\caption{\footnotesize The closed path $C_{+}+C_{-}$ for integration of $K(-Q^2) \Pi_V(Q^2)/Q^2$. The limit $R \to \infty$ is understood.}
\label{Figcontamu}
\end{figure}

The theoretical evaluation of $a_{\mu}^{\rm had(1)}$ is given by
\be
a_{\mu}^{\rm had(1)} = \frac{\alpha_{em}^2}{3 \pi^2} \int_0^{\infty} \frac{ds}{s} K(s) R_{\gamma,{\rm data}}(s),
\label{ahada} \ee
where
\bes
\label{KRkap}
\bea
K(s) &=&  \int_0^1 dx \frac{x^2 (1-x)}{x^2 + \frac{s}{m_{\mu}^2} (1-x)},
\label{Ks} \\
R_{\gamma,{\rm data}}(s) &=& 4 \pi k_f \; {\rm Im} \Pi_{\rm V}(-s-i \epsilon),
\label{Rs} \\
k_f &=& 3 \sum_{f} Q_f^2.
\label{kappa}
\eea \ees
We use $N_f=3$, which implies that $k_f=2$. When can apply the Cauchy theorem to the integrand $K(-Q^2) \Pi_{\rm V}(Q^2)/Q^2$
\be
\oint_{C_{+}+C_{-}} \frac{d Q^2}{Q^2} K(-Q^2) \Pi_{\rm V}(Q^2) = 0,
\label{CauchyKP}
\ee
where the closed path $C_{+}+C_{-}$ is made of two disconnected parts and is shown in Fig.~\ref{Figcontamu} ($R \to \infty$).
We note that $\Pi_{\rm V}(Q^2)$ is holomorphic with a cut along the negative semiaxis in the complex $Q^2$-plane, and $K(-Q^2)$ is holomorphic with a cut along the positive semiaxis. The relation (\ref{CauchyKP}), in conjunction with Eqs.~(\ref{ahada})-(\ref{KRkap}), then implies for $a_{\mu}^{\rm had(1)}$ the following expression in terms of the V-channel $\Pi_V(Q^2)$ correlation function:
 \be
a_{\mu}^{\rm had(1)} = \frac{8}{3} \alpha_{em}^2\int_0^1 dx (1-x) \left[ \Pi_V(Q^2=0) - \Pi_V \left( Q^2\!=\!m^2_{\mu} \frac{x^2}{(1-x)} \right) \right].
\label{ahadb0}
\ee  
When applying integration by parts, this can be expressed in terms of the (full) Adler function (\ref{DV})
\be
a_{\mu}^{\rm had(1)} =  \frac{\alpha_{em}^2}{3 \pi^2} \! \int_0^{1} \! \frac{dx}{x} (1-x) (2-x) {\cal D}_{\rm V} \left( Q^2\!=\!m^2_{\mu} \frac{x^2}{(1-x)} \right).
\label{ahadb}
\ee
We can see from this expression that the main contribution to this quantity comes from the deep IR-regime $Q^2 \lesssim m_{\mu}^2$ ($\sim 0.01 \ {\rm GeV}^2$). The Adler function $D_V$ that appears in this integration is in principle represented by the OPE expansion (\ref{DV}), where the $D=0$ contribution is the renormalon-motivated resummation Eq.~(\ref{dresA}), cf.~also Appendix \ref{app:renmod}.

On the other hand, the terms with $D$ ($\equiv 2 n$) $\geq 4$ in Eq.~(\ref{DV}) require the knowledge of the V-channel condensates $\langle O_{2n} \rangle_{\rm V}$. 
In the previous Section, we extracted the (V+A)-channel condensates, cf. Eqs.~(\ref{O11770200}) and Table \ref{tabO4O6}.

For $D=4$, the V-channel condensate values are simply connected with those of the (V+A)-channel
\bea
\langle O_4 \rangle_{{\rm V}} & = & \langle O_4 \rangle_{{\rm A}} = \frac{1}{2}  \langle O_4 \rangle_{{\rm V+A}},
\label{O4V} \eea
which gives us the values of $\langle O_4 \rangle_{{\rm V}}$ immediately from Eq.~(\ref{O411770200}) for the input parameters $\alpha_s(M_Z^2;\MSbar)=0.1177$ and $r_{\tau,{\rm th}}^{(D=0)}=0.200$
\be
\langle O_4 \rangle_{\rm V} = (+0.00014 \pm 0.00008) \ {\rm GeV}^4.
\label{O4Vres}
\ee
On the other hand, for the $D=6$ condensates, a sum rule analysis of the (V-A)-channel \cite{GPRS} gives for $\langle O_6 \rangle_{{\rm V-A}}$
\bes
\label{GPRSOPAL}
\bea
\langle O_6 \rangle_{{\rm V-A}} & = & (-0.00570 \pm 0.00120) \ {\rm GeV}^6 \; {\rm (OPAL)},
\label{O6VmAOP} \\
\langle O_6 \rangle_{{\rm V-A}} & = & (-0.00360 \pm 0.00070) \ {\rm GeV}^6 \; {\rm (ALEPH)},
\label{O6VmAAL} \eea \ees
When we average these results over OPAL and ALEPH in the same way as we averaged $\langle O_D \rangle_{{\rm V+A}}$ ($D=4, 6$) in Sec.~\ref{subs:num}, Eqs.~(\ref{avO})-(\ref{sigavO}), the result is\footnote{\textcolor{black}{
Similar results can be obtained from the earlier analyses \cite{BoiOP} [$(-0.00660 \pm 0.00110) \ {\rm GeV}^6$ (OPAL)] and \cite{BoiAL} [$(-0.00316 \pm 0.00091) \ {\rm GeV}^6$ (ALEPH)] which give the average $\langle O_6 \rangle_{{\rm V-A}} = (-0.00488 \pm 0.00186) \ {\rm GeV}^6$.}} 
\be
\langle O_6 \rangle_{{\rm V-A}} = (-0.00465 \pm 0.00126) \ {\rm GeV}^6 .
\label{O6VmA}
\ee
When we now take into account that $\langle O_6 \rangle_{\rm V} = (1/2) ( \langle O_6 \rangle_{{\rm V+A}} +  \langle O_6 \rangle_{{\rm V-A}})$, we obtain from Eq.~(\ref{O6VmA}) and Eq.~(\ref{O611770200}) [i.e., for $\alpha_s(M_Z^2;\MSbar)=0.1177$ and $r_{\tau,{\rm th}}^{(D=0)}=0.200$]
\be
\langle O_6 \rangle_{\rm V} = (-0.00196 \pm 0.00064) \ {\rm GeV}^6,  
\label{O6Vres} 
\ee
For other cases of the input parameters, we use for the central values of $\langle O_D \rangle_{\rm V+A}$ the corresponding values obtained from arithmetic averages of the OPAL and ALEPH values in Table \ref{tabO4O6}, which changes then the corresponding central values in Eqs.~(\ref{O4Vres}) and (\ref{O6Vres}), but the uncertainties ($\pm 0.00008 \ {\rm GeV}^4$;  $\pm 0.00064 \ {\rm GeV}^6$) remain the same.

Having now the condensate values of the first two higher-dimension terms ($D=4, 6$) in the OPE expansion (\ref{DV}), we still do not have a workable expression for ${\cal D}_V(Q^2)$ at very low momenta $Q^2 \lesssim m_{\mu}^2$ which we need in the integral (\ref{ahadb}). The problem does not reside in the $D=0$ term of the OPE expansion (\ref{DV}), because this term is evaluated in $\A$QCD in its resummed form (\ref{dresA}) and thus goes to zero at $Q^2 \to 0$ [because $\A(Q^{'2}) \sim Q^{'2} \to 0$ when $Q^{'2} \to 0$], cf.~Fig.~\ref{FigdposQ2}. The problem resides in the higher-dimension terms ($D \geq 4$) which diverge in the deep IR-regime. This is a general (and problematic) feature of the OPE expansions. We have to regularize these terms in the deep IR-regime. The $D=4, 6$ terms will be regularized here with regularization mass parameters\footnote{In Refs.~\cite{KTG}, and later in \cite{APTappl1b}, similar regularizations of higher-twist OPE terms were applied in the analysis of the Bjorken Sum Rule.} $\cM_2, \cM_3$, by replacing in the $D=4$ term in the denominator $Q^2 \mapsto (Q^2+\cM_2^2)$, and by replacing in the $D=6$ term in the denominator $Q^2 \mapsto (Q^2 + \cM_3^2)$. The residue of the corresponding $D=6$ term must also be modified so that the expansion for large $Q^2$ reproduces the OPE expansion in Eq.~(\ref{DV}) up to $D=6$. This leads to the following ansatz for the (IR regularized) part ${\cal D}(Q^2)^{\rm (NP)}$ in Eq.~(\ref{DV}):
\bes
\bea
\lefteqn{
  {\cal D}_{\rm V}(Q^2)^{\rm (NP)}  =  1 + 2 \pi^2 \sum_{n \geq 2}
\frac{ n 2 \langle O_{2n} \rangle_{\rm V}}{(Q^2)^n}
\label{NPOPE}
}
\\
& = &
1 + 4 \pi^2
\left[ \frac{2 \langle O_4 \rangle_{\rm V}}{ (Q^2+ \cM_2^2)^2 } 
            + \frac{\left( 3 \langle O_6 \rangle_{\rm V} + 4 \cM_2^2 \langle O_4 \rangle_{\rm V} \right)}{(Q^2 + \cM_3^2)^3} \right]
\label{NPOPEreg}
\eea \ees
We know that ${\cal D}_V(Q^2) \to 0$ when $Q^2 \to 0$. Since the $D=0$ term in (\ref{DV}) already has this property, as mentioned earlier, this then implies
\be
  {\cal D}_{\rm V}(0) = 0 \; \Rightarrow  {\cal D}_{\rm V}(0)^{\rm (NP)} =0.
  \label{deepIR1}
  \ee
  Consequently, $\cM_3$ is not independent, but is a function of the mass $\cM_2$ and of the two condensate values
  \be
\cM_3^2  = \left[ \frac{(-3) \langle O_6 \rangle_{\rm V} - 4 \cM_2^2 \langle O_4 \rangle_{\rm V}}{\frac{1}{4 \pi^2} + 2   \langle O_4 \rangle_{\rm V}/\cM_2^4} \right]^{1/3}.
\label{cM32}
 \ee
 At this point, the only free parameter is the mass $\cM_2$. In principle, this parameter can now be determined in the following way: We require the reproduction of the experimental value (\ref{amuhad1}) of $a_{\mu}^{\rm had(1)}$ by the integration (\ref{ahadb}), where in the latter integral we use for ${\cal D}_V(Q^2)$ the expression
\be
{\cal D}_{\rm V}(Q^2; {\cal M}_2^2) = d(Q^2)_{D=0} + {\cal D}_{\rm V}(Q^2; {\cal M}_2^2)^{\rm (NP)},
\label{DVb}
\ee
and where ${\cal D}_{\rm V}(Q^2; {\cal M}_2^2)^{\rm (NP)}$ is given in Eq.~(\ref{NPOPEreg}) in conjunction with Eq.~(\ref{cM32}).

Here: (I) for the $D=0$ contribution $d(Q^2)_{(D=0)}$ we use the renormalon-motivated resummation Eq.~(\ref{dresA}), where the coupling $\A(Q^{'2})$ is the $3\delta$ $\A$QCD coupling obtained in Sec.~\ref{sec:Ares}; (II) the $D>0$ contribution ${\cal D}_{\rm V}(Q^2)^{\rm (NP)}$ is represented by the the IR-regularized higher-twist terms Eq.~(\ref{NPOPEreg}) where $\cM_3$ is determined by Eq.~(\ref{cM32}); (III) the condensate values appearing in these Eqs.~(\ref{NPOPEreg}) and (\ref{cM32}) are those of Eqs.~(\ref{O4Vres}) and (\ref{O6Vres}) when the input parameters for the coupling $\A(Q^2)$ are  taken to be $\alpha_s(M_Z^2;\MSbar)=0.1177$ and $r_{\tau,{\rm th}}^{(D=0)}=0.200$; for other values of the input parameters, the condensate values are those as explained in the text after Eq.~(\ref{O6Vres}).

If the described QCD framework is consistent in the entire IR-regime, the described procedure should give us for the IR-regularizing mass parameters $\cM_2$ and $\cM_3$ real values (i.e., real positive values for the squares $\cM_2^2$ and $\cM_3^2$), and these values are expected to be in the typical regime of the nonperturbative QCD $\cM_2 \sim \cM_3 \lesssim 1 \ {\rm GeV}^2$. The aforedescribed numerical procedure of determination of the mass parameter $\cM_2$ indeed gives such values. Namely, when the input parameters for the coupling have the values $\alpha_s(M_Z^2;\MSbar)=0.1177$ and $r_{\tau,{\rm th}}^{(D=0)}=0.200$, we obtain
\bes
\label{cM23}
\bea
\cM_2 &= & \left[ 0.384^{+0.019}_{-0.040} (O_4)^{-0.019}_{+0.014} (O_6) \right] \ {\rm GeV},
\label{cM2} 
\\
\cM_3 & = & \left[ 0.730^{-0.012}_{+0.016} (O_4)^{-0.055}_{+0.042} (O_6) \right] \ {\rm GeV}.
\label{cM3} 
\eea \ees
The uncertainties of the extracted masses due to the uncertainties $\delta \langle O_4 \rangle_V = \pm 0.00008 \ {\rm GeV}^4$ and $\delta \langle O_6 \rangle_V = \pm 0.00064 \ {\rm GeV}^4$ are included in Eqs.~(\ref{cM23}). 
These values of ${\cal M}_2$ and ${\cal M}_3$ are both real (i.e, their squares are positive) and lie in the region around or below $1$ GeV, as expected for IR-regularizing masses in QCD.
The variation $\pm 4$ of the experimental value $10^{10} \times a_{\mu; {\rm exp}}^{\rm had(1)}$, Eq.~(\ref{amuhad1}), affects only slightly the extracted mass values: $\delta \cM_2 =\pm 0.65$ MeV and $\delta \cM_3 =\pm 0.27$ MeV. If we used the higher value $10^{10} \times a_{\mu; {\rm exp}}^{\rm had(1)} = 712.4$ (which is the central value of the prediction in Ref.~\cite{Borsanyi:2020mff}), the extracted values would not increase much: $\delta \cM_2 = 3.1$ MeV and $\delta \cM_3=1.3$ MeV.\footnote{\textcolor{black}{This mild dependence of the mass parameters $\cM_2$ and $\cM_3$ on the variation of $a_{\mu; {\rm exp}}^{\rm had(1)}$ means that our approach can accommodate equally well any values in the range $690 < 10^{10} \times a_{\mu; {\rm exp}}^{\rm had(1)} < 725$; this implies that our approach by itself cannot provide a solution of the problem of discrepancy between the (total) SM value $a_{\mu}^{({\rm SM})}$ and the directly measured value $a_{\mu}^{{\rm dir.exp}}$ mentioned in footnote \ref{amuSM}.}} 

For other values of the input parameters, the extracted central values of $\cM_2$ and $\cM_3$ are given in Table \ref{tabM23}, where also the central values of $\langle O_4 \rangle_V$ and $\langle O_6 \rangle_V$ are included.
 \begin{table}
   \caption{The extracted central values of the V-channel condensates $\langle O_{D} \rangle$ ($D=2,3$), in units of ${\rm GeV}^{D}$; the extracted values of the IR-regularizing masses $\cM_n$ (in GeV), and the variation of these masses when the $V$ channel condensates vary around their central values by their uncertainties $\delta \langle O_{4} \rangle = \pm 0.0008 \ {\rm GeV}^4$ and  $\delta \langle O_{6} \rangle = \pm 0.0064 \ {\rm GeV}^6$. The various cases of the $\A$-coupling input parameters are given in the same order as in Table \ref{tabres3d}.}
   \label{tabM23}
 \renewcommand{\arraystretch}{1.2}
\begin{ruledtabular}
  \begin{tabular}{ll|rr|rrr|rrr}
  ${\overline \alpha}_s(M_Z^2)$ &  $r^{(D=0)}_{\tau, {\rm th}}$ & $\langle O_4 \rangle_V$ & $\langle O_6 \rangle_V$ & $\cM_2$ & $\delta \cM_2 (O_4)$ & $\delta \cM_2 (O_6)$ & $\cM_3$ & $\delta \cM_3 (O_4)$ & $\delta \cM_3 (O_6)$
\\
\hline
$0.1184$ & 0.203 & -0.00051 & -0.00161 & 3.241 & $^{+0.290}_{-0.229}$ & $^{+0.143}_{-0.149}$ & 1.006 & 0.000 & 0.000
 \\
$0.1181$ & 0.201 & -0.00032 & -0.00174 & 4.389 & $^{+0.679}_{-0.464}$ & $^{+0.168}_{-0.175}$ & 1.028 & 0.000 & 0.000
 \\
$0.1181$ & 0.203 &  -0.00014 & -0.00175 & 9.096 & $^{+4.799}_{-1.840}$ & $^{+0.187}_{-0.190}$ & 1.126 & 0.000 & 0.000
\\
\hline
$0.1180$ & 0.202 &  -0.00010 & -0.00179 & 10.512 & $^{+12.993}_{-2.677}$ & $^{+0.226}_{-0.231}$ & 1.118 & 0.000 & 0.000
 \\
$0.1180$ & 0.203 &  +0.00007 & -0.00184 & 0.346 & $^{+0.035}_{+43.904}$ & $^{-0.017}_{+0.012}$ & 0.734 & $^{-0.016}_{+0.487}$ & $^{-0.057}_{+0.042}$ 
 \\
$0.1179$ & 0.201 &  -0.00006 & -0.00184 & 13.243 &  $^{-12.947}_{-4.573}$ & $^{+2.738}_{-3.473}$ & 1.1109 & $^{-0.359}_{+0.000}$ & 0.000 
 \\
$0.1179$ & 0.202 & +0.00009 & -0.00188 & 0.359 & $^{+0.029}_{-0.099}$ & $^{-0.017}_{+0.013}$ & 0.733 & $^{-0.015}_{+0.025}$ & $^{-0.057}_{+0.042}$ 
 \\
$0.1178$ & 0.200 & -0.00003 & -0.00188 & 18.274 & $^{-17.935}_{-8.731}$ & $^{+0.432}_{-0.444}$ & 1.103 & $^{-0.358}_{+0.000}$ & 0.000
\\
$0.1178$ & 0.201 & +0.00011 & -0.00192 & 0.371 & $^{+0.024}_{-0.061}$ & $^{-0.018}_{+0.013}$ & 0.732 & $^{-0.013}_{+0.020}$ & $^{-0.056}_{+0.042}$ 
\\
$0.1177$ & 0.199 & +0.00001 &  -0.00193 & 0.267 & $^{+0.101}_{-11.394}$ & $^{-0.013}_{+0.010}$ & 0.763 & $^{-0.024}_{+0.332}$ & $^{-0.053}_{+0.039}$ 
\\ 
$0.1177$ & 0.200 & +0.00014 & -0.00196 & 0.384 & $^{+0.019}_{-0.040}$ & $^{-0.019}_{+0.014}$ & 0.730 & $^{-0.012}_{+0.016}$ & $^{-0.055}_{+0.042}$ 
\\
$0.1176$ & 0.199 & +0.00016 & -0.00200 & 0.392 & $^{+0.017}_{-0.033}$ & $^{-0.020}_{+0.014}$ & 0.730 & $^{-0.012}_{+0.015}$ & $^{-0.054}_{+0.042}$ 
\\ 
$0.1175$ & 0.197 & +0.00008 & -0.00201 & 0.367 & $^{+0.032}_{--}$ & $^{-0.018}_{+0.015}$ & 0.748 & $^{-0.014}_{--}$ & $^{-0.053}_{+0.041}$ 
\\
$0.1175$ & 0.198 & +0.00019 & -0.00203 & 0.401 & $^{+0.014}_{-0.025}$ & $^{-0.020}_{+0.015}$ & 0.728 & $^{-0.011}_{+0.013}$ & $^{-0.054}_{+0.041}$ 
\\
$0.1175$ & 0.1988 & +0.00041 & -0.00213 & 0.425 & $^{+0.005}_{-0.008}$ & $^{-0.021}_{+0.016}$ & 0.703 & $^{-0.009}_{+0.009}$ & $^{-0.054}_{+0.041}$

\\
$0.1174$ & 0.197 & +0.00022 & -0.00207 & 0.409 & $^{+0.013}_{-0.020}$ & $^{-0.020}_{+0.016}$ & 0.727 & $^{-0.011}_{+0.012}$ & $^{-0.054}_{+0.041}$ 
\\
$0.1173$ & 0.196 & +0.00025 & -0.00211 & 0.417 & $^{+0.010}_{-0.017}$ & $^{-0.021}_{+0.016}$ & 0.726 & $^{-0.010}_{+0.012}$ & $^{-0.053}_{+0.041}$ 
\\
$0.1173$ & 0.197 & +0.00046 & -0.00219 & 0.432 & $^{+0.004}_{-0.007}$ & $^{-0.022}_{+0.016}$ & 0.702 & $^{-0.009}_{+0.009}$ & $^{-0.053}_{+0.041}$
\\
$0.1172$ & 0.195 &  +0.00028 & -0.00216 & 0.423 & $^{+0.010}_{-0.013}$ & $^{-0.021}_{+0.017}$ & 0.726 & $^{-0.009}_{+0.011}$ & $^{-0.052}_{+0.041}$ 
\\
$0.1172$ & 0.196 & +0.00045 & -0.00221 & 0.434 & $^{+0.004}_{-0.007}$ & $^{-0.022}_{+0.016}$ & 0.706 & $^{-0.009}_{+0.009}$ & $^{-0.053}_{+0.041}$
\\
$0.1171$ & 0.195 & +0.00046 & -0.00224 & 0.437 & $^{+0.004}_{-0.007}$ & $^{-0.022}_{+0.017}$ & 0.708 & $^{-0.009}_{+0.009}$ & $^{-0.052}_{+0.040}$
\\
$0.1170$ & 0.194 & +0.00047 & -0.00227 & 0.440 & $^{+0.004}_{-0.006}$ & $^{-0.023}_{+0.017}$ & 0.710 & $^{-0.009}_{+0.008}$ & $^{-0.052}_{+0.040}$
\\
$0.1169$ & 0.193 &  +0.00049 & -0.00231 & 0.443 & $^{+0.004}_{-0.005}$ & $^{-0.022}_{+0.018}$ & 0.711 & $^{-0.008}_{+0.009}$ & $^{-0.051}_{+0.040}$
\\
\hline
$0.1181$ & 0.199 & -0.00043 & -0.00177 & 3.003 & $^{+0.327}_{-0.248}$ & $^{+0.181}_{-0.193}$ & 0.968 & 0.000 & 0.000
\\
$0.1184$ & 0.201 & -0.00063 & -0.00164 & 2.553 & $^{+0.182}_{-0.151}$ & $^{+0.147}_{-0.156}$ & 0.972 & 0.000 & 0.000
\\
$0.1184$ & 0.199 & -0.00074 & -0.00168 & 2.437 & $^{+0.147}_{-0.125}$ & $^{+0.132}_{-0.139}$ & 0.982 & $^{+0.000}_{-0.001}$ & $^{+0.000}_{-0.001}$ 
\\
$0.1189$ & 0.201 & -0.00114 & -0.00148 & 2.255 & $^{+0.087}_{-0.079}$ & $^{+0.093}_{-0.099}$ & 1.015 & 0.000 & 0.000
\end{tabular}
\end{ruledtabular}
\end{table}
We see that the values of $\cM_2$ and $\cM_3$ are both below $1$ GeV only when $\langle O_4 \rangle_V$ is positive.\footnote{In such cases, also the value of the gluon condensate $\langle a GG \rangle$ is positive, due to the relation $\langle a GG \rangle =12 \langle O_4 \rangle_V + 6 f^2_{\pi} m^2_{\pi}$ (where $ 6 f^2_{\pi} m^2_{\pi} \approx 0.00199 \ {\rm GeV}^4$).} Whenever the value of the condensate $\langle O_4 \rangle_V$ is negative, the extracted value of the mass ${\cal M}_2$ becomes at least several GeV and often $\sim 10$ GeV, which is not expected for an IR-regularizing mass in QCD. When the value of  $\langle O_4 \rangle_V$ approaches zero from the negative side, the extracted value of $\cM_2$ increases to very large values.

Therefore, we conclude that in our construction of the $\A$-coupling, at a given input value of $\alpha_s(M_Z^2; \MSbar)$, the value of the other input parameter $r_{{\tau}, {\rm th}}^{(D=0)}$  ($\equiv r_{{\tau}, {\rm th}}^{(D=0, m_{\tau}^2)}$) should be such that the Borel-Laplace sum rules yield a positive value of the extracted condensate $\langle O_4 \rangle_{V+A}$ ($= 2 \langle O_4 \rangle_{V}$). From Table \ref{tabM23} we can see that this then severely restricts the possible values of $r_{{\tau}, {\rm th}}^{(D=0)}$ toward the higher values. On the other hand, we have also several other restrictions to be fulfilled for a our constructed $\A$-coupling, namely that it should pass the $r_{\tau}$-consistency check for ALEPH and OPAL (cf.~Table \ref{tabrtau}), e.g., $|{\rm dev}_{r_{\tau}}| < 1.5$; and that the fit quality parameters in the Borel-Laplace sum rules $\chi^2(\Psi)$ ($\Psi=\pi/6, \pi/4, 0$) should be reasonably small, e.g.. $\chi^2(\Psi) < 2 \ \chi^2_{\rm exp}(\Psi)$, i.e., ${\rm rat}_{\chi^2}(\Psi) < 2$, cf.~Tables \ref{tabchi2OP}-\ref{tabchi2AL}. In Table \ref{tabsel} we summarize these results for all the considered cases, in order to see the mentioned trends more clearly. 
 \begin{table}
   \caption{The parameters ${\rm rat}_{\chi^2}(\Psi){\rm (ALEPH)}$ ($\Psi=\pi/6, \pi/4, 0$), ${\rm dev}_{r_{\tau}}{\rm (ALEPH)}$, $\langle O_4 \rangle_V$ (in ${\rm GeV}^4$) and $\cM_2$ (in GeV), for the considered $\A$-coupling cases. These six parameters are critical in the sense of selection of the acceptable input parameter cases, see the text. The various cases of the $\A$-coupling input parameters  $\alpha_s(M_Z^2; \MSbar)$ and $r_{{\tau}, {\rm th}}^{(D=0)}$ are presented in the same order as in Table \ref{tabres3d}.}
   \label{tabsel}
   \renewcommand{\arraystretch}{1.2}
\begin{ruledtabular}
\begin{tabular}{ll|ccc|c|cc}
${\overline \alpha}_s(M_Z^2)$ &  $r^{(D=0)}_{\tau, {\rm th}}$ & ${\rm rat}_{\chi^2}^{\rm (AL)}\left(\frac{\pi}{6} \right)$ & ${\rm rat}_{\chi^2}^{\rm (AL)}\left(\frac{\pi}{4} \right)$ & ${\rm rat}_{\chi^2}^{\rm (AL)}(0)$ & ${\rm dev}_{r_{\tau}}^{\rm (AL)}$ & $\langle O_4 \rangle_V$ & $\cM_2$ 
\\
\hline
$0.1184$ & 0.203 & 0.58 & 1.15 & 2.14 & 1.24 & $-0.00051\pm 0.00008$ & $3.241^{+0.290}_{-0.229}(O_4)^{+0.143}_{-0.149}(O_6)$ 
 \\
$0.1181$ & 0.201 & 0.80& 1.15 & 1.83 & 0.89 & $-0.00032\pm 0.00008$ & $4.389^{+0.679}_{-0.464}(O_4)^{+0.168}_{-0.175}(O_6)$ 
 \\
$0.1181$ & 0.203 & 0.39 & 0.76 & 1.90 & 1.52  & $-0.00014\pm 0.00008$ & $9.096^{+4.799}_{-1.840}(O_4)^{+0.187}_{-0.190}(O_6)$
\\
\hline
$0.1180$ & 0.202 & 0.50 & 0.95 & 1.76 & 1.29 & $-0.00010\pm 0.00008$ & $10.512^{+12.993}_{-2.677}(O_4)^{+0.226}_{-0.231}(O_6)$
 \\
$0.1180$ & 0.203 & 0.26 & 0.52 & 1.64 & 1.65 & $+0.00007\pm 0.00008$ & $0.346^{+0.035}_{+43.904}(O_4)^{-0.017}_{+0.012}(O_6)$ 
 \\
$0.1179$ & 0.201 & 0.62 & 1.17 & 1.65 & 1.07 & $-0.00006\pm 0.00008$ & $13.243^{-12.947}_{-4.573}(O_4)^{+2.738}_{-3.473}(O_6)$
 \\
$0.1179$ & 0.202 & 0.37 & 0.70 & 1.54 & 1.42 & $+0.00009\pm 0.00008$ & $0.359^{+0.029}_{-0.099}(O_4)^{-0.017}_{+0.013}(O_6)$
 \\
$0.1178$ & 0.200 & 0.76 & 1.41 & 1.55 & 0.85 & $-0.00003\pm 0.00008$ & $18.274^{-17.935}_{-8.731}(O_4)^{+0.432}_{-0.444}(O_6)$
\\
$0.1178$ & 0.201 &  0.49 & 0.90 & 1.45 & 1.19 & $+0.00011\pm 0.00008$ & $0.371^{+0.024}_{-0.061}(O_4)^{-0.018}_{+0.013}(O_6)$
\\
$0.1177$ & 0.199 & 0.93 & 1.68 & 1.47 & 0.63 & $+0.00001\pm 0.00008$ & $0.267^{+0.101}_{-11.394}(O_4)^{-0.013}_{+0.010}(O_6)$
\\ 
$0.1177$ & 0.200 & 0.63 & 1.14 & 1.37 & 0.96 & $+0.00014\pm 0.00008$ & $0.384^{+0.019}_{-0.040}(O_4)^{-0.019}_{+0.014}(O_6)$
\\
$0.1176$ & 0.199 & 0.78 & 1.40 & 1.31 & 0.74 & $+0.00016\pm 0.00008$ & $0.392^{+0.017}_{-0.033}(O_4)^{-0.020}_{+0.014}(O_6)$
\\ 
$0.1175$ & 0.197 & 1.31 & 2.30 & 1.37 & 0.18 & $+0.00008\pm 0.00008$ & $0.367^{+0.032}_{--}(O_4)^{-0.018}_{+0.015}(O_6)$
\\
$0.1175$ & 0.198 & 0.96 & 1.69 & 1.27 & 0.51 & $+0.00019\pm 0.00008$ & $0.401^{+0.014}_{-0.025}(O_4)^{-0.020}_{+0.015}(O_6)$
\\
$0.1175$ & 0.1988 & 0.56 & 0.98 & 0.97 & 0.84 & $+0.00041\pm 0.00008$ & $0.425^{+0.005}_{-0.008}(O_4)^{-0.021}_{+0.016}(O_6)$
\\
$0.1174$ & 0.197 & 1.15 & 2.01 & 1.23 & 0.28 & $+0.00022\pm 0.00008$ & $0.409^{+0.013}_{-0.020}(O_4)^{-0.020}_{+0.016}(O_6)$
\\
$0.1173$ & 0.196 & 1.36 & 2.36 & 1.22 & 0.06 & $+0.00025\pm 0.00008$ & $0.417^{+0.010}_{-0.017}(O_4)^{-0.021}_{+0.016}(O_6)$
\\
$0.1173$ & 0.197 & 0.85 & 1.46 & 0.93 & 0.44 & $+0.00046\pm 0.00008$ & $0.432^{+0.004}_{-0.007}(O_4)^{-0.022}_{+0.016}(O_6)$
\\
$0.1172$ & 0.195 & 1.60 & 2.73 & 1.22 & -0.17 & $+0.00028\pm 0.00008$ & $0.423^{+0.010}_{-0.013}(O_4)^{-0.021}_{+0.017}(O_6)$
\\
$0.1172$ & 0.196 & 1.10 & 1.87 & 0.98 & 0.19 & $+0.00045\pm 0.00008$ & $0.434^{+0.004}_{-0.007}(O_4)^{-0.022}_{+0.016}(O_6)$
\\
$0.1171$ & 0.195 & 1.35 & 2.28 & 1.02 & -0.05 & $+0.00046\pm 0.00008$ & $0.437^{+0.004}_{-0.007}(O_4)^{-0.022}_{+0.017}(O_6)$
\\
$0.1170$ & 0.194 & 1.61 & 2.70 & 1.06 & -0.29 & $+0.00047\pm 0.00008$ & $0.440^{+0.004}_{-0.006}(O_4)^{-0.023}_{+0.017}(O_6)$
\\
$0.1169$ & 0.193 & 1.89 & 3.15 & 1.11 & -0.52 & $+0.00049\pm 0.00008$ & $0.443^{+0.004}_{-0.005}(O_4)^{-0.022}_{+0.018}(O_6)$
\\
\hline
$0.1181$ & 0.199 & 1.11 & 2.06 & 1.54 & 0.26 & $-0.00043\pm 0.00008$ & $3.003^{+0.327}_{-0.248}(O_4)^{+0.181}_{-0.193}(O_6)$
\\
$0.1184$ & 0.201 & 0.81 & 1.57 & 1.72 & 0.59 & $-0.00063\pm 0.00008$ & $2.553^{+0.182}_{-0.151}(O_4)^{+0.147}_{-0.156}(O_6)$
\\
$0.1184$ & 0.199 & 1.11 & 2.10 & 1.42 & -0.08 & $-0.00074\pm 0.00008$ & $2.437^{+0.147}_{-0.125}(O_4)^{+0.132}_{-0.139}(O_6)$
\\
$0.1189$ & 0.201 & 0.84 & 1.67 & 1.49 & 0.01 & $-0.00114\pm 0.00008$ & $2.255^{+0.087}_{-0.079}(O_4)^{+0.093}_{-0.099}(O_6)$
\end{tabular}
\end{ruledtabular}
 \end{table}
In the Table we present the five obtained critical parameters, namely ${\rm rat}_{\chi^2}(\psi){\rm (ALEPH)}$ ($\Psi=\pi/6, \pi/4, 0$), ${\rm dev}_{r_{\tau}}{\rm (ALEPH)}$, and $\langle O_4 \rangle_V$.\footnote{The fit-quality parameters for the OPAL data, ${\rm rat}_{\chi^2}(\psi){\rm (OPAL)}$ ($\Psi=\pi/6, \pi/4, 0$) and ${\rm dev}_{r_{\tau}}{\rm (OPAL)}$, are in the considered cases significantly smaller and thus acceptable (cf.~Tables \ref{tabchi2OP} and \ref{tabrtau}), due to the significantly higher experimental uncertainty of OPAL data.}  The last column has the extracted values of the IR-regularizing mass $\cM_2$, to remind us that $\langle O_4 \rangle_V$ must be positive in order to have acceptable values of $\cM_2$ ($\lesssim 1$ GeV). We see that the input values $\alpha_s(M_Z^2;\MSbar) \geq 0.1180$ are not in the preferred regime, because in such cases the positivity of $\langle O_4 \rangle_V$ requires to use a relatively large value of the other input parameter, $r^{(D=0)}_{\tau, {\rm th}} > 0.202$, which then implies that the $r_{\tau}$-consistency condition for ALEPH data, say $|{\rm dev}_{r_{\tau}}{\rm (ALEPH)}| < 1.5$, is not fulfilled. We recall that in Table \ref{tabsel} (as well as in other Tables \ref{tabres3d}-\ref{tabO4O6}, \ref{tabchi2OP}-\ref{tabM23}), all the input data with $\alpha_s(M_Z^2;\MSbar) \leq 0.1180$ have the values of the input parameter $r^{(D=0)}_{\tau, {\rm th}}$ presented up to its maximum possible value at that $\alpha_s(M_Z^2;\MSbar)$ within three digits (i.e., by increasing the maximal presented value of $r^{(D=0)}_{\tau, {\rm th}}$ by 0.001 we would surpass the maximal possible value).\footnote{Except for the case of $\alpha_s(M_Z^2;\MSbar)=0.1175$ where $r^{(D=0)}_{\tau, {\rm th}}=0.1988$ is at the maximally admissible value within four digits, i.e., $r^{(D=0)}_{\tau, {\rm th}}=0.1989$ is already above the maximum value achievable in our $3 \delta$ $\A$QCD framework at that value of $\alpha_s(M_Z^2;\MSbar)$.} If requiring ${\rm rat}_{\chi^2}(\Psi) < 2$ ($\Psi=\pi/6, \pi/4, 0$) and $|{\rm dev}_{r_{\tau}}| < 1.5$, and $\langle O_4 \rangle_V > 0$, then the results summarized in Table \ref{tabsel} suggest that, within the considered $3 \delta$ $\A$QCD framework, the preferred values of the input parameters are: $0.1172 \leq \alpha(M_Z^2;\MSbar) \leq 0.1179$. In these cases, the corresponding values of $r^{(D=0)}_{\tau, {\rm th}}$ are in a narrow interval of width $0.001$-$0.002$, which includes the maximal possible value of $r^{(D=0)}_{\tau, {\rm th}}$ as achievable in the considered $3 \delta$ $\A$QCD famework [at the considered value of $\alpha(M_Z^2;\MSbar)$]. Among the Borel-Laplace sum rule quality-fit parameters ${\rm rat}_{\chi^2}(\Psi)$ for ALEPH data, the parameter with $\Psi=\pi/4$ is often the most stringent. In that sense, among the specific choices of the values of the output parameters in Table \ref{tabsel}, the most attractive appear to be the choices of ($\alpha(M_Z^2;\MSbar)$, $r^{(D=0)}_{\tau, {\rm th}}$): (0.1178, 0.201); (0.1177, 0.200); (0.1175, 0.1988); (0.1173, 0.197).

When $\alpha_s(M_Z^2; \MSbar) > 0.1179$, it turns out that the positive value of $\langle O_4 \rangle_V$ (and thus $\cM_2 \lesssim 1$ GeV) is achieved at a price of not passing clearly the $r_{\tau}$-test for the ALEPH data [${\rm dev}_{\tau}({\rm ALEPH}) > 1.5$]. On the other hand, when $\alpha_s(M_Z^2; \MSbar) < 0.1173$, the coupling gives unattractive values $\chi^2(\pi/4) > 2 \ \chi^2(\pi/4)_{\rm exp}$ for ALEPH data, i.e., ${\rm rat}_{\chi^2}^{\rm (AL)}(\pi/4) > 2$, even when $r^{(D=0)}_{\tau, {\rm th}}$ is very close to its maximal possible value [at the considered value of $\alpha(M_Z^2;\MSbar)$]. 

\textcolor{black}{In all the considered cases, it turned out that our $D=0$ contribution of the Adler function $d(Q^2)_{D=0}$ contributes to $a_{\mu;{\rm exp}}^{\rm had(1)}$ at least 60 percent of its total value Eq.~(\ref{amuhad1}), and the NP-part the rest [cf.~also Eqs.~(\ref{DV}), (\ref{ahadb}) and (\ref{deepIR1})]. In general, if we had $\A(Q^2) \to a_0$ ($>0$) when $Q^2 \to 0$, this would not be so, since the $D=0$ contribution would then lead in Eq.~(\ref{ahadb}) to an integral diverging at $x \to 0$.}

All the solutions presented here are mutually related in the way that we come from one solution to another by continuously varying the imput parameters, i.e., they represent one class of solutions. However, for given input values $\alpha_s(M_Z^2; \MSbar)$ and $r_{{\tau}, {\rm th}}^{(D=0)}$, there is in general at least one other class of solutions, which gives higher $s_0$ values: $s_0 > 1000$, thus $M_0^2 > 12 \ {\rm GeV}^2$. These solutions are less attractive, because at $Q^2 \to 0$ they behave as $\A(Q^2) = k Q^2$ with $k > 20 \ {\rm GeV}^{-2}$, which makes them significantly higher than the lattice value $\A_{\rm latt}(Q^2)$ at low positive $Q^2$ and significantly lower at the local maximum. 

\section{Summary}
 \label{sec:summ}

 In this work we analysed a QCD-variant  whose coupling $\A(Q^2)$ [the analog of $\alpha_s(Q^2)/\pi$] has a spectral function $\rho_{\A}(\sigma) \equiv {\rm Im} \A (-\sigma - i \epsilon)$ such that at large scales $\sigma \geq M_0^2$ ($ \gg \Lambda^2_{\rm QCD}$) it coincides with the underlying pQCD spectral function, and at scales $\sigma < M_0^2$ it is parametrized in terms of three Dirac-delta functions. Such a spectral function $\rho_{\A}(\sigma)$ contains seven parameters.

 Four of the parameters are used (adjusted) as a ``precision tool'' to achieve that at large scales ($|Q^2| > 1 {\rm GeV}^2$) the considered coupling $\A(Q^2)$ approaches the underlying pQCD coupling $a(Q^2)$ ($\equiv \alpha_s(Q^2)/\pi$) to an increasingly large precision (when $|Q^2|$ increases), $\A(Q^2) - a(Q^2) \sim (\Lambda^2_{\rm QCD}/Q^2)^5$. One of the remaining three parameters is used (adjusted) to reproduce the values of the semihadronic strangeless $\tau$ decay rate ratio $r_{\tau}$ consistent with experiments (the physics of ``intermediate'' scales $|Q^2| \sim 1 \ {\rm GeV}^2$). The two remaining parameters are used to achieve the behaviour of $\A(Q^2)$ at $|Q^2| < 1 \ {\rm GeV}^2$ as suggested by large-volume lattice calculations \cite{LattcoupNf02a,LattcoupNf02b}, namely that $\A(Q^2) \sim Q^2$ at $Q^2 \to 0$ and that $\A(Q^2)$ at positive $Q^2$ has a local maximum at $Q^2 \approx 0.135 \ {\rm GeV}^2$ (when using the $\MSbar$-scaling).

 The construction of the coupling is performed in a renormalization scheme which coincides up to the four-loop level with the MM scheme\footnote{We recall that the MM scheme is used in the large-volume lattice calculations of the Landau gauge gluon and ghost propagators \cite{LattcoupNf02a,LattcoupNf02b}, and the scheme is theoretically known at present up to the four-loop level \cite{MiniMOM,BoucaudMM,CheRet}.} but rescaled to the usual $\MSbar$ scaling ($\Lambda_{\rm MM} \mapsto \Lambda_{\MSbar}$). In addition to the mentioned seven parameters, there is an additonal implicit parameter involved, namely the strength of the underlying pQCD coupling in the considered low-momenta regime $N_f=3$; we use for this parameter the widely used $\alpha_s(M_Z^2; \MSbar)$ (at $N_f=5$). In the constructed framework, we varied in practice, within the relatively narrow phenomenologically allowed intervals, the values of $\alpha_s(M_Z^2; \MSbar)$ and of the leading-twist (i.e., dimension zero, $D=0$) contribution $r_{\tau,{\rm th}}^{(D=0)}$ of the semihadronic $\tau$-decay ratio. As a result of the numerical implementation of the mentioned conditions, the obtained parameters in the spectral function $\rho_{\A}(\sigma)$ turned out to be such that  $\rho_{\A}(\sigma) =0$ for $\sigma \leq 0$, i.e., the obtained QCD coupling $\A(Q^2)$ is free of Landau singularities.
 
We then applied the Borel-Laplace sum rules to the OPAL and ALEPH data of the (strangeless) semihadronic $\tau$ decays [in the (V+A)-channel], in order to extract the condensate values of $D=4$ and $D=6$ condensates $\langle O_D \rangle_{\rm V+A}$. Some of the applied input parameter values of $\alpha_s(M_Z^2; \MSbar)$ and $r_{\tau,{\rm th}}^{(D=0)}$ did not give us good fit quality of the Borel-Laplace sum rules, especially for the (more precise) ALEPH data. Furthermore, with the obtained values of the condensates, we then compared the theoretical values of $r_{\tau}^{(D=0, \sigma_{\rm max})}$ with the corresponding experimental values (where $\sigma_{\rm max}$ is the upper bound of squared energy of the spectral functions, different for OPAL and ALEPH data). This $r_{\tau}$-consistency test acts as yet another filter for the acceptable range of the input parameter values $\alpha_s(M_Z^2; \MSbar)$ and $r_{\tau,{\rm th}}^{(D=0)}$  of the $\A$-coupling.

Such a construction and its evaluations have been performed already in our previous work \cite{3dAQCD}. But this time, besides some refined numerical aspects [such as the five-loop RGE-evolution of $a(Q^2)$ from $Q^2=M_Z^2$ down to $N_f=3$ regime], the main difference is that for the evaluation of the Adler function (needed in the calculation of $r_{\tau,{\rm th}}^{(D=0)}$ and of the Borel-Laplace sum rules) we now used a renormalon-motivated resummation \cite{renmod}, in contrast to Ref.~\cite{3dAQCD} where the truncated perturbation series (adjusted to the $\A$-coupling formalism) was used. This aspect changes appreciably the obtained numerical values of the parameters of $\A$-coupling.

We finally evaluated the hadronic vacuum polarization (HVP) contribution to muon anomalous magnetic moment,  $(g_{\mu}/2-1)^{\rm had(1)} \equiv a_{\mu}^{\rm had(1)}$, with our obtained coupling, which involves the Adler function in the V-channel. We used the V-channel condensate values $\langle O_D \rangle_{\rm V}$ ($D=4, 6$), obtained from the aforementioned values of the (V+A)-channel condensates $\langle O_D \rangle_{\rm V+A}$ (which were obtained from Borel-Laplace sum rules) and the (V-A)-channel condensate value $\langle O_6 \rangle_{\rm V-A}$ \cite{GPRS}. In the V-channel Adler function ${\cal D}_{\rm V}(Q^2)$ we wrote the OPE $D=4, 6$ contributions in a form which involves the IR-regularizing masses ${\cal M}_D$ ($D=4,6$), in order to apply the obtained ${\cal D}_{\rm V}(Q^2)$ at very low $Q^2 \sim m_{\mu}^2$ as needed for the evaluation of $a_{\mu}^{\rm had(1)}$. These two masses ${\cal M}_D$ are interrelated via the condition ${\cal D}_{\rm V}(0)=0$. Therefore, the adjustment of the value of the mass ${\cal M}_2$ can lead in general to the correct experimental value of $a_{\mu}^{\rm had(1)}$ ($\approx 7 \times 10^{-8}$). Since the masses ${\cal M}_D$ reflect nonperturbative physics, we expect them to be around or below $1$ GeV. We varied the mentioned input parameters of the $\A$-coupling, $\alpha_s(M_Z^2; \MSbar)$ and $r_{\tau,{\rm th}}^{(D=0)}$, within the phenomenologically acceptable range, and extracted the values of the IR-regularizing masses ${\cal M}_D$ ($D=4,6$). It turned out that these two masses were in the acceptable range $0 < {\cal M}_D < 1$ GeV for narrow intervals of values of the $\A$-coupling input parameters  $\alpha_s(M_Z^2; \MSbar)$ and $r_{\tau,{\rm th}}^{(D=0)}$. In particular, the values of  $r_{\tau,{\rm th}}^{(D=0)}$ have to lie in a narrow interval around its possible maximum (as allowed, at given value of $\alpha_s(M_Z^2; \MSbar)$, by the construction of the $\A$-coupling), namely those which gave positive values of $\langle O_4 \rangle_{\rm V+A} >0$ (and  thus $\langle O_4 \rangle_{\rm V} >0$).\footnote{
  These correspond also to positive value of the gluon condensate, $\langle a GG\rangle > 0$.}

When we take into account, in addition to the mentioned condition  $0 < {\cal M}_D < 1$ GeV, also the requirements of the acceptably good quality of fits in the Borel-Laplace sum rules and of the acceptable quality of the mentioned $r_{\tau}$-consistency test, the described framework of $\A$-coupling turns out to prefer the values of $0.1172 \leq \alpha_s(M_Z^2;\MSbar) \leq 0.1179$, and, for such values of $\alpha_s(M_Z^2;\MSbar)$, the values of $r_{\tau,{\rm th}}^{(D=0)}$ close to the allowed maximal values, cf.~Table \ref{tabsel} and the discussion at the end of Sec.~\ref{sec:amu}.

Ref.~\cite{shortv} represents a short version, which summarizes some of the main features of this work.

\begin{acknowledgments}
G.C.~acknowledges the support by FONDECYT (Chile) Grant No.~1180344.
\end{acknowledgments}

\appendix

\section{Renormalon-motivated summation; the case of Adler function}
\label{app:renmod}

The method \cite{renmod} is based on the following observations. We consider a spacelike physical observable ${\cal D}(Q^2)$ whose perturbation expansion is written in two (equivalent) ways, cf.~Eqs.~(\ref{dD0TPS})
\bes
\label{Dptlpt}
\bea
{\cal D}_{\rm pt}(Q^2) & = & a(\kappa Q^2) + \sum_{n=1}^{\infty} d_n(\kappa) a(\kappa Q^2)^{n+1}
\label{Dpt}\\
& = &  a(\kappa Q^2) + \sum_{n=1}^{\infty} \td_n(\kappa) \ta_{n+1}(\kappa Q^2),
\label{Dlpt} \eea \ees
where $Q^2$ is the physical spacelike scale of the process, and $\kappa$ is the renormalization scale parameter ($\kappa \sim 1$), i.e., $\mu^2 = \kappa Q^2$ is the renormalization scale. The logarithmic derivatives $\ta_{n+1}$ and the coefficients $\td_n$ of the ``reorganized'' series (\ref{Dlpt}) are given in Eqs.~(\ref{tan}) and (\ref{tdndn}), respectively.\footnote{The coefficients $k_m(p)$ and $\tk_m(p)$ in Eqs.~(\ref{tananrel})-(\ref{tdndn}) are independent of any scale, i.e., independent of $Q^2$ and independent of $\kappa$. Thus, for example: $\td_n(\kappa) = d_n(\kappa) + \sum_{s=1}^{n-1} \tk(n+1-s) \; d_{n-s}(\kappa)$.}
In the series Eq.~(\ref{Dlpt}) of ${\cal D}$ in logarithmic derivatives, we replace the logarithmic derivatives $\ta_{n+1}$ by the powers $a^{n+1}$, and obtain thus an auxiliary quantity $\tD$
\be
\tD(Q^2;\kappa) \equiv a(\kappa Q^2) + \sum_{n=1}^{\infty} \td_n(\kappa) a(\kappa Q^2)^{n+1}.
\label{tD} \ee
This quantity is renormalization scale dependent ($\kappa$-dependent) at the level beyond one-loop (at the one-loop level, we have $\ta_{n+1}=a^{n+1}$, and $\tD$ then coincides with ${\cal D}$). It si possible to show \cite{renmod} that the Borel transform ${\rm B}[\tD](u;\kappa)$ of $\tD$
\be
{\rm B}[\tD](u;\kappa) \equiv 1 + \sum_{n=1}^{\infty} \frac{\td_n(\kappa)}{n! \beta_0^n} u^n
\label{BtD} \ee
has a simple $\kappa$-dependence
\be
{\rm B}[\tD](u;\kappa) = \kappa^u {\rm B}[\tD](u), \label{kapBtD}. \ee
This dependence is exact, and is equal to the one-loop level (or: large-$\beta_0$ level) $\kappa$-dependence of Borel transforms of QCD observables such as ${\rm B}[D](u;\kappa)$  [the full $\kappa$-dependence of ${\rm B}[D](u;\kappa)$ is more complicated].

This indicates strongly that the structure of Borel transforms ${\rm B}[\tD](u)$ is the same as, or very similar to, the simple structure of the large-$\beta_0$ (LB) Borel transform ${\rm B}[{\cal D}](u;\kappa)^{\rm (LB)}$ of the observable $\cal D$, i.e., its renormalon poles are simple or multiple poles: $1/(p - u)^{n_p}$, $1/(p + u)^{m_p}$  ($n_p, m_p=1,2,\ldots$). In addition, the limiting ``zero'' multiplicity poles $\ln(1 \mp u/p)$ can be included in ${\rm B}[\tD](u)$ (corresponding to $n_p \to 0$ or $m_p \to 0$).

It was shown numerically in \cite{renmod} that such (simple) structures of ${\rm B}[\tD](u)$ then imply for ${\rm B}[D](u)$ the theoretically expected renormalon structures of the poles with specific fractional (noninteger) multiplicities $1/(p - u)^{{\widetilde \gamma}_p+n_p-k}$, $1/(p + u)^{{\overline \gamma}_p+m_p-k}$, where: ${\widetilde \gamma}_p= 1 + p {\beta}_1/{\beta}_0^2$, ${\overline \gamma}_p=1 - p {\beta}_1/{\beta}_0^2$, and $k=1,2,\ldots$.

For the massless Adler function ${\cal D}(Q^2) = d_{(D=0)}(Q^2)$, discussed in Sec.~\ref{sec:constr}, the large-$\beta_0$ (LB) Borel transform  ${\rm B}[d](u)^{\rm (LB)}$ was calculated in the literature \cite{LBAdl1,LBAdl2,BenekeRev}, and has the IR renormalon poles at $u=2$ ($p=2$, simple pole $n_p=1$), $u=3$ ($p=3$, double pole $n_p=2$), etc.; the UV renormalon poles are at $u=-1$ ($p=1$, double pole $m_p=2$), $u=-2$ ($p=2$, $m_p=2$), etc. This then suggests that we can write down for the Borel  transform ${\rm B}[\td](u)$ of the auxiliary (to Adler) quantity $\td(Q^2;\kappa)_{(D=0)}$ the following ansatz
\bea
{\rm B}[\td](u)^{\rm (4 P)} & = &
\exp \left( \tK u \right) \pi
{\Big \{}
\td_{2,1}^{\rm IR} \left[ \frac{1}{(2-u)} + \tal (-1) \ln \left( 1 - \frac{u}{2} \right) \right] + \frac{ \td_{3,2}^{\rm IR} }{(3 - u)^2} + \frac{ \td_{1,2}^{\rm UV} }{(1 + u)^2}
{\Big \}},
\label{BtD4P}
\eea
which has four adjustable parameters: the renormalon residue parameters $\td_{2,1}^{\rm IR}$, $\td_{3,2}^{\rm IR}$, $\td_{1,2}^{\rm UV}$ and the ``scaling'' parameter\footnote{Note that, according to Eq.~(\ref{kapBtD}), the factor $\exp(\tK u)$ corresponds to a change of the renormalization scale $\mu^2_{\rm new} = \mu^2_{\rm old} \exp(\tK)$.} $\tK$. As we see, the ansatz (\ref{BtD4P}) contains information about the three leading renormalons (at $u=2,3;-1$). It turns out that the parameter $\tal$ in Eq.~(\ref{BtD4P}) is not free, but is determined by the knowledge of a subleading coefficient of the $p=2$ IR renormalon of the Adler function \cite{renmod}. The four mentioned parameters are fixed by the knowledge of the first four coefficients in the perturbation expansion of the Adler function (\ref{dD0TPS}). In the LMM scheme, this then yields the values of the parameters as given after Eq.~(\ref{Gds}) (cf.~also Table V of Ref.~\cite{renmod}). When transformed into the $\MSbar$ scheme, the obtained result then gives $d_4(\MSbar)=338.2$.

Once the parameters of the Borel transform ${\rm B}[\tD](u)^{\rm (4 P)} $ of the auxiliary quantity (\ref{tD}) are fixed (in a chosen scheme), this generates all the higher-order coefficients $\td_n$ (and thus $d_n$) ($n=4,5,\ldots$) of the perturbation series (\ref{Dptlpt}). In principle, this could be regarded as sufficient to evaluate the perturbation series (\ref{Dptlpt}) or its $\A$QCD analog as explained in Eqs.~(\ref{tatA})-(\ref{dD0TPSA}). However, these series are asymptotically divergent [even in the $\A$QCD version Eq.~(\ref{dD0TPSA})], due to factorial divergence of the coefficients $\td_n \sim n!$ at large $n$. Nevertheless, the summation with the generated coefficients $\td_n$ can be performed, in terms of an integral which involves the coupling and a characteristic function $F_D(t)$
\be
{\cal D}_{\rm res.}(Q^2) =
\int_0^{+\infty} \frac{dt}{t} F_D(t) a(t Q^2).
\label{Dres1}
\ee
It turns out that the characteristic function $F_D(t)$ exists and it is the inverse Mellin transform of the Borel transform ${\rm B}[\tD](u)$ of the mentioned auxiliary quantity $\tD$
\be
F_D(t) =
\frac{1}{2 \pi i}
\int_{u_0- i \infty}^{u_0 + i \infty} du {\rm B}[\tD](u) t^u,
\label{FDinvMell}
\ee
where $u_0$ is any real value for which the Mellin transform
\be
{\rm B}[\tD](u) =
\int_0^{+\infty} \frac{dt}{t} F_D(t) t^{-u}
\label{BtDMell}
\ee
exists. For the specific case of the Adler function $d(Q^2)_{(D=0)}$, the integrals in the inverse Mellin (\ref{FDinvMell}) of ${\rm B}[\td](u)$ [see Eq.~(\ref{BtD4P})] can be explicitly evaluated and yield for the corresponding characteristic function the result Eqs.~(\ref{Gds}), with the resummation of the type of (\ref{Dres1}) given in Eq.~(\ref{drespQCD}) in Sec.~\ref{sec:constr}. There, we also explained that in pQCD for $Q^2>0$ there is an ambiguity of evaluation of such a resummation, but that in $\A$QCD the corresponding evaluation Eq.~(\ref{dresA}) is without ambiguities if $\A(Q^{'2})$ has no Landau singularities.

When we apply in the resummation (\ref{dresA}) [or: (\ref{drespQCD})] the Taylor expansion of $\A(t \exp(-\tK) Q^2)$ around $\A(\kappa Q^2)$
\be
\A(t e^{-\tK} Q^2) = \A(\kappa Q^2) + \sum_{n=1}^{\infty} (-\beta_0)^n \ln^n\left( \frac{t}{\kappa} e^{-\tK} \right) \tA_{n+1}(\kappa Q^2),
\label{TaylorA} \ee
and exchange the summation and the integration, we can check that the resummation (\ref{dresA}) can be written formally as the expansion in logarithmic derivatives [cf.~Eq.~(\ref{dD0TPSA}) for the truncated version]
\be
d(Q^2)_{(D=0)}^{\A{\rm QCD}} =  \A(\kappa Q^2) + \sum_{n=1}^{\infty} \td_n(\kappa) \; \tA_{n+1}(\kappa Q^2),
\label{dD0A}
\ee
with the coefficients $\td_n(\kappa)$ as generated by the Borel transform ${\rm B}[\td](u;\kappa)^{\rm (4 P)}= \kappa^u {\rm B}[\td](u)^{\rm (4 P)}$ where ${\rm B}[\td](u)^{\rm (4 P)}$ is from Eq.~(\ref{BtD4P}).\footnote{We recall that the series (\ref{dD0A}) is asymptotically divergent, while the resummed form Eq.~(\ref{dresA}) is unambiguous and convergent (if $\A$ has no Landau singularities).} This is so because it turns out that the obtained characteristic functions $G_d^{(\pm)}(t)$ and $G_d^{\rm (SL)}(t)$ given in Eqs.~(\ref{Gds}) fulfill the (necessary by construction) ``sum rules''
\bea
\td_n(\kappa) & = & (- \beta_0)^n {\Bigg \{} \int_0^{1} \frac{dt}{t} G_d^{(-)}(t) \ln^n \left( \frac{t}{\kappa} e^{-\tK} \right) + \int_1^{+\infty} \frac{dt}{t} G_d^{(+)}(t) \ln^n \left( \frac{t}{\kappa} e^{-\tK} \right)
\nonumber\\ &&
+ \int_0^{\infty}  \frac{dt}{t} G_d^{\rm (SL)}(t) \left[ \ln^n \left( \frac{t}{\kappa} e^{-\tK} \right) -  \ln^n \left( \frac{1}{\kappa} e^{-\tK} \right) \right] {\Bigg \}},
\label{GDLSLSR}
\eea
for $n=0,1,2,\ldots$.

It is interesting that the characteristic function $F_D$ (or: $G_d^{(\pm)}$ and $G_d^{\rm (SL)}$) of the {\it full\/} leading-twist contribution ${\cal D}(Q^2)$ Eq.~(\ref{Dptlpt}) is the inverse Mellin of the Borel transform of the {\it auxiliary\/} quantity $\tD(Q^2;\kappa)$ Eq.~(\ref{tD}). On the other hand, it is the Borel transform ${\rm B}[\tD](u)$ of the auxiliary quantity $\tD$ that is considerably simpler than the Borel transform ${\rm B}[D](u)$ of the full quantity. These two aspects combine in a very beneficial way towards the construction of the characteristic function and thus towards the described renormalon-motivated resummation of the spacelike QCD observables.

We want to point out yet another interesting aspect of the described method: it leads to the resummation of the full leading-twist quantity (and not just of the large-$\beta_0$ part of it), because the coefficients $\td_n$ contain the full information of the leading-twist contribution ${\cal D}(Q^2)$. This is despite the fact that the auxiliary quantity $\tD$ (which contains all $\td_n$'s) behaves under the variation of the renormalization scale as a large-$\beta_0$ part of an observable.\footnote{This approach can be regarded as an extension of the Neubert method of characteristic functions \cite{Neubert}. Neubert worked with the large-$\beta_0$ quantities ${\cal D}^{\rm (LB)}(Q^2)$ and with the corresponding one-loop pQCD coupling; at that level (one-loop), there is the coincidence $\ta^{n+1}=a^{n+1}$ and $\tD^{\rm (LB)}(Q^2)={\cal D}^{\rm (LB)}(Q^2)$. The ambiguity of integration (with the one-loop pQCD coupling) was fixed in \cite{Neubert} by adopting the Principal Value convention.}


\begin{thebibliography}{99}

\bibitem{3dAQCD}
  C.~Ayala, G.~Cveti\v{c}, R.~K\"ogerler and I.~Kondrashuk,
  ``Nearly perturbative lattice-motivated QCD coupling with zero IR limit,''
 J. Phys. G \textbf{45} (2018) no.3, 035001
[arXiv:1703.01321 [hep-ph]].

\bibitem{APT}  
  D.~V.~Shirkov and I.~L.~Solovtsov,
  ``Analytic QCD running coupling with finite IR behaviour and universal ${\bar \alpha}_s(0)$ value,''
{\it JINR Rapid Commun.\/} {\bf 2}[{\bf 76}] (1996), 5-10, 
 [arXiv:hep-ph/9604363 [hep-ph]];
``Analytic model for the QCD running coupling with universal  alpha(s)-bar(0) value,''
Phys.\ Rev.\ Lett.\  {\bf 79} (1997), 1209
  [hep-ph/9704333];
K.~A.~Milton and I.~L.~Solovtsov,
``Analytic perturbation theory in QCD and Schwinger's connection between the beta function and the spectral density,''
Phys. Rev. D \textbf{55} (1997), 5295-5298
[arXiv:hep-ph/9611438 [hep-ph]];
  D.~V.~Shirkov,
  ``Analytic perturbation theory for QCD observables,''
  {\it Theor.\ Math.\ Phys.\/}  {\bf 127} (2001), 409
  [hep-ph/0012283];
``Analytic perturbation theory in analyzing some QCD observables,''
Eur. Phys. J. C \textbf{22} (2001), 331-340
[arXiv:hep-ph/0107282 [hep-ph]].
  
\bibitem{KS}
  A.~I.~Karanikas and N.~G.~Stefanis,
  ``Analyticity and power corrections in hard scattering hadronic functions,''
Phys.\ Lett.\ B {\bf 504} (2001), 225
  Erratum: [Phys.\ Lett.\ B {\bf 636} (2006), 330]
  [hep-ph/0101031].
  
\bibitem{BMS}
  A.~P.~Bakulev, S.~V.~Mikhailov and N.~G.~Stefanis,
  ``QCD analytic perturbation theory: From integer powers to any power of the running coupling,''
 Phys.\ Rev.\ D {\bf 72} (2005), 074014
  [Phys.\ Rev.\ D {\bf 72} (2005), 119908]
  [hep-ph/0506311];
  ``Fractional Analytic Perturbation Theory in Minkowski space and application to Higgs boson decay into a b anti-b pair,''
 Phys.\ Rev.\ D {\bf 75} (2007), 056005
  Erratum: [Phys.\ Rev.\ D {\bf 77} (2008), 079901]
  [hep-ph/0607040];
 ``Higher-order QCD perturbation theory in different schemes: From FOPT to CIPT to FAPT,''
JHEP {\bf 1006} (2010), 085
  [arXiv:1004.4125 [hep-ph]].

\bibitem{revAPT}
  G.~M.~Prosperi, M.~Raciti and C.~Simolo,
  ``On the running coupling constant in QCD,''
 Prog.\ Part.\ Nucl.\ Phys.\  {\bf 58} (2007), 387
  [hep-ph/0607209];
  D.~V.~Shirkov and I.~L.~Solovtsov,
  ``Ten years of the analytic perturbation theory in QCD,''
Theor.\ Math.\ Phys.\  {\bf 150} (2007), 132
  [hep-ph/0611229];
A.~P.~Bakulev,
  ``Global Fractional Analytic Perturbation Theory in QCD with Selected Applications,''
  Phys.\ Part.\ Nucl.\  {\bf 40} (2009), 715
  doi:10.1134/S1063779609050050
  [arXiv:0805.0829 [hep-ph]] (arXiv preprint in Russian);
  N.~G.~Stefanis,
  ``Taming Landau singularities in QCD perturbation theory: The Analytic approach,''
Phys.\ Part.\ Nucl.\  {\bf 44} (2013), 494
  [Phys.\ Part.\ Nucl.\  {\bf 44} (2013), 494]
  [arXiv:0902.4805 [hep-ph]].

 \bibitem{APTappl1}
  K.~A.~Milton, I.~L.~Solovtsov and O.~P.~Solovtsova,
  ``The Bjorken sum rule in the analytic approach to perturbative QCD,''
  Phys.\ Lett.\ B {\bf 439} (1998), 421
  [hep-ph/9809510];
  R.~S.~Pasechnik, D.~V.~Shirkov, O.~V.~Teryaev, O.~P.~Solovtsova and V.~L.~Khandramai,
  ``Nucleon spin structure and pQCD frontier on the move,''
  Phys.\ Rev.\ D {\bf 81} (2010), 016010
  [arXiv:0911.3297 [hep-ph]];
  R.~S.~Pasechnik, J.~Soffer and O.~V.~Teryaev,
  ``Nucleon spin structure at low momentum transfers,''
  Phys.\ Rev.\ D {\bf 82} (2010), 076007
  [arXiv:1009.3355 [hep-ph]];
  V.~L.~Khandramai, R.~S.~Pasechnik, D.~V.~Shirkov, O.~P.~Solovtsova and O.~V.~Teryaev,
  ``Four-loop QCD analysis of the Bjorken sum rule vs data,''
  Phys.\ Lett.\ B {\bf 706} (2012), 340
  [arXiv:1106.6352 [hep-ph]].
C.~Ayala, G.~Cveti\v{c}, A.~V.~Kotikov and B.~G.~Shaikhatdenov,
``Bjorken sum rule in QCD frameworks with analytic (holomorphic) coupling,''
Int. J. Mod. Phys. A \textbf{33} (2018) no.18n19, 1850112
[arXiv:1708.06284 [hep-ph]].


\bibitem{APTappl1b}
C.~Ayala, G.~Cveti\v{c}, A.~V.~Kotikov and B.~G.~Shaikhatdenov,
``Bjorken polarized sum rule and infrared-safe QCD couplings,''
Eur. Phys. J. C \textbf{78} (2018) no.12, 1002
[arXiv:1812.01030 [hep-ph]].

\bibitem{APTappl2}
  G.~Cveti\v{c}, A.~Y.~Illarionov, B.~A.~Kniehl and A.~V.~Kotikov,
  ``Small-$x$ behavior of the structure function $F_2$ and its slope $\partial \ln F_2 / \partial \ln(1/x)$ for 'frozen' and analytic strong-coupling constants,''
  Phys.\ Lett.\ B {\bf 679} (2009), 350
  [arXiv:0906.1925 [hep-ph]];
  A.~V.~Kotikov, V.~G.~Krivokhizhin and B.~G.~Shaikhatdenov,
  ``Analytic and 'frozen' QCD coupling constants up to NNLO from DIS data,''
  Phys.\ Atom.\ Nucl.\  {\bf 75} (2012), 507
  [arXiv:1008.0545 [hep-ph]];
  C.~Ayala and S.~V.~Mikhailov,
  ``How to perform a QCD analysis of DIS in analytic perturbation theory,''
  Phys.\ Rev.\ D {\bf 92} (2015), 014028
  [arXiv:1503.00541 [hep-ph]];
  A.~V.~Sidorov and O.~P.~Solovtsova,
  ``The QCD analysis of $xF_3$ structure function based on the analytic approach,''
  Nonlin.\ Phenom.\ Complex Syst.\  {\bf 16} (2013), 397
  [arXiv:1312.3082 [hep-ph]];
  ``The QCD analysis of the combined set for the $F_3$ structure function data based on the analytic approach,''
  Mod.\ Phys.\ Lett.\ A {\bf 29} (2014) no. 36, 1450194
  [arXiv:1407.6858 [hep-ph]];
  ``QCD analysis of the F$_{3}$ structure function based on inverse Mellin transform in analytic perturbation theory,''
  Phys.\ Part.\ Nucl.\ Lett.\  {\bf 14} (2017) no. 1, 1
  ``Non-singlet $Q^2$-evolution and the analytic approach to Quantum Chromodynamics,''
  Nonlin.\ Phenom.\ Complex Syst.\  {\bf 18} (2015), 222;
  L.~Ghasemzadeh, A.~Mirjalili and S.~Atashbar Tehrani,
 ``Nonsinglet polarized nucleon structure function in infrared-safe QCD,''
  Phys.\ Rev.\ D {\bf 100} (2019), 114017
  [arXiv:1906.01606 [hep-ph]].
  
  \bibitem{APTappl3}
  P.~Allendes, C.~Ayala and G.~Cveti\v{c},
  ``Gluon Propagator in Fractional Analytic Perturbation Theory,''
 Phys.\ Rev.\ D {\bf 89} (2014), 054016
  [arXiv:1401.1192 [hep-ph]].
  
\bibitem{Nest2}
  A.~V.~Nesterenko and J.~Papavassiliou,
  ``The massive analytic invariant charge in QCD,''
 Phys.\ Rev.\ D {\bf 71} (2005), 016009
  [hep-ph/0410406].

\bibitem{Webber} 
  B.~R.~Webber,
  ``QCD power corrections from a simple model for the running coupling,''
 JHEP {\bf 9810} (1998), 012
  [hep-ph/9805484].

\bibitem{Boucaud}
P.~Boucaud, F.~De Soto, A.~Le Yaouanc, J.~P.~Leroy, J.~Micheli, H.~Moutarde, O.~Pene and J.~Rodr\'{\i}guez-Quintero,
  ``The strong coupling constant at small momentum as an instanton detector,''
  JHEP {\bf 0304} (2003), 005
  [hep-ph/0212192];
  P.~Boucaud, F.~De Soto, A.~Le Yaouanc, J.~P.~Leroy, J.~Micheli, O.~Pene and J.~Rodr\'{\i}guez-Quintero,
  ``Modified instanton profile effects from lattice Green functions,''
  Phys.\ Rev.\ D {\bf 70} (2004), 114503
  [hep-ph/0312332].

\bibitem{Alekseev}
  A.~I.~Alekseev and B.~A.~Arbuzov,
  ``An invariant charge model for all $q^2 > 0$ in QCD and gluon condensate,''
 Mod. Phys. Lett. A \textbf{20} (2005), 103-116
[arXiv:hep-ph/0411339 [hep-ph]].
  A.~I.~Alekseev,
  ``Analytic invariant charge in QCD with suppression of nonperturbative contributions at large $Q^2$,''
Theor. Math. Phys. \textbf{145} (2005), 1559-1575
[Teor.\ Mat.\ Fiz.\  {\bf 145} (2005), 221]
  ``Synthetic running coupling of QCD,''
 Few Body Syst. \textbf{40} (2006), 57-70
   [hep-ph/0503242].

\bibitem{mes2}
  M.~Baldicchi, A.~V.~Nesterenko, G.~M.~Prosperi, D.~V.~Shirkov and C.~Simolo,
  ``Bound state approach to the QCD coupling at low energy scales,''
 Phys.\ Rev.\ Lett.\  {\bf 99} (2007), 242001
  [arXiv:0705.0329 [hep-ph]];
M.~Baldicchi, A.~V.~Nesterenko, G.~M.~Prosperi and C.~Simolo,
  ``QCD coupling below 1 GeV from quarkonium spectrum,''
   Phys.\ Rev.\ D {\bf 77} (2008), 034013
  [arXiv:0705.1695 [hep-ph]].

\bibitem{CV12}
  G.~Cveti\v{c} and C.~Valenzuela,
  ``An approach for evaluation of observables in analytic versions of QCD,''
 J.\ Phys.\ G {\bf 32} (2006), L27
  [hep-ph/0601050];
  ``Various versions of analytic QCD and skeleton-motivated evaluation of
 observables,''
Phys. Rev. D \textbf{74} (2006), 114030
[erratum: Phys. Rev. D \textbf{84} (2011), 019902]
[arXiv:hep-ph/0608256 [hep-ph]].


\bibitem{1danQCD}
  C.~Contreras, G.~Cveti\v{c}, O.~Espinosa and H.~E.~Mart\'{\i}nez,
  ``Simple analytic QCD model with perturbative QCD behavior at high momenta,''
Phys. Rev. D \textbf{82} (2010), 074005
[arXiv:1006.5050 [hep-ph]].
Phys.\ Rev.\ D {\bf 82} (2010), 074005
 
\bibitem{2danQCD} 
  C.~Ayala, C.~Contreras and G.~Cveti\v{c},
  ``Extended analytic QCD model with perturbative QCD behavior at high momenta,''
Phys. Rev. D \textbf{85} (2012), 114043
[arXiv:1203.6897 [hep-ph]].

  
\bibitem{anOPE} 
  G.~Cveti\v{c} and C.~Villavicencio,
  ``Operator Product Expansion with analytic QCD in tau decay physics,''
Phys.\ Rev.\ D {\bf 86} (2012), 116001
  [arXiv:1209.2953 [hep-ph]].

 \bibitem{anOPE2}
  C.~Ayala and G.~Cveti\v{c},
  ``Calculation of binding energies and masses of quarkonia in analytic QCD models,''
Phys. Rev. D \textbf{87} (2013) no.5, 054008
[arXiv:1210.6117 [hep-ph]];
C.~Ayala, G.~Cveti\v{c} and L.~Gonz\'alez,
``Evaluation of neutrinoless double beta decay: QCD running to sub-GeV scales,''
Phys. Rev. D \textbf{101} (2020) no.9, 094003
[arXiv:2001.04000 [hep-ph]].


\bibitem{Brod}
  S.~J.~Brodsky, G.~F.~de Teramond and A.~Deur,
  ``Nonperturbative QCD coupling and its $\beta$-function from Light-Front Holography,''
  Phys.\ Rev.\ D {\bf 81} (2010), 096010
  [arXiv:1002.3948 [hep-ph]];
  T.~Gutsche, V.~E.~Lyubovitskij, I.~Schmidt and A.~Vega,
  ``Dilaton in a soft-wall holographic approach to mesons and baryons,''
  Phys.\ Rev.\ D {\bf 85} (2012), 076003
  [arXiv:1108.0346 [hep-ph]].

\bibitem{Brod2} 
  A.~Deur, S.~J.~Brodsky and G.~F.~de Teramond,
  ``On the interface between perturbative and nonperturbative QCD,''
  Phys.\ Lett.\ B {\bf 757} (2016), 275
  [arXiv:1601.06568 [hep-ph]].


\bibitem{ArbZaits}
  B.~A.~Arbuzov and I.~V.~Zaitsev,
  ``Elimination of the Landau pole in QCD with the spontaneously generated anomalous three-gluon interaction,''
  arXiv:1303.0622 [hep-th].

\bibitem{Shirkovmass}
  D.~V.~Shirkov,
  ``'Massive' Perturbative QCD, regular in the IR limit,''
 Phys.\ Part.\ Nucl.\ Lett.\  {\bf 10} (2013), 186
  [arXiv:1208.2103 [hep-th]].
  
\bibitem{KKS}
  A.~V.~Kotikov, V.~G.~Krivokhizhin and B.~G.~Shaikhatdenov,
  ``Analytic and 'frozen' QCD coupling constants up to NNLO from DIS data,''
  Phys.\ Atom.\ Nucl.\  {\bf 75} (2012), 507
  [arXiv:1008.0545 [hep-ph]].

\bibitem{Luna1}
  E.~G.~S.~Luna, A.~L.~dos Santos and A.~A.~Natale,
  ``QCD effective charge and the structure function $F_{2}$ at small-$x$,''
  Phys.\ Lett.\ B {\bf 698} (2011), 52
  [arXiv:1012.4443 [hep-ph]];
  D.~A.~Fagundes, E.~G.~S.~Luna, M.~J.~Menon and A.~A.~Natale,
  ``Aspects of a Dynamical Gluon Mass Approach to elastic hadron scattering at LHC,''
  Nucl.\ Phys.\ A {\bf 886} (2012), 48
  [arXiv:1112.4680 [hep-ph]];
  C.~A.~S.~Bahia, M.~Broilo and E.~G.~S.~Luna,
  ``Energy-dependent dipole form factor in a QCD-inspired model,''
  J.\ Phys.\ Conf.\ Ser.\  {\bf 706} (2016), 052006
  [arXiv:1508.07359 [hep-ph]];
  ``Nonperturbative QCD effects in forward scattering at the LHC,''
  Phys.\ Rev.\ D {\bf 92} (2015), 074039
  [arXiv:1510.00727 [hep-ph]].

\bibitem{Luna2}
D.~Hadjimichef, E.~Luna and M.~Pel\'aez,
``QCD effective charges and the structure function $F_{2}$ at small-$x$: Higher twist effects,''
Phys. Lett. B \textbf{804} (2020), 135350
[arXiv:1907.07577 [hep-ph]].

\bibitem{Nest1}
  A.~V.~Nesterenko,
  ``Quark antiquark potential in the analytic approach to QCD,''
 Phys.\ Rev.\ D {\bf 62} (2000), 094028
  [hep-ph/9912351];
  ``New analytic running coupling in spacelike and timelike regions,''
 Phys.\ Rev.\ D {\bf 64} (2001), 116009
  [hep-ph/0102124];
  ``Analytic invariant charge in QCD,''
 Int.\ J.\ Mod.\ Phys.\ A {\bf 18} (2003), 5475
  [hep-ph/0308288];
  A.~C.~Aguilar, A.~V.~Nesterenko and J.~Papavassiliou,
  ``Infrared enhanced analytic coupling and chiral symmetry breaking in QCD,''
 J.\ Phys.\ G {\bf 31} (2005), 997
  [hep-ph/0504195].

\bibitem{Pelaez}
  M.~Pel\'aez, U.~Reinosa, J.~Serreau, M.~Tissier and N.~Wschebor,
  ``Small parameters in infrared quantum chromodynamics,''
  Phys.\ Rev.\ D {\bf 96} (2017), 114011
  [arXiv:1703.10288 [hep-th]];
  J.~A.~Gracey, M.~Pel\'aez, U.~Reinosa and M.~Tissier,
  ``Two loop calculation of Yang-Mills propagators in the Curci-Ferrari model,''
  Phys.\ Rev.\ D {\bf 100} (2019), 034023
  [arXiv:1905.07262 [hep-th]].

\bibitem{Siringo}
  F.~Siringo,
  ``Calculation of the nonperturbative strong coupling from first principles,''
  Phys.\ Rev.\ D {\bf 100} (2019), 074014
  [arXiv:1902.04110 [hep-ph]].

\bibitem{NestBook}
  A.~V.~Nesterenko,
  ``Strong interactions in spacelike and timelike domains: dispersive approach,'' Elsevier, Amsterdam, 2016, eBook ISBN: 9780128034484.

\bibitem{GCrev}
  G.~Cveti\v{c} and C.~Valenzuela,
  ``Analytic QCD: a short review,''
  Braz.\ J.\ Phys.\  {\bf 38} (2008), 371
  [arXiv:0804.0872 [hep-ph]].
 
\bibitem{Brodrev}
A.~Deur, S.~J.~Brodsky and G.~F.~de Teramond,
``The QCD running coupling,''
Nucl. Phys. \textbf{90} (2016), 1
doi:10.1016/j.ppnp.2016.04.003
[arXiv:1604.08082 [hep-ph]].
  
\bibitem{BK}
  A.~V.~Nesterenko and C.~Simolo,
  ``QCDMAPT: Program package for Analytic approach to QCD,''
Comput.\ Phys.\ Commun.\  {\bf 181} (2010), 1769
  [arXiv:1001.0901 [hep-ph]];
  ``${\rm QCDMAPT}_F$: Fortran version of QCDMAPT package,''
 Comput.\ Phys.\ Commun.\  {\bf 182} (2011), 2303
  [arXiv:1107.1045 [hep-ph]].
  A.~P.~Bakulev and V.~L.~Khandramai,
  ``FAPT: a Mathematica package for calculations in QCD Fractional Analytic Perturbation Theory,''
Comput.\ Phys.\ Commun.\  {\bf 184} (2013), 183
  [arXiv:1204.2679 [hep-ph]].


\bibitem{ACprogr}
  C.~Ayala and G.~Cveti\v{c},
  ``anQCD: a Mathematica package for calculations in general analytic QCD models,''
  Comput.\ Phys.\ Commun.\  {\bf 190} (2015), 182
  [arXiv:1408.6868 [hep-ph]];
  ``anQCD: Fortran programs for couplings at complex momenta in various analytic QCD models,''
  Comput.\ Phys.\ Commun.\  {\bf 199} (2016), 114
  [arXiv:1506.07201 [hep-ph]].
 
\bibitem{MSS}
  I.~L.~Solovtsov and D.~V.~Shirkov,
  ``Analytic approach to perturbative QCD and renormalization scheme dependence,''
 Phys.\ Lett.\ B {\bf 442} (1998), 344
  [hep-ph/9711251].
  K.~A.~Milton, I.~L.~Solovtsov and O.~P.~Solovtsova,
  ``Analytic perturbation theory and inclusive tau decay,''
   Phys.\ Lett.\ B {\bf 415} (1997), 104
  [hep-ph/9706409];
  ``The Adler function for light quarks in analytic perturbation theory,''
 Phys.\ Rev.\ D {\bf 64} (2001), 016005
  [hep-ph/0102254].

\bibitem{MagrGl} 
  B.~A.~Magradze,
  ``The gluon propagator in analytic perturbation theory,''
Conf.\ Proc.\ C {\bf 980518} (1999), 158
  [hep-ph/9808247].
  
\bibitem{DeRafael}
  S.~Peris, M.~Perrottet and E.~de Rafael,
  ``Matching long and short distances in large-$N_c$ QCD,''
 JHEP {\bf 9805} (1998), 011
  [hep-ph/9805442].


\bibitem{MagrTau} 
  B.~A.~Magradze,
  ``Testing the concept of quark-hadron duality with the ALEPH $\tau$ decay data,''
 Few Body Syst.\  {\bf 48} (2010), 143
  Erratum: [Few Body Syst.\  {\bf 53} (2012), 365]
  [arXiv:1005.2674 [hep-ph]];
  ``Strong coupling constant from $\tau$ decay within a dispersive approach to perturbative QCD,''
  {\it Proceedings of A. Razmadze Mathematical Institute\/} 160 (2012) 91-111
  [arXiv:1112.5958 [hep-ph]].

\bibitem{Nest3a}
A.~V.~Nesterenko and J.~Papavassiliou,
  J.\ Phys.\ G {\bf 32} (2006), 1025
  [hep-ph/0511215].
  
\bibitem{Nest3b}
 A.~V.~Nesterenko,
 Phys.\ Rev.\ D {\bf 88} (2013), 056009
  [arXiv:1306.4970 [hep-ph]];
  J.\ Phys.\ G {\bf 42} (2015), 085004
  [arXiv:1411.2554 [hep-ph]].

 \bibitem{renmod}
  G.~Cveti\v{c},
  ``Renormalon-motivated evaluation of QCD observables,''
  Phys.\ Rev.\ D {\bf 99} (2019) no. 1, 014028
  [arXiv:1812.01580 [hep-ph]].

\bibitem{shortv}
  G.~Cveti\v{c} and R.~K\"ogerler,
  ``Infrared-suppressed QCD coupling and the hadronic contribution to muon g-2,''
 J. Phys. G \textbf{47} (2020) no.10, 10LT01
 [arXiv:2007.05584 [hep-ph]].

 
  \bibitem{Peris}
  S.~Peris,
  ``Large-$N_c$ QCD and Pad\'e approximant theory,''
  Phys.\ Rev.\ D {\bf 74} (2006), 054013
  [hep-ph/0603190].

 \bibitem{LattcoupNf02a}
  I.~L.~Bogolubsky, E.-M.~Ilgenfritz, M.~M\"uller-Preussker and A.~Sternbeck, 
``Lattice gluodynamics computation of Landau gauge Green's functions in the deep infrared,''
  Phys.\ Lett.\ B {\bf 676} (2009), 69
  [arXiv:0901.0736 [hep-lat]]; 

\bibitem{LattcoupNf02b}
  E.-M.~Ilgenfritz, M.~M\"uller-Preussker, A.~Sternbeck and A.~Schiller,  
``Gauge-variant propagators and the running coupling from lattice QCD,''
  hep-lat/0601027.

\bibitem{Lattcoupb}
  A.~G.~Duarte, O.~Oliveira and P.~J.~Silva,
  `Lattice gluon and ghost propagators, and the strong coupling in pure SU(3) Yang-Mills theory: finite lattice spacing and volume effects,''
  Phys.\ Rev.\ D {\bf 94} (2016) no. 1, 014502
  [arXiv:1605.00594 [hep-lat]].

 \bibitem{Lattcoupc} 
  B.~Blossier {\it et al.},
  ``The Strong running coupling at $\tau$ and $Z_0$ mass scales from lattice QCD,''
  Phys.\ Rev.\ Lett.\  {\bf 108} (2012), 262002
  [arXiv:1201.5770 [hep-ph]];
  ``Ghost-gluon coupling, power corrections and $\Lambda_{\bar{\rm MS}}$ from lattice QCD with a dynamical charm,''
  Phys.\ Rev.\ D {\bf 85} (2012), 034503
  [arXiv:1110.5829 [hep-lat]].

\bibitem{Taylor} 
  J.~C.~Taylor,
  ``Ward Identities and Charge Renormalization of the Yang-Mills Field,''
 Nucl.\ Phys.\ B {\bf 33} (1971), 436.

 \bibitem{Latt3gluon}
  A.~Athenodorou, P.~Boucaud, F.~De Soto, J.~Rodr\'{\i}guez-Quintero and S.~Zafeiropoulos,
  ``Gluon Green functions free of quantum fluctuations,''
  Phys.\ Lett.\ B {\bf 760} (2016), 354
  [arXiv:1604.08887 [hep-ph]];
  A.~Athenodorou, D.~Binosi, P.~Boucaud, F.~De Soto, J.~Papavassiliou, J.~Rodr\'{\i}guez-Quintero and S.~Zafeiropoulos,
  ``On the zero crossing of the three-gluon vertex,''
  Phys.\ Lett.\ B {\bf 761} (2016), 444
  [arXiv:1607.01278 [hep-ph]];
  P.~Boucaud, F.~De Soto, J.~Rodr\'{\i}guez-Quintero and S.~Zafeiropoulos,
  ``Refining the detection of the zero crossing for the symmetric and asymmetric three-gluon vertices,''
Phys. Rev. D \textbf{95} (2017) no.11, 114503
[arXiv:1701.07390 [hep-lat]];
A.~Athenodorou, P.~Boucaud, F.~De Soto, J.~Rodr\'{\i}guez-Quintero and S.~Zafeiropoulos,
``Instanton liquid properties from lattice QCD,''
JHEP \textbf{02} (2018), 140
[arXiv:1801.10155 [hep-lat]].
  
\bibitem{MiniMOM}
  L.~von Smekal, K.~Maltman and A.~Sternbeck,
  ``The Strong coupling and its running to four loops in a minimal MOM scheme,''
  Phys.\ Lett.\ B {\bf 681} (2009), 336
  [arXiv:0903.1696 [hep-ph]].

\bibitem{BoucaudMM}
  P.~Boucaud, F.~De Soto, J.~P.~Leroy, A.~Le Yaouanc, J.~Micheli, O.~Pene and J.~Rodr\'{\i}guez-Quintero,
  ``Ghost-gluon running coupling, power corrections and the determination of Lambda(MS-bar),''
  Phys.\ Rev.\ D {\bf 79} (2009), 014508
  [arXiv:0811.2059 [hep-ph]];
S.~Zafeiropoulos, P.~Boucaud, F.~De Soto, J.~Rodr\'{\i}guez-Quintero and J.~Segovia,
``Strong running coupling from the gauge sector of domain wall Lattice QCD with physical quark masses,''
Phys. Rev. Lett. \textbf{122} (2019) no.16, 162002
[arXiv:1902.08148 [hep-ph]];


\bibitem{CheRet}
  K.~G.~Chetyrkin and A.~Retey,
  ``Three loop three linear vertices and four loop similar to MOM beta functions in massless QCD,''
  hep-ph/0007088.

  \bibitem{AKGCR}
A.~V.~Garkusha, A.~L.~Kataev and V.~S.~Molokoedov,
``Renormalization scheme and gauge (in)dependence of the generalized Crewther relation: what are the real grounds of the $\beta$-factorization property?,''
JHEP \textbf{02} (2018), 161
[arXiv:1801.06231 [hep-ph]].

\begingroup \color{black}
\bibitem{DSEdecoup}
A.~C.~Aguilar and J.~Papavassiliou,
``Gluon mass generation in the PT-BFM scheme,''
JHEP \textbf{12} (2006), 012
doi:10.1088/1126-6708/2006/12/012
[arXiv:hep-ph/0610040 [hep-ph]];
A.~C.~Aguilar, D.~Binosi and J.~Papavassiliou,
``Gluon and ghost propagators in the Landau gauge: Deriving lattice results from Schwinger-Dyson equations,''
Phys. Rev. D \textbf{78} (2008), 025010
doi:10.1103/PhysRevD.78.025010
[arXiv:0802.1870 [hep-ph]];
P.~Boucaud, J.~P.~Leroy, A.~Le Yaouanc, J.~Micheli, O.~Pene and J.~Rodr\'{\i}guez-Quintero,
``On the IR behaviour of the Landau-gauge ghost propagator,''
JHEP \textbf{06} (2008), 099
doi:10.1088/1126-6708/2008/06/099
[arXiv:0803.2161 [hep-ph]];
D.~Binosi and J.~Papavassiliou,
``Pinch Technique: theory and applications,''
Phys. Rept. \textbf{479} (2009), 1-152
doi:10.1016/j.physrep.2009.05.001
[arXiv:0909.2536 [hep-ph]].
\endgroup
  
\bibitem{PDG2019}
  M.~Tanabashi {\it et al.} [Particle Data Group],
  Phys.\ Rev.\ D {\bf 98} (2018) no. 3, 030001 and 2019 update (http://pdg.lbl.gov/index.html).

\bibitem{GCIK}
  G.~Cveti\v{c} and I.~Kondrashuk,
  ``Explicit solutions for effective four- and five-loop QCD running coupling,''
  JHEP {\bf 1112} (2011), 019
  [arXiv:1110.2545 [hep-ph]].

 \bibitem{5lMSbarbeta}
  P.~A.~Baikov, K.~G.~Chetyrkin and J.~H.~K\"uhn,
  ``Five-loop running of the QCD coupling constant,''
 Phys.\ Rev.\ Lett.\  {\bf 118} (2017) no. 8, 082002
  [arXiv:1606.08659 [hep-ph]].
  

\bibitem{4lquarkthresh}
  Y.~Schr\"oder and M.~Steinhauser,
  ``Four-loop decoupling relations for the strong coupling,''
  JHEP {\bf 0601} (2006), 051
  doi:10.1088/1126-6708/2006/01/051
  [hep-ph/0512058];
  B.~A.~Kniehl, A.~V.~Kotikov, A.~I.~Onishchenko and O.~L.~Veretin,
  ``Strong-coupling constant with flavor thresholds at five loops in the anti-MS scheme,''
  Phys.\ Rev.\ Lett.\  {\bf 97} (2006), 042001
  [hep-ph/0607202].
  
\bibitem{4lMSbarbeta}
  T.~van Ritbergen, J.~A.~M.~Vermaseren and S.~A.~Larin,
  ``The Four loop beta function in quantum chromodynamics,''
  Phys.\ Lett.\ B {\bf 400} (1997), 379
  doi:10.1016/S0370-2693(97)00370-5
  [hep-ph/9701390].
  
\bibitem{3lquarkthresh}
  K.~G.~Chetyrkin, B.~A.~Kniehl and M.~Steinhauser,
  ``Strong coupling constant with flavour thresholds at four loops in the
  MSbar scheme,''
  Phys.\ Rev.\ Lett.\  {\bf 79} (1997), 2184
  [arXiv:hep-ph/9706430].

\bibitem{d1}
  K.~G.~Chetyrkin, A.~L.~Kataev and F.~V.~Tkachov,
``Higher order corrections to $\sigma_T$ ($e^+ e^- \to$ Hadrons) 
in Quantum Chromodynamics,''
 Phys.\ Lett.\ B {\bf 85} (1979), 277;
  M.~Dine and J.~R.~Sapirstein,
  ``Higher order QCD corrections in $e^+e^-$ annihilation,''
  Phys.\ Rev.\ Lett.\  {\bf 43} (1979), 668;
  W.~Celmaster and R.~J.~Gonsalves,
``An analytic calculation of higher order Quantum Chromodynamic 
corrections in  $e^+e^-$ annihilation,''
  Phys.\ Rev.\ Lett.\  {\bf 44} (1980), 560.
 
\bibitem{d2}
  S.~G.~Gorishnii, A.~L.~Kataev and S.~A.~Larin,
``The ${\cal O}(\alpha_s^3)$ corrections to $\sigma_{tot} (e^+ e^- \to$ hadrons) and 
$\Gamma(\tau^- \to \nu_{\tau} + {\rm hadrons})$ in QCD,''
 Phys.\ Lett.\ B {\bf 259} (1991), 144;
  L.~R.~Surguladze and M.~A.~Samuel,
``Total hadronic cross-section in $e^+e^-$ annihilation 
at the four loop level of perturbative QCD,''
 Phys.\ Rev.\ Lett.\  {\bf 66} (1991), 560
  Erratum: [Phys.\ Rev.\ Lett.\  {\bf 66} (1991), 2416].


\bibitem{d3}
  P.~A.~Baikov, K.~G.~Chetyrkin and J.~H.~K\"uhn,
  ``Order $\alpha^4_s$ QCD corrections to $Z$ and $\tau$ Decays,''
 Phys.\ Rev.\ Lett.\  {\bf 101} (2008), 012002
  [arXiv:0801.1821 [hep-ph]].


\bibitem{BenekeRev}
M.~Beneke,
``Renormalons,''
Phys. Rept. \textbf{317} (1999), 1-142
[arXiv:hep-ph/9807443 [hep-ph]].

\bibitem{Rens1}
A.~Maiezza and J.~C.~Vasquez,
``Non-local Lagrangians from Renormalons and Analyzable Functions,''
Annals Phys. \textbf{407} (2019), 78-91
[arXiv:1902.05847 [hep-th]];
J.~Bersini, A.~Maiezza and J.~C.~Vasquez,
``Resurgence of the renormalization group equation,''
Annals Phys. \textbf{415} (2020), 168126
[arXiv:1910.14507 [hep-th]].

\bibitem{Rens2}
C.~Ayala, X.~Lobregat and A.~Pineda,
``Superasymptotic and hyperasymptotic approximation to the operator product expansion,''
Phys. Rev. D \textbf{99} (2019) no.7, 074019
[arXiv:1902.07736 [hep-th]];
``Hyperasymptotic approximation to the top, bottom and charm pole mass,''
Phys. Rev. D \textbf{101} (2020) no.3, 034002
[arXiv:1909.01370 [hep-ph]];
``Determination of $\alpha(M_z)$ from an hyperasymptotic approximation to the energy of a static quark-antiquark pair,''
JHEP \textbf{09} (2020), 016
[arXiv:2005.12301 [hep-ph]].

\bibitem{Rens3}
D.~Boito, P.~Masjuan and F.~Oliani,
``Higher-order QCD corrections to hadronic $\tau$ decays from Pad\'e approximants,''
JHEP \textbf{08} (2018), 075
[arXiv:1807.01567 [hep-ph]];
D.~Boito and F.~Oliani,
``Renormalons in integrated spectral function moments and $\alpha_s$ extractions,''
Phys. Rev. D \textbf{101} (2020) no.7, 074003
[arXiv:2002.12419 [hep-ph]].

\bibitem{Rens4}
  H.~Takaura,
``Formulation for renormalon-free perturbative predictions beyond large-$\beta_0$ approximation,''
[arXiv:2002.00428 [hep-ph]].


\bibitem{Sternb}
 A.~Sternbeck, private communication.
 
\bibitem{OPAL}
  K.~Ackerstaff {\it et al.} [OPAL Collaboration],
  ``Measurement of the strong coupling constant $\alpha_s$ and the vector and axial vector spectral functions in hadronic tau decays,''
  Eur.\ Phys.\ J.\ C {\bf 7} (1999), 571
  [hep-ex/9808019].
 
\bibitem{PerisPC1}
  D.~Boito, M.~Golterman, M.~Jamin, A.~Mahdavi, K.~Maltman, J.~Osborne and S.~Peris,
  ``An updated determination of $\alpha_s$ from $\tau$ decays,''
  Phys.\ Rev.\ D {\bf 85} (2012), 093015
  [arXiv:1203.3146 [hep-ph]].

\bibitem{PerisPC2}
  We are grateful to S.~Peris for providing us with the measured spectral functions and covariance matrices of OPAL Collaboration; these data are the update, made by the authors of Ref.~\cite{PerisPC1}, of the OPAL data, based on the older OPAL data given to them by S.~Menke.
  
 \bibitem{ALEPH2}
  S.~Schael {\it et al.}  [ALEPH Collaboration],
``Branching ratios and spectral functions of tau decays: final ALEPH measurements and physics implications,''
Phys.\ Rept.\  {\bf 421} (2005), 191
  [hep-ex/0506072];
  M.~Davier, A.~H\"ocker and Z.~Zhang,
  ``The Physics of hadronic tau decays,''
 Rev.\ Mod.\ Phys.\  {\bf 78} (2006), 1043
  [hep-ph/0507078].
  
\bibitem{DDHMZ}
  M.~Davier, S.~Descotes-Genon, A.~H\"ocker, B.~Malaescu and Z.~Zhang,
  ``The Determination of $\alpha_s$ from $\tau$ decays revisited,''
 Eur.\ Phys.\ J.\ C {\bf 56} (2008), 305
  [arXiv:0803.0979 [hep-ph]].
  
   \bibitem{ALEPHfin}
  M.~Davier, A.~H\"ocker, B.~Malaescu, C.~Z.~Yuan and Z.~Zhang,
  ``Update of the ALEPH non-strange spectral functions from hadronic $\tau$ decays,''
  Eur.\ Phys.\ J.\ C {\bf 74} (2014) no. 3, 2803
  [arXiv:1312.1501 [hep-ex]].

\bibitem{ALEPHwww}
  The measured data of ALEPH Collaboration, with covariance matrix corrections described in Ref.~\cite{ALEPHfin}, are available on the following web page:
http://aleph.web.lal.in2p3.fr/tau/specfun13.html

\bibitem{Davier:2019can}
M.~Davier, A.~Hoecker, B.~Malaescu and Z.~Zhang,
``A new evaluation of the hadronic vacuum polarisation contributions to the muon anomalous magnetic moment and to $\mathbf{\boldsymbol\alpha(m_Z^2)}$,''
Eur. Phys. J. C \textbf{80} (2020) no.3, 241
[arXiv:1908.00921 [hep-ph]].

\bibitem{AKR}
S.~Eidelman, F.~Jegerlehner, A.~L.~Kataev and O.~Veretin,
``Testing nonperturbative strong interaction effects via the Adler function,''
Phys. Lett. B \textbf{454} (1999), 369-380
[arXiv:hep-ph/9812521 [hep-ph]].

\bibitem{ANR}
A.~V.~Nesterenko,
``Explicit form of the R-ratio of electron–positron annihilation into hadrons,''
J. Phys. G \textbf{46} (2019) no.11, 115006
[arXiv:1902.06504 [hep-ph]];
``Recurrent form of the renormalization group relations for the higher-order hadronic vacuum polarization function perturbative expansion coefficients,''
[arXiv:2004.00609 [hep-ph]].

\bibitem{amurev}
T.~Aoyama \textit{et al.},
``The anomalous magnetic moment of the muon in the Standard Model,''
[arXiv:2006.04822 [hep-ph]].

\bibitem{Brook}
G.~Bennett \textit{et al.} [Muon g-2],
Phys. Rev. D \textbf{73} (2006), 072003
[arXiv:hep-ex/0602035 [hep-ex]].

\bibitem{Borsanyi:2020mff}
S.~Borsanyi \textit{et al.},
``Leading-order hadronic vacuum polarization contribution to the muon magnetic momentfrom lattice QCD,''
arXiv:2002.12347 [hep-lat].

\bibitem{Lehner:2020crt}
C.~Lehner and A.~S.~Meyer,
``Consistency of hadronic vacuum polarization between lattice QCD and the R-ratio,''
Phys. Rev. D \textbf{101} (2020), 074515
[arXiv:2003.04177 [hep-lat]].


\bibitem{amuHT}
H.~Terazawa,
``All the hadronic contributions to the anomalous magnetic moment of the muon and the Lamb shift in the hydrogen atom,''
Prog. Theor. Phys. \textbf{39} (1968), 1326-1332;
``Spectral function of the photon propagator-mass spectrum and timelike form-factors of particles,''
Phys. Rev. \textbf{177} (1969), 2159-2166.


\bibitem{amuBR}
J.~S.~Bell and E.~de Rafael,
``Hadronic vacuum polarization and g(mu)-2,''
Nucl. Phys. B \textbf{11} (1969), 611-620.

\bibitem{LPQ}
M.~Lindner, M.~Platscher and F.~S.~Queiroz,
``A call for new physics: the muon anomalous magnetic moment and Lepton Flavor Violation,''
Phys. Rept. \textbf{731} (2018), 1-82
[arXiv:1610.06587 [hep-ph]].
  
\bibitem{SF}
F.~Correia and S.~Fajfer,
``Restrained dark $U(1)_d$ at low energies,''
Phys. Rev. D \textbf{94} (2016) no.11, 115023
[arXiv:1609.00860 [hep-ph]];
``Light mediators in anomaly free U (1)$_{X}$ models. Part I. Theoretical framework,''
JHEP \textbf{10} (2019), 278
[arXiv:1905.03867 [hep-ph]];
JHEP \textbf{10} (2019), 279
[arXiv:1905.03872 [hep-ph]];
I.~Dor\v{s}ner, S.~Fajfer and S.~Saad,
``$\mu \to e \gamma$ selecting scalar leptoquark solutions for the $(g-2)_{e,\mu}$ puzzles,''
[arXiv:2006.11624 [hep-ph]].

\bibitem{CSK}
G.~Cveti\v{c}, C.~Kim, D.~Lee and D.~Sahoo,
``Probing new physics scenarios of muon $g-2$ via $J/\psi$ decay at BESIII,''
[arXiv:2004.03124 [hep-ph]], to appear in JHEP.  

\begingroup \color{black}
\bibitem{Pass}
M.~Passera, W.~J.~Marciano and A.~Sirlin,
``The muon $g-2$ and the bounds on the Higgs boson mass,''
Phys. Rev. D \textbf{78} (2008), 013009
doi:10.1103/PhysRevD.78.013009
[arXiv:0804.1142 [hep-ph]];
A.~Keshavarzi, W.~J.~Marciano, M.~Passera and A.~Sirlin,
``Muon $g-2$ and $\Delta \alpha$ connection,''
Phys. Rev. D \textbf{102} (2020) no.3, 033002
doi:10.1103/PhysRevD.102.033002
[arXiv:2006.12666 [hep-ph]].
\endgroup

\bibitem{CHMM}
A.~Crivellin, M.~Hoferichter, C.~A.~Manzari and M.~Montull,
``Hadronic vacuum polarization: $(g-2)_\mu$ versus global electroweak fits,''
Phys. Rev. Lett. \textbf{125} (2020) no.9, 091801
[arXiv:2003.04886 [hep-ph]].

  
\bibitem{GPRS}
  M.~Gonz\'alez-Alonso, A.~Pich and A.~Rodr\'{\i}guez-S\'anchez,
  ``Updated determination of chiral couplings and vacuum condensates from hadronic $\tau$ decay data,''
  Phys.\ Rev.\ D {\bf 94} (2016), 014017
  [arXiv:1602.06112 [hep-ph]].


\bibitem{BoiOP}
D.~Boito, M.~Golterman, M.~Jamin, K.~Maltman and S.~Peris,
``Low-energy constants and condensates from the $\tau$ hadronic spectral functions,''
Phys. Rev. D \textbf{87} (2013) no.9, 094008
[arXiv:1212.4471 [hep-ph]].

\bibitem{BoiAL}
D.~Boito, A.~Francis, M.~Golterman, R.~Hudspith, R.~Lewis, K.~Maltman and S.~Peris,
``Low-energy constants and condensates from ALEPH hadronic $\tau$ decay data,''
Phys. Rev. D \textbf{92} (2015) no.11, 114501
[arXiv:1503.03450 [hep-ph]]. 

  
 \bibitem{KTG}
  O.~Teryaev,
  ``Analyticity and higher twists,''
  Nucl.\ Phys.\ Proc.\ Suppl.\  {\bf 245} (2013), 195
  [arXiv:1309.1985 [hep-ph]];
  V.~L.~Khandramai, O.~V.~Teryaev and I.~R.~Gabdrakhmanov,
  ``Infrared modified QCD couplings and Bjorken sum rule,''
  J.\ Phys.\ Conf.\ Ser.\  {\bf 678} (2016), 012018;
  I.~R.~Gabdrakhmanov, O.~V.~Teryaev and V.~L.~Khandramai,
  ``Infrared models for the Bjorken sum rule in the APT approach,''
  J.\ Phys.\ Conf.\ Ser.\  {\bf 938} (2017) no. 1, 012046.

  
\bibitem{LBAdl1}
  D.~J.~Broadhurst,
  ``Large N expansion of QED: asymptotic photon propagator and contributions to the muon anomaly, for any number of loops,''
  Z.\ Phys.\ C {\bf 58} (1993), 339.

\bibitem{LBAdl2}
  D.~J.~Broadhurst and A.~L.~Kataev,
  ``Connections between deep inelastic and annihilation processes at next to next-to-leading order and beyond,''
  Phys.\ Lett.\ B {\bf 315} (1993), 179
  [hep-ph/9308274].


\bibitem{Neubert}
  M.~Neubert,
  ``Scale setting in QCD and the momentum flow in Feynman diagrams,''
  Phys.\ Rev.\ D {\bf 51} (1995), 5924
  doi:10.1103/PhysRevD.51.5924
  [hep-ph/9412265].


  
\end{thebibliography}
\end{document}